\documentclass[a4paper,12pt]{article}

\usepackage{amsmath}
\numberwithin{equation}{section}
\usepackage{amssymb}
\usepackage{textcomp}
\usepackage{hyperref}
\usepackage{cite}
\usepackage{multirow}
\usepackage{accents} 
\usepackage{slashed}

\title{Harmonic Superspace for  Ali-Ilahi's ADHM Instanton Sigma Model}
\author{Abbas Ali\footnote{Email:aali.ph@amu.ac.in}, Mohsin Ilahi, P.P. Abdul Salih \\and Shafeeq Rahman Thottoli\footnote{Present Address : Department of Physical Sciences, Physics Division, College of Science,
Jazan University, Jazan 45142, Kingdom of Saudi Arabia}\\
		Physics Department, Aligarh Muslim University,\\ Aligarh, India}
\date{}
\begin{document}

\maketitle

\begin{abstract}
ADHM Yang-Mills instantons are extended field theoretical objects. These are more general than the more familiar 't Hooft Yang-Mills instantons. Their counter parts exist in string theory in terms of sigma models. Nearly three decades ago Witten constructed a (0,4) supersymmetric linear sigma model incorporating ADHM instantons.  Witten's construction was in component form. Galperin and Sokatchev constructed the corresponding off-shell supersymmetric version using the harmonic superspace. Recently Ali and Ilahi constructed an ADHM instanton linear sigma model that is complementary, in the sense of being dual, to the original model constructed by Witten. Full (0,4) supersymmetric off-shell harmonic superspace formalism for Ali-Ilahi's complementary ADHM instanton sigma model is developed in this note.
\end{abstract}

\newpage

\tableofcontents

\newpage 

\section{Introduction}

To study strings in background fields we use two dimensional (supersymmetric) sigma model formalism that incorporate gravitational $G_{\mu\nu}(X)$, anti-symmetric tensor field $B_{\mu\nu}(X)$, dilaton $\Phi(X)$ and Yang-Mills $A^a_\mu(X)$ background fields. One example of Yang-Mills backgrounds is given by instantons. Instantons, like solitons, are
extended classical self-dual Yang-Mills field configurations to field theoretical equations that complement the perturbation theory in an essential way.  Simplest Yang-Mills instanton solutions are the 't Hooft instantons but these contain a restricted set of $5k+3$ parameters as compared to the full possible set of $8k$ for $SU(2)$ gauge  group. Supersymmetric sigma models, some incorporating 't Hooft Yang-Mills instantons,  were studies in the Refs.\cite{Fradkin:1985ys, Callan:1985ia,  Alvarez-Gaume:1981exv, Strominger:1986uh, Strominger:1990et, Callan:1991dj, Callan:1991ky, Callan:1991at}.

Apart from 't Hooft self-dual Yang-Mills instanton there is the ADHM construction of Yang-Mills instantons from 1978 \cite{Atiyah:1978ri}. This incorporates the full set of parameters and is generalizable to general Lie groups \cite{Witten:1978qe, Christ:1978jy, Corrigan:1978ce, Weinberg:2012pjx, Tong:2005un}. 

To incorporate the ADHM instanton as a background for heterotic string theory Witten constructed a (0,4) supersymmetric linear, that is massive, instanton sigma model in Ref.\cite{Witten:1994tz} in 1995. ADHM construction is a complex formalism and it is closely related to another sophisticated construction - the twistor space. Witten's construction, on the other hand, was a lucid field theoretic construction. The ADHM form of the self-dual Yang-Mills instanton comes out naturally in Witten's construction. 

Witten also investigated corresponding moduli space of the ADHM instanton sigma model and the structure of the moduli space was found to be mysterious. In this model an additional branch of the moduli space appeared in the small instanton limit. There is a $Z_2$ symmetry between the original moduli space and the new branch. In the finite instanton case the two branches are at infinite distance from each other but these meet each other in the small instanton case. Witten revisited the issue of the moduli space mystery in Ref.\cite{Witten:1995zh} but the issue remained essentially unsolved.  

In Witten's construction there were two $SU(2)$ symmetries, $F=SU(2)$ and $F'=SU(2)'$ with a $Z_2$ symmetry between them. He broke this symmetry and used the $F'$ symmetry for his construction. He also suggested that it would be interesting to do a complementary construction using the $F$ symmetry. In Ref.\cite{Ali:2023csc}  an ADHM instanton sigma model was constructed by Ali and Ilahi  that is complementary to the one constructed by Witten in the sense of being dual to it. This complementary construction uses the $F$ symmetry, as suggested by Witten. The moduli space that results in this construction is such that it clarifies the  nearly three decades old mystery of the moduli space of Witten's original construction. There is a  duality between the moduli spaces of Witten's original and Ali-Ilahi's complementary or dual ADHM instanton sigma models. In the small instanton limit the two models coalesce and the duality becomes a $Z_2$ symmetry. That is why the additional branch of the moduli space appears in the small instanton limit. This happens in both cases - the Witten's Original Model and Ali-Ilahi's Complementary Model.

There have been following  further developments along the line of these investigations. Witten in his original construction had suggested one more construction that uses both $F$ and $F'$ $SU(2)$ symmetries for a construction. Corresponding construction was done in Ref.\cite{Ali:2023icn} by Ali and Salih and called the complete ADHM instanton sigma model. Quantization of complementary ADHM instanton sigma model was done in Ref.\cite{Ali:2023zxt}) by Ali, Ilahi and Thottoli along the lines of Refs.\cite{Howe:1987qv, Howe:1992tg, Howe:1988cj, Lambert:1995dp}.

The insights obtained from these investigations have helped in clarification of several other long standing problems. In Ref.\cite{Giveon:1998ns} Giveon, Kutasov and Seiberg investigated $AdS_3$ superstrings and come out with a number of insights. Like Witten's paper on ADHM instanton sigma model there was a mystery in their paper too. This was the mystery of the doubling of Ramond superalgebra. Like the appearance of an additional branch of moduli space in Witten's original construction the Giveon-Kutasov-Seiberg Ramond superalgebra doubling too remained unsolved for quarter of a century. This was solved in Ref.\cite{Ali:2023xov} by Ali and this was a result of insights gained in the construction of the complementary and complete ADHM instanton sigma models. The mystery was related to the appearance of a $Z_2$ symmetry and doubling of the instanton moduli space in the small instanton limit. The resolution of this in Ref.\cite{Ali:2023xov} consisted of the fact that in the small instanton limit the moduli spaces of the original and complementary models coalesce and hence the appearance of  $Z_2$  symmetry because of the doubling of the moduli space. In general case there are two independent branches of the moduli space in case of the ADHM instanton sigma models. In case of Ramond superalgebra doubling there are two independent $AdS_3$ superstrings that are independent of each other and related to each other by a duality.

The presence of two branches of the  moduli space and the $Z_2$ symmetry between them in the small instanton limit also helped in resolving another mystery in $AdS_3$ superstrings in Ref.\cite{Ali:2023kkf}. In Ref.\cite{Gukov:2004ym} the authors surveyed all proposals for superconformal field theory dual to superstrings moving on $AdS_3\times S^3\times S^3\times S^1$ backgrounds and rejected all of them. They investigated the connection of the problem with $N=4$ superconformal algebras \cite{Ademollo:1975an, Ademollo:1976pp, Ademollo:1976wv, Sevrin:1988ew, Ali:2000zu, Ali:2003aa, Ali:2000we, Ali:1993sd, Ivanov:1987mz, Ivanov:1988rt}. To investigate the issue they needed a free field realization of the large $N=4$ superconformal algebra that is more general then the most commonly used free field realization by Sevrin, Troost, van Proeyen. They tried to find the needed free field realization  by taking a tensor product of two Sevrin, Troost, van Proeyen free field realizations but gave up because of a technical problem. The tensor product of the two Sevrin-Troost-van-Proeyen realizations turned out not to be associative. This hurdle was overcome in Ref.\cite{Ali:2023kkf} using a mapping to duality between the two branches of the moduli space of the ADHM instanton sigma models. The non-associativity of two free field realizations goes away for particular values of the parameters.

The supersymmetry structure of  ADHM instanton linear sigma models, $N=4$ superconformal algebras and superconformal field theories dual to superstrings moving in $AdS_3$ backgrounds in view of above findings were clarified in Ref.\cite{Ali:2024amc}.  First of all the information about the (0, 4) supersymmetries of Witten's original, Ali-Ilahi's complementary and Ali-Salih's complete ADHM instanton linear sigma models were collected together. Then the information about large, small and middle $N=4$ superconformal algebras was collected and   their interrelations  with the ADHM instanton linear sigma models were summarized.  It was  then argued that the former, that is the supersymmetry of the ADHM instanton linear sigma models, flows to latter, that is the small, middle and large $N=4$ superconformal symmetries, in the infrared limit. Finally the mapping of these $N=4$ superconformal symmetries onto the superconformal symmetries dual to superstrings moving on $AdS_3\times S^3\times {\mathcal M}^4$ backgrounds with ${\mathcal M}^=K3, T^4$ and $S^3\times S^1$ was summarized.

In short the explorations that started with the investigations of the complementary and complete ADHM instanton sigma models have resulted in a very rich tapestry of connections. 

Our objective in this note is to develop harmonic superspace covariant formalism \cite{Galperin:1984av} for Ali-Ilahi's ADHM instanton sigma model \cite{Ali:2023csc} along the lines that was done for Witten's original ADHM instanton sigma model \cite{Witten:1994tz}  by Galperin and Sokatchev in Ref.\cite{Galperin:1994qn} (in this note the phrase {\it the authors} refers to this paper). 

In Witten's  construction out of the full (0,4) supersymmetry only the (0,1) part  was off-shell. Clearly it was desirable to have off-shell formalism for the full (0, 4) supersymmetry. This difficult task was taken up by Galperin and Sokatchev (GS) in Ref.\cite{Galperin:1994qn}. They used the harmonic superspace formalism  to give the full off-shell supersymmetric description. The harmonic superspace formalism for 't Hooft instantons was done in Ref.\cite{Galperin:1995pq} by the same authors.

In \cite{Ali:2023csc}  the original construction was closely followed and hence the resulting construction inherits not only the strengths of the original construction but also its weaknesses. Thus in the resulting construction, just like the original construction, only (0,1) part of the full (0,4) supersymmetry was off-shell. Thus it is desirable to have the corresponding off-shell formalism of the full (0,4) supersymmetry. That is the burden of the present note.

This can be done in a few  different ways. One way is to use the duality between the original and the complementary constructions. Another possibility is to do the harmonic superspace construction in an \textit{ab initio} manner. The third possibility is to improvise on the route followed by  Galperin and Sokatchev. We shall choose the last option.

In either case the construction will be straightforward in principle but very involved in practice. The reason is two fold. The task of developing the harmonic superspace formalism for the complementary ADHM sigma model might bring us face to face with two very technical topics - ADHM instanton sigma models and the harmonic superspace. Secondly the duality between Witten's Original Model and Ali-Ilahi's Complementary Model is very simple in principle but very complex to implement in the construction.

In their narrative Galperin and Sokatchev uncovered a relation of the ADHM construction with twistor construction. We shall not pay attention to that connection. We shall also avoid related topic of integrability conditions for self-dual Yang-Mills theories. There are two other topics that Galperin and Sokatchev pay attention to but will be overlooked in our narrative. These are the questions of infinite number auxiliary fields and reduction of $N=2$ superspace from $D=4$ to $D=2.$

The plan of the rest of this note is as follows.

To formulate Ali-Ilahi's complementary ADHM instanton sigma model in harmonic superspace in an off-shell manner we need corresponding complementary or dual harmonic superspace. It should be dual to the harmonic superspace used by Galperin and Sokatchev. This complementary or dual harmonic superspace is constructed in Section \ref{dual}. Before doing the harmonic superspace construction for Witten's original ADHM instanton sigma model Galperin and Sokatchev gave their own perspective about it. To do the harmonic superspace construction for Ali-Ilahi's model we need similar perspective on it. In Section \ref{perspective} we look at Ali-Ilahi's construction from the perspective of Galperin and Sokatchev. In Section \ref{free0} we describe the super-multiplets that are needed for the off-shell superspace formalism of the Ali-Ilahi's ADHM instanton sigma model. In Section \ref{dual-int} we discuss the interaction  of the Ali-Ilahi's complementary ADHM instanton sigma model and the structure of the corresponding instanton gauge field. We conclude in Section \ref{conclusions}.

In this note we have covered a lot of review material in the appendices. This extensive coverage of the background material is necessary to follow the narrative of this note because it deals with complex issues and subsumes very extensive background. In appendices we have tried to collect representative background.

In Appendix \ref{a} we begin  by giving some background on Yang-Mills instantons. Aspects of harmonic superspace are reviewed in Appendix \ref{b}. In Appendix \ref{c} we review Witten's original ADHM instanton sigma model construction. In turn we do the same for Ali-Ilahi's complementary ADHM instanton sigma model construction in Appendix \ref{d}. The harmonic superspace off-shell formalism for Witten's ADHM instanton sigma model is summarized in Appendix \ref{e}.

To read the rest of this note one should start with the appendices because the narrative subsumes the corresponding background.

\section{Dual Harmonic Superspace}\label{dual}

Purpose of this note is to formulate Ali-Ilahi's complementary ADHM instanton linear sigma model in harmonic superspace in an off-shell manner. Corresponding harmonic superspace will be complementary or dual to the harmonic superspace employed by Galperin and Sokatchev to formulate Witten's original model in an off-shell manner. Galperin and Sokatchev's formalism is reviewed in Appendix \ref{b}.

Witten's original ADHM instanton sigma model and Ali-Ilahi's complementary model are dual to each other and are independent of each other. These exist in independent harmonic superspaces. Thus to formulate Ali-Ilahi's complementary ADHM instanton sigma model in an off-shell manner we need corresponding harmonic superspace that is independent of and dual to the harmonic superspace employed by Galperin and Sokatchev. We shall construct it in this section.

Our construction of the dual harmonic superspace will be constructed by applying duality transformations on the original harmonic superspace described in Appendix \ref{b}. Since we have included liberal amount of background material in that appendix our introduction in this section will be brief.

In Witten's construction the $SO(4)$ automorphism of the $(0, 4)$ supersymmetry is broken as $SO(4)\sim SU(2)\times SU(2)'=F\times F'.$ Clearly there is a $Z_2$ automorphism between the two $SU(2)$s. The resulting Lagrangian in Witten's construction is invariant only under $F'=SU(2)'$. The Lagrangian of the Ali-Ilahi model is invariant only under $F=SU(2)$.

The duality between Witten's original and Ali-Ilahi's complementary model is the duality $F\leftrightarrow F'$ or $SU(2)\leftrightarrow SU(2)'$. This entails, among other things, the duality between the indices of two $SU(2)$s : $A\leftrightarrow A'$. Since the original and complementary models are independent of each other this means that for the dual harmonic space we have to introduce independent harmonic variables $\hat{u}_{A'}$. We shall do further specification of the duality between Witten's Original Model and Ali-Ilahi's Complementary Model later on as the need arises.

With this choice the equation corresponding to Eqn.(\ref{zweib}) is
\begin{equation}\label{zweib0}
 \hat{u}^{+A'}\hat{u}^{-}_{A'}=1,~~~~~~   \begin{pmatrix}
\hat{u}^{-}_{1'} & \hat{u}^{+}_{1'} \\
\hat{u}^{-}_{2'} & \hat{u}^{+}_{2'} 
\end{pmatrix} \in SU(2)'.
\end{equation}
Similarly equation corresponding to Eqn.(\ref{uone}) becomes
\begin{equation}\label{uone0}
\hat\theta^{+}_{\alpha}=\hat\theta^{A'}_{\alpha}\hat{u}^{+}_{A'}, ~~~~\hat\theta^{-}_{\alpha}=\hat\theta^{A'}_{\alpha}\hat{u}^{-}_{A'}
\end{equation}
and Eqn.(\ref{conuone}) takes the following form
\begin{equation}\label{conuone0}
\hat\theta^{A'}_{\alpha}=\hat{u}^{+A'}\hat\theta^{-}_{\alpha}-\hat{u}^{-A'}\hat\theta^{+}_{\alpha}.
\end{equation}
Clearly the $SL(2, C)$ group here is different from Witten's case and should be called $SL(2, C)'$ and the indices $\alpha, \beta$ should be called $\alpha', \beta'$ but to avoid cluttering of notation we are not adopting this notation because we do not foresee much room for confusion.

The dual real (or the central) basis given by
\begin{equation}
   (x^\mu, \hat\theta_{\alpha A'}, \bar{\hat\theta}^{A'}_{\alpha'}).\label{basis1}
\end{equation}
Corresponding supersymmetry transformations are
\begin{eqnarray}\label{n=2ss0}
    \delta x^\mu&=&i(\epsilon^{A'}\sigma^{\mu}\bar{\hat\theta}_{A'}-\hat\theta^{A'}\sigma^{\mu}\bar{\epsilon}_{A'}),\nonumber\\~~~\delta\theta_{\alpha A'}&=&\epsilon_{\alpha A'},\nonumber\\~~~\delta\bar{\hat\theta}^{A'}_{\dot{\alpha}}&=&\bar{\epsilon}^{A'}_{\dot{\alpha}}.
\end{eqnarray}

The analytic subspace of $N=2$ superspace in the dual case looks like following:
\begin{equation}\label{subset0}
     \hat{x}^{\mu}_S = x^\mu-2i\hat\theta^{(A'}\sigma^\mu\bar{\hat\theta}^{B')}\hat{u}^{+}_{A'}u^{-}_{B'},~~~\hat\theta^{+}_{\alpha},~~~\bar{\hat\theta}^{+}_{\dot{\alpha}},~~~\hat{u}^{\pm}_{A'}.
\end{equation}

The $N=2$ supersymmetry transformations Eqn.(\ref{neq2susytr}) become
\begin{eqnarray}\label{neq2susytr0}
 \delta\hat{x}^{\mu}_S &=&-2i(\hat\epsilon^{A'}\sigma^\mu\bar{\hat\theta}^{+}+\hat\theta^{+}\sigma^\mu\bar{\hat\epsilon}^{A'})\hat{u}^{-}_{A'},\nonumber\\
\delta\hat\theta^{+}_{\alpha}&=&\hat\epsilon^{A'}_{\alpha}\hat{u}^{+}_{A'},~~~\delta \bar{\hat\theta}^{+}_{\dot{\alpha}}=\bar{\hat\epsilon}^{A'}_{\dot{\alpha}}\hat{u}^{+}_{A'},~~~\delta \hat{u}^{\pm}_{A'}=0.    \end{eqnarray}

The superfield in the dual harmonic superspace corresponding to the superfield of original harmonic superspace given by Eqn.(\ref{anasf}) becomes:
\begin{eqnarray}\label{anasf0}
\hat\phi^{(q)}(\hat{x}_S,\hat\theta^+, \bar{\hat\theta}^+, \hat{u}^\pm)&=&\hat{F}^{(q)}(\hat{x}_S, \hat{u}^{\pm})+\hat\theta^{\alpha+}\hat\psi^{(q-1)}_{\alpha}(\hat{x}_S, \hat{u}^{\pm})\nonumber \\
&+&\bar{\hat\theta}^{+}_{\dot{\alpha}}\bar{\hat\varphi}^{\dot{\alpha}(q-1)}(\hat{x}_S, \hat{u}^{\pm})
+\bar\theta^{+}\bar\theta^{+}\hat{M}^{(q-2)}(\hat{x}_S, \hat{u}^{\pm})\nonumber \\
&+&\bar{\hat\theta}^{-}\bar{\hat\theta}^{+}\bar{N}^{(q-2)}(\hat{x}_S, \hat{u}^{\pm}) + \hat\theta^{+}\sigma^a\bar{\hat\theta}^{+}\hat{A}_{a}^{(q-2)}(\hat{x}_S, \hat{u}^{\pm}) \nonumber \\
&+& \bar{\hat\theta}^{+}\bar{\hat\theta}^{+}{\hat\theta}^{\alpha+}\hat{\xi}^{q-3}_{\alpha}(\hat{x}_S, u^{\pm})+\hat\theta^+\hat\theta^+\bar{\hat\theta}^+_{\dot{\alpha}}\bar{\hat\chi}^{\dot{\alpha}(q-3)}(\hat{x}_S, \hat{u}^{\pm}) \nonumber \\
&+& \hat\theta^+\hat\theta^+\bar{\hat\theta}^+\bar{\theta}^{+}\hat{D}^{(q-4)}          (\hat{x}_S, \hat{u}^{\pm}).
\end{eqnarray}
Here the $U(1)$ charge $q$ should be called $q'$ because this is a different $U(1)$ that should be called $U(1)'$. But we are using $q$ in place of $q'$ to avoid cluttering of notation and we do not see any scope for confusion.  Since henceforth we shall be talking about dual harmonic space and Ali-Ilahi's complementary or dual ADHM instanton sigma model we could ignore all the carets. That would give us some respite from cluttering. But then there is the complete ADHM instanton linear sigma model construction in Ref.\cite{Ali:2023icn} where Witten's and Ali-Ilahi's constructions occur simultaneously. Because of that we have to use the notation with carets over some symbols. For the harmonic superspace formulation of that complete ADHM instanton sigma model we shall need both superspaces and hence it is worthwhile to bear the cluttering of the notation here.

Every individual field in the superfield in the Eqn.(\ref{anasf0}) has an expansion like follows:
\begin{equation}\label{fnewcor0}
\hat{F}^{(q)}(\hat{x}_S, u^{\pm})=\sum_{n=0}^{\infty} \hat{f}^{A'_1\cdots A'_{n+q}B'_1\cdots B'_n}(\hat{x}_S)\hat{u}^{+}_{(A'_1}\cdots \hat{u}^{+}_{A'_{n+q}}\hat{u}^{-}_{B'_1}\cdots \hat{u}^{-}_{B'_n)}
\end{equation}
for $q\geq 1$. This is the counterpart of Eqn.(\ref{fnewcor}).

The gauge transformation of Eqn.(\ref{gaugetran}) becomes
\begin{equation}\label{gaugetran0}
\delta\hat{V}^{++} = \hat{D}^{++}\hat\lambda 
\end{equation}
with $\hat\lambda=\hat\lambda(\hat{x}, \hat\theta^+, \bar{\hat\theta}^-, \hat{u}^\pm)).$
The derivative Eqn.(\ref{deransf}), when applied to an analytic superfield in the dual harmonic superspace, takes the following expression: 
\begin{equation}\label{deransf0}
\hat{D}^{++} = \hat{u}^{+A'}\frac{\partial}{\partial \hat{u}^{-A'}} -2 i\hat\theta^+\sigma^\mu\bar{\hat\theta}^+
\frac{\partial}{\partial\hat{x}_S^\mu}.
\end{equation}
The Wess-Zumino-like gauge for $\hat{V}^{++}$ in dual harmonic superspace, earlier given in Eqn.(\ref{wzlike}) for original harmonic superspace,  is now given by the following expression:
\begin{eqnarray}\label{wzlike0}
\hat{V}^{++}(\hat{x}_S, \hat\theta^+,\bar{\hat\theta}^-,\hat{u}^{\pm})&=&\hat\theta^+\hat\theta^+[\hat{M}(\hat{x}_S)+i\hat{N}(\hat{x}_S)] + \bar{\hat\theta}^+\bar{\hat\theta}^+ [\hat{M}(\hat{x}_S)-i\hat{N}(\hat{x}_S)] \nonumber \\
&+&i{\hat\theta}^+\sigma^{a}\bar{\hat\theta}^+\hat{A}_{a}(\hat{x}_S)
+\bar{\hat\theta}^+\bar{\hat\theta}^+{\hat\theta}^{+\alpha}\hat{\psi}^{A'}_{\alpha}(\hat{x}_S)\hat{u}^{-}_{A'}\nonumber \\
&+&\hat\theta^+\hat\theta^+\bar{\hat\theta}^+_{\dot{\alpha}}\bar{\hat\psi}^{\dot{\alpha}A}(\hat{x}_S)\hat{u}^{-}_{A'}+  \hat\theta^+\hat\theta^+\bar{\hat\theta}^+\bar{\hat\theta}^+\hat{D}^{(A'B')}   (\hat{x}_S)\hat{u}^{-}_{A'}\hat{u}^{-}_{B'}.\nonumber\\
\end{eqnarray}

The harmonic variables of the dual harmonic superspace have the following properties
\begin{eqnarray}
\hat{u}^{+A'}\hat{u}^-_{A'}&=&1,\nonumber\\
\hat{u}^{+A}\hat{u}^+_{A'}&\equiv&\hat{u}^{+A'}\epsilon_{A'B'}\hat{u}^{+B'}=0,\nonumber\\
\hat{u}^{-A'}\hat{u}^{-}_{A'}&=&0.\label{harv3}
\end{eqnarray}
and
\begin{equation}
(\hat{u}^{+A})^*=\hat{u}^{-A},~~(\hat{u}^-_{A'})^*=-\hat{u}^+_{A'}.\label{harv4}
\end{equation}

The three derivative operators in dual harmonic superspace preserving the relations given by Eqn.(\ref{harv3}) are given by 
\begin{equation}\label{threedefop0}
\hat{D}^{++} = \hat{u}^{+A'}\frac{\partial}{\partial \hat{u}^{-A'}},~~~    \hat{D}^{--} = \overline{\hat{D}^{++}},~~~ \hat{D}^{0} = \hat{u}^{+A'}\frac{\partial}{\partial \hat{u}^{+A'}} - \hat{u}^{-A'}\frac{\partial}{\partial\hat{u}^{-A'}}. 
\end{equation}
These obey an $SU(2)$ algebra.

The  eigen-functions of the charge operator for dual harmonic superspace are defined as follows:

\begin{equation}\label{harmfn0}
\hat{D}^{0}\hat{f}^{q}(\hat{u}) = q\hat{f}^{q}(\hat{u}) ,~~~ q=0,\pm 1, \pm 2,\cdots.
\end{equation}
Corresponding harmonic expansion is :
\begin{equation}\label{harmexp0}
\hat{f}^{q}({u})=\sum_{n=0}^{\infty}\hat{f}^{A'_1\cdots A'_{n+q}B'_1\cdots B'_n}\hat{u}^{+}_{(A_1}\cdots \hat{u}^{+}_{A'_{n+q}}\hat{u}^{-}_{B'_1}\cdots\hat{u}^{-}_{B'_n)}.
\end{equation}

The reality conditions on harmonic functions for dual harmonic superspace are
 \begin{equation}\label{partconj0}
\widetilde{\hat{f}^{A'B'\cdots}}=\overline{\hat{f}^{A'B'\cdots}},~~~\widetilde{\hat{u}^{\pm {A'}}}=\hat{u}^{\pm}_{A'},~~~\widetilde{\hat{u}^{\pm}_{A'}}=-\hat{u}^{\pm A'}.
 \end{equation}

If we define
\begin{equation}
    \hat{D}^3=\frac{1}{2}\hat{D}^0\label{def0}
\end{equation}
then we get the $SU(2)$ algebra of derivatives in dual harmonic superspace as:
\begin{equation}\label{def1}
   [\hat{D}^{++}, \hat{D}^{--}]=2\hat{D}^3,\nonumber 
\end{equation}
\begin{equation}   
  [\hat{D}^3, \hat{D}^{++}]=\hat{D}^{++}, [\hat{D}^3, \hat{D}^{--}]=-\hat{D}^{--}.\label{dalgebra0}  
\end{equation} 

Action of the derivatives on the harmonic variables in dual harmonic superspace is given below.
\begin{equation}
\hat{D}^{++}\hat{u}^{+A'}=0,~~~\hat{D}^{++}\hat{u}^{-A'}=\hat{u}^{+A'}  \label{def4}
\end{equation}
and
\begin{equation}
\hat{D}^{-}\hat{u}^{+A'}=\hat{u}^{-A'},~~~\hat{D}^{++}\hat{u}^{-A'}=0\label{def2}  
\end{equation}
as well as
\begin{equation}
    \hat{D}^0\hat{u}^{\pm {A'}}=\pm \hat{u}^{\pm {A'}}.\label{def3}
\end{equation}

We can discuss the problem of solution to the differential equation, like we did for Eqn.(\ref{diffe}), in dual harmonic superspace 
\begin{equation}
\hat{D}^{++}\hat{X}^{(q)}=\hat{F}^{(q+2)}\label{diffe1}
\end{equation}
in $\hat{u}^\pm$-calculus for some unknown $\hat{X}$.

We begin with the homogeneous problem
\begin{equation}
 \hat{D}^{++}\hat{X}_0^{(q)}=0. \label{diffe2}
\end{equation}
Corresponding solution is
\begin{eqnarray}
\hat{X}_0^{(q)}&=&\hat{X}^{(A'_1\cdots A'_q)}\hat{u}_{A'_1}\cdots \hat{u}_{A'_q},\;\;q\geq 0, \nonumber\\
\hat{X}^{(q)}_0&=&0,\;\;{q< 0}.\label{soln2}
\end{eqnarray}

The solution to the inhomogeneous equation now can be written as
\begin{eqnarray}
    \hat{X}^{(q)}&=&(1/\hat{D}^{++})\hat{F}^{(q+2)}+\hat{X}_0^{(q)},q\geq 0\nonumber\\
    \hat{X}^-&=&(1/\hat{D}^{++})\hat{F}^{(1)},\;\;q=-1.\label{soln4}
\end{eqnarray}
For $q\leq -1$ the solution exists only for special $\hat{F}^{(q+2)}$s. Here we have used the dual version of the following partial solution to Eqn.(\ref{diffe})
\begin{equation}
\frac{1}{\hat{D}^{++}}\hat{F}^{(q+2)}=\sum_{n=0}^{\infty}\frac{1}{n+1}\hat{f}^{\cdots 2n+q+2\cdots}(\hat{u}^+)^{n+q+1} (\hat{u}^-)^{n+1}.\label{soln3}   
\end{equation}

Here are two formulas in the dual harmonic superspace that will be used in the following to wrap up our brief introduction to the harmonic calculus. The simple rule gives the definition of a harmonic integral.

\begin{equation}\label{harmint1}
\int d\hat{u} 1=1,~~\int d\hat{u} \hat{u}^{+}_{(A'_1}\cdots\hat{u}^{+}_{A'_{p}}\hat{u}^{-}_{B_1}\cdots\hat{u}^{-}_{B_r)} ~~~\text{for} ~p+r>0.
\end{equation}

In the expansion of the integrand, the harmonic integral essentially projects out the singlet portion just like the original harmonic superspace. Integration by parts for the derivatives (\ref{threedefop0}) is compliant with the aforementioned integration rule. Additionally, one can can prove the following identity:

\begin{equation}\label{identity0}
\hat{D}^{++}_1 \frac{1}{\hat{u}^{+}_1\hat{u}^{+}_2}=\delta^{+,-}(\hat{u}_1,\hat{u}_2),
\end{equation}
where $\hat{u}^{+}_1\hat{u}^{+}_2 \equiv \hat{u}^{+A'}_1\hat{u}^{+}_{2A'}$ and $\delta^{+,-}(\hat{u}_1, \hat{u}_2)$ is a harmonic delta function. This is, as stated in Sub-appendix \ref{calculus},  equivalent to
\begin{equation}\label{delta-rule0}
    \frac{\partial}{\partial \bar z}=\pi\delta(z).
\end{equation}

In central basis the dual $N=2$ harmonic superspace is parametrized by the coordinates
\begin{equation}
    \hat{z}^M=(x^\mu, \hat{\theta}_{\alpha A}, \bar{\hat\theta}^{A'}_{\dot\alpha}=\overline{(\theta_{\alpha A'})}).
    \label{cb0}
\end{equation}
Corresponding supersymmetry transformations are given by the Eqn.(\ref{n=2ss}).

The coordinates of the dual analytic basis include the harmonic variables
\begin{equation}
    \hat{Z}^M_S=(\hat{x}^\mu_S, \hat\theta^+_\alpha, \bar{\hat\theta}^+_{\dot\alpha}, \hat\theta^-_\alpha, \bar{\hat\theta}^-_{\dot\alpha},\hat{u}^\pm_{A'}).\label{ab0}
\end{equation}
Where  we have defined
\begin{eqnarray}
    \label{def5}
    \hat{x}^\mu_S&=&x^\mu-2i\hat\theta^{(A'}\sigma^\mu{\bar\theta}^{B')}\hat{u}^+_{A'}\hat{u}^-_{B'},\nonumber\\
    \hat\theta^\pm_\alpha&=&\hat\theta^{A'}_\alpha u^\pm_{A'},\nonumber\\
    \bar{\hat\theta}^\pm_\alpha&=&\bar{\hat\theta}^{A'}_\alpha \hat{u}^\pm_{A'}.
\end{eqnarray}
Corresponding supersymmetry transformations in the dual harmonic superspace are given below:

\begin{eqnarray}\label{neq2susytr2}
 \delta\hat{x}^{\mu}_S &=&-2i(\epsilon^{A}\sigma^\mu\bar{\hat\theta}^{+}+\hat\theta^{+}\sigma^\mu\bar{\epsilon}^{A'})\hat{u}^{-}_{A'},\nonumber\\
\delta\hat\theta^+_{\alpha}&=&\epsilon^{A'}_{\alpha}\hat{u}^{+}_{A'},~~~\delta \bar{\hat\theta}^{+}_{\dot{\alpha}}=\bar{\epsilon}^{A'}_{\dot{\alpha}}\hat{u}^{+}_{A'},\nonumber\\   
\delta\hat\theta^{-}_{\alpha}&=&\epsilon^{A'}_{\alpha}\hat{u}^{-}_{A'},~~~\delta \bar{\hat\theta}^{-}_{\dot{\alpha}}=\bar{\epsilon}^{A'}_{\dot{\alpha}}\hat{u}^{-}_{A'},~~~\delta \hat{u}^{\pm}_{A'}=0.
\end{eqnarray}

The rest of the narrative will be slightly terse and brief owing to the fact that we have included very extensive background before describing our construction that we begin in the next section.

\section{Ali-Ilahi Model : Galperin-Sokatchev Perspective}\label{perspective}

In this section we give the Galperin-Sokatchev perspective of Ali-Ilahi's ADHM instanton sigma model. This perspective is needed for the main task of this note - off-shell, that is, harmonic superspace formulation of Ali-Ilahi's complementary ADHM instanton linear sigma model.

In providing Galperin and Sokatchev's perspective on Witten's ADHM instanton sigma model we did not use their notation but a notation that is more close to Witten's original notation. In this section too we shall again use a notation closer to Witten's notation rather than Galperin-Sokatchev notation. 

This time, as different from the case of Witten's Original Model, the initial step in the ADHM construction for the case of $SO(n')$ involves the rectangular matrix $A^{a'}_{A'Y}$ with indices $a' = 1, \cdots, n'+4k$, $Y = 1, \cdots, 2k$ (where $n'$ and $k$ are positive numbers) and $A'$ being the same $Sp(1)$ index as mentioned earlier in the last section.  

The tensor $A^{a'}_{A'Y}$ should obey the reality condition :
\begin{equation}\label{real3}
\overline{A^{a'}_{A'Y}}=\epsilon^{A'B'}\epsilon^{YZ}A^{a'}_{B'Z}.
\end{equation}

As per Eqn.(\ref{linear2}) it consists of tensors
$\hat N^{a'}_{A'Y}$ and $\hat E^{a'}_{YY'}$. In this case, $A^{a'}_{A'Y}(\hat\phi)$ is linear in $\hat\phi$ 
\begin{equation}
 A^{a'}_{A'Y}(\hat\phi)=\hat N^{a'}_{A'Y}+\hat E^{a'}_{YY'}\hat\phi_{A'}^{Y'}\label{linear5}   
\end{equation}
since $\hat N^{a'}_{A'Y}$ and $\hat E^{a'}_{YY'}$ are constant.

The tensor $A^{a'}_{A'Y}(\hat\phi)$ must satisfy the algebraic constraint:
\begin{equation}\label{real5}
A^{a'}_{A'Y}(\hat\phi)A^{a'}_{B'Z}(\hat\phi)=\epsilon^{A'B'}\hat R_{YZ}(\hat\phi),
\end{equation}
here $\hat R_{YZ}$ is an invertible antisymmetric $2k\times 2k$ matrix.

The matrices $\hat N^{a'}_{A'Y}$ and $\hat  E^{a'}_{YY'}$ should have maximal rank. These conditions are essential for the successful implementation of the ADHM construction in the case of $SO(n')$.

To obtain the instanton field the real rectangular matrix denoted as $\hat v^{a'}_{i'}$, introduced in Appendix \ref{d}, is required, where $i'$ is the $SO(n')$ index. The matrix satisfies two important properties - (\ref{ortho2}) and (\ref{ortho3}).

By employing these properties, the $SO(n')$ gauge field, Eqn.(\ref{aijay4}),  can be straightforwardly expressed as follows:
\begin{equation}\label{aijay5}
\hat A_{i'j'A'Y'}= \hat v^{a'}_{i'}\partial_{A'Y'}\hat v^{a'}_{j'}.
\end{equation}
Here
\begin{equation}
   \partial_{A'Y'}=\frac{\partial}{\partial \phi^{A'Y'}}. 
\end{equation}

It also corresponds to an instanton solution with finite action and instanton number $k$. 
 
Just like the Original Model in the present case too the matrices $A^{a'}_{A'Y}$ and $\hat v^{a'}_{i'}$ have lots of arbitrariness. The index $a'$ has  global $SO(n'+4k)$ transformation freedom that leaves (\ref{aijay4}) invariant. Also the index $Y$ has  global $GL(2k, C)$ transformation invariance.  The reality condition, Eqn.(\ref{real3}), reduces it to $GL(k, C)$ invariance. The index $i'$ is the $SO(n')$ Yang-Mills local gauge invariance index.

The $SO(n'+4k) \times GL(k,Q)$ freedom is the one that is relevant for counting the instanton parameters. This can be used to put the tensors $\hat N^{a'}_{A'Y}$ and $\hat E^{a'}_{YY'}$ into canonical form 
 
\begin{equation}\label{canon2}
 \hat N^{a'}_{A'Y}\ \rightarrow \ \left(\begin{array}{ c} \hat b_{4k \times n'}
\\ \hat d_{4k \times 4k }
\end{array}\right), \ \ \
\hat E^{a'}_{YY'} \ \rightarrow \ \left(\begin{array}{c} 0_{4k\times n'} \\
1_{4k\times 4k}
\end{array} \right).
\end{equation}
The matrices $b$ and $d$ are subject to algebraic constraints derived from (\ref{real5}). Together with the remaining elements in the instanton solutions, these symmetry transformations result in the appropriate count of independent parameters.  Again the references  \cite{Christ:1978jy, Weinberg:2012pjx} tell us more about the demanding task of counting the instanton parameters.

\section{Case of Dual (0,4) Supersymmetry}\label{04susy0}

In Appendix \ref{b} we have covered the background on harmonic superspace in some detail. That formalism was for $N=2$ supersymmetry. To formulate Witten's ADHM instanton sigma models in off-shell superspace we need harmonic superspace for $(0, 4)$ superspace. Galperin and Sokatchev discussed this in Section 2 of their narrative. We have reviewed that in Sub-appendix \ref{04susy}. In this section we take up corresponding task for the dual formalism.

Harmonic variables are useful for the analysis of $(0,4)$ supersymmetric theories because of the existence of three complex structures.

The $(0,4)$ world sheet superspace in case of the dual harmonic superspace, relevant for off-shell formulation of Ali-Ilahi's complementary model, is described using the coordinates $\hat{x}_{S++}$, $x_{--}$, and $\hat{\theta}^{AA'}_+$. To avoid confusion, as in the original case, the harmonic $U(1)$ charges are expressed as upper indices while the Lorentz ($SO(1,1)$) weights as lower indices. The Grassmann variables $\hat{\theta}^{AA'}_+$ carry doublet indices of 
the  automorphism group of $(0,4)$ supersymmetry,
\begin{equation}
    SO(4)\sim SU(2)\times SU(2)'
\end{equation}
and they also obey the reality condition 
\begin{equation}
\overline{\hat{\theta}^{AA'}} =
\epsilon_{AB}\epsilon_{A'B'} \hat{\theta}^{BB'}.
\end{equation}
 
The counterparts of the supersymmetry generators, the spinor covariant derivatives in the dual harmonic superspace are defined as
\begin{equation}\label{D0}
\hat{D}_{-AA'} = \frac{\partial}{\partial\hat{\theta}^{AA'}_+} + i\hat{\theta}_{+AA'}
\frac{\partial}{\partial x_{++}}
\end{equation}
which satisfy the superalgebra with $(0,4)$ supersymmetry
\begin{equation}\label{susy9}
\{ \hat{D}_{-AA'}, \hat{D}_{-BB'}\} = 2i\epsilon_{AB}\epsilon_{A'B'} \partial_{--}.
\end{equation} 

Defining  
\begin{equation}\label{D+0}
\hat{D}^+_{-A} \equiv \hat{u}^{+A'}\hat{D}_{-AA'},
\end{equation}
we can rewrite Eqn.(\ref{susy9}) as follows
\begin{equation}\label{++0}
\{ \hat{D}^+_{-A}, \hat{D}^+_{-B}\} = 0.
\end{equation}
The torsion term from the right side of Eqn.(\ref{susy9}) disappears after this projection.

The Grassmann analytic 	superfields can be defined by the equation
\begin{equation}\label{ga0}
\hat{D}^+_{-A}\Phi(\hat{x_S},\hat{\theta},\hat{u}) = 0.
\end{equation}
We now define the {\it analytic basis} for the dual harmonic superspace, which has been enlarged by the addition of harmonic variables.
\begin{equation}\label{bas0}
\hat{x}_{S++} = x_{++} + i\hat{\theta}^{AA'}_+\hat{\theta}^{B'}_{+A} \hat{u}^+_{(A'}\hat{u}^-_{B')},
\ \ x_{--}, \ \ \hat{\theta}^{\pm A}_{+}= \hat{u}^\pm_{A'} \hat{\theta}^{AA'}_+, \ \ \hat{u}^\pm.\end{equation}

The derivative $\hat{D}^+_{-A'}$ in the dual harmonic space is just a partial one
\begin{equation}
    \hat{D}^+_{-A} =
\frac{\partial}{\partial\hat\theta^{-A}_+}.
\end{equation}
The Grassmann analyticity criterion,  Eqn.(\ref{ga0}),  can thus be resolved in the following manner:
\begin{equation}\label{asf0}
\hat{D}^+_{-A}\hat\Phi(\hat{x},{\hat\theta},\hat{u}) = 0 \ \ \Rightarrow \ \ \hat\Phi = \hat\Phi(\hat{x}_{S++},
x_{--}, {\hat\theta}^+_+, \hat{u}).
\end{equation}
In the meantime the dual harmonic derivative ${D}^{++}$ receives a vielbein term:
\begin{equation}\label{der0}
\hat{D}^{++} = \hat{u}^{+A}\frac{\partial}{\partial\hat{u}^{-A'}} + i{\hat\theta}^{+A'}_+{\hat\theta}^+_{+A'}
\frac{\partial}{\partial\hat{x}_{S++}}.
\end{equation}
We also have
\begin{equation}\label{hgr0}
[\hat{D}^{++}, \hat{D}^+_{-A}] = 0.
\end{equation}

The analytic superfields defined in Eqn.(\ref{asf0}) has a non-vanishing $U(1)$ harmonic charge $q$ short Grassmann expansion,
\begin{equation}\label{expa0}
\hat\Phi^q(\hat{x},{\hat\theta}^+,\hat{u}) = \hat\phi^q(\hat{x}, \hat{u}) + {\hat\theta}^{+A}_+\hat\xi^{q-1}_{-A'}(\hat{x}, \hat{u})
+({\hat\theta}^+_+)^2 \hat{f}^{q-2}_{--}(\hat{x}, \hat{u}),
\end{equation}
where $({\theta}^+_+)^2 \equiv {\theta}^{+A}_+{\theta}^+_{+A}\;$. The coefficients in Eqn.(\ref{expa0}) are harmonic-dependent fields (remember that all of the terms in Eqn.(\ref{expa0}) conserve the total $U(1)$ charge $q$). In addition, we should note that the harmonic analytic superfields Eqn.(\ref{expa0}) can occasionally be made real in the sense of the special conjugation Eqn.(\ref{partconj0}).

Here Galperin and Sokatchev remind us, in case of the Original Model, that they have harmonized only one $SU(2)$ subgroup of the $SO(4)$ automorphism group and left the other, $SU(2)'$, unaffected. This is because  the Original Model has only $SU(2)'$ symmetry and $SU(2)$ invariance is broken.  

In case of Ali-Ilahi's complementary ADHM instanton sigma model the role of the $SU(2)$ and $SU(2)'$ subgroups of the $SO(4)$ automorphism group of $(0, 4)$ supersymmetry are interchanged. This is what has been constructed in this section. The Complementary Model has $SU(2)$ invariance and $SU(2)'$ invariance is broken.
 
\section{Dual Free Supermultiplets}\label{free0}

To formulate Ali-Ilahi's ADHM instanton sigma model in harmonic superspace we need corresponding super-multiplets in the superfield form. In his construction of the original ADHM instanton sigma model Witten used three supermultiplets in component form. First one is the scalar supermultiplet based on bosonic coordinates $X^{AY}$ of the $R^{4k}$ target space and their fermionic partners $\psi^{A'Y}$. This is called the fundamental scalar supermultiplet. Then there is a right moving supermultiplet.  Galperin and Sokatchev call this supermultiplet containing the right moving fermions $\lambda^a_-$ and its superpartners (denoted by $F^a$ by Witten) as the  chiral fermion supermultiplet. The third supermultiplet consists of the $4k'$ scalars $\phi^{A'Y'}$ and corresponding supersymmetric partners $\chi^{AY'}$. Corresponding supermultiplet is termed as twisted scalar multiplet by Witten.

Corresponding harmonic superspace superfields were described in their Section 3.1 of their narrative by Galperin and Sokatchev. We covered the summary of their construction in Sub-appendix \ref{free}.

In this section we shall describe the super-multiplets needed for the off-shell supersymmetric formulation of Ali-Ilahi's complementary ADHM instanton sigma model \cite{Ali:2023csc} using dual harmonic superspace.

Ali-Ilahi's complementary ADHM instanton sigma model is related to Witten's original model by a simple duality that includes the transformations
$A\leftrightarrow A'$, $Y\leftrightarrow Y'$, $X\leftrightarrow\phi$, $\psi\leftrightarrow\chi$. The Original and Complementary Models are independent of each other. The free supermultiplets of the two models are related by duality. We shall denote the fields of the complementary model with a caret. Thus the chiral fermion superfield for Ali-Ilahi's complementary model will be denoted by $(\hat\lambda^{a'}, \hat F^{a'}).$ We shall use carets in case of scalar and twisted scalar supermultiplets also because they are independent from each other by definition.

Because of the duality between the original and complementary models the supermultiplet containing the fields $(\hat\phi, \hat\chi)$ will be taken as fundamental scalar and the supermultiplet containing the fields$( \hat{X}, \hat\psi)$ as twisted scalar supermultiplet. $(\hat\lambda^{a'}, \hat{F}^{a'})$ will be the right moving chiral fermion supermultiplet for the complementary model.

Here, we'll discuss their superspace equivalents of the three supermultiplets needed for the construction of Ali-Ilahi's complementary or dual ADHM instanton sigma model. Galperin and Sokatchev thought of these as the result of trivial dimensional reduction and truncation from the two fundamental $N=2, D=4$ superfields. We shall not go into that digression.

We begin with the construction of the dual fundamental scalar superfield based on scalar supermultiplet $(\hat\phi, \hat\chi)$. Here $\hat\phi^{A'Y'}$ are the coordinates of the target space $R^{4k'}$ because $A'$ takes two values and $Y'$ takes $2k'$ values. Corresponding supermultiplet in harmonic superspace is defined as :

 \begin{equation}\label{X0}
\hat\Phi^{+Y'}(x, {\hat\theta}^+, \hat{u}) = \hat\phi^{+Y'}(x, \hat{u}) + i{\hat\theta}^{+A}_+\hat\chi^{Y'}_{-A}(x, \hat{u})
+({\hat\theta}^+_+)^2 \hat{f}^{-Y'}_{--}(x, \hat{u}).
\end{equation}

These superfields are real in the sense of the conjugation defined in Eq.(\ref{real}), that is,
\begin{equation}\label{Xreal0}
\widetilde{\hat\Phi^{+Y'}} = \epsilon_{Y'Z'}\hat\Phi^{+Z'}.
\end{equation}
Every harmonic field has an infinite expansion on $S^2$. The multiplet we are searching for, though, ought to have a finite number of physical fields. It turns out that by enforcing the following harmonic irreducibility condition, we can truncate the harmonic expansions
\begin{equation}\label{irr0}
\hat{D}^{++}\hat\Phi^{+Y'} = 0.
\end{equation}
Finding the component solution to this constraint is not difficult, as explained in Sub-appendix \ref{free}.  Using the
analytic basis form of $\hat{D}^{++}$ (\ref{der0}), for $\phi^{+Y'}(x, \hat{u})$ we find
\begin{equation}\label{eqX0}
\partial^{++} \phi^{+Y'}(x,\hat{u}) = 0 \ \ \Rightarrow \ \ \phi^{+Y'}(x,\hat{u}) =
\hat{u}^+_{A'} \phi^{A'Y'}(x),
\end{equation}
as follows from the harmonic expansion  for $q=+1$. Here $\partial^{++}$ denotes the purely harmonic part of (\ref{der0}) and following Galperin and Sokatchev we are using $x$ in place of $\hat{x}$. Similarly, for
the other two fields in Eqn.(\ref{X0}) we get
\begin{equation}\label{solu0}
\chi^{Y'}_{-A}(x,\hat{u}) = \chi^{Y'}_{-A}(x), \ \ \ \hat{f}^{-Y'}_{--}(x,{u}) = -i \hat{u}^-_{A'}
\partial_{--}\phi^{A'Y'}(x).
\end{equation}
Here we would like to clarify once again that, for example, in field $\hat{f}^{-Y'}_{--}$ the $Y'$ index is the $Sp(2k')$ index, upper $-$ index is the $U(1)$ charge and the lower indices $--$ are the Lorentz ones. The component fields are real because of (\ref{Xreal0}),
$\overline{\phi^{A'Y'}}=
\epsilon_{A'B'}\epsilon_{Y'Z'}\phi^{B'Z'}, \; \overline{\chi_-^{AY'}}=
\epsilon_{AB}\epsilon_{Y'Z'}\chi_-^{BZ'}$. The harmonic dependence of the fields  was therefore changed to a linear dependence as a result of the constraint. We infer that the fields is in  the form an off-shell $(0,4)$ supersymmetric multiplet since the constraint  is clearly supersymmetric and does not need equations of motion for the component fields. It matches with the field content in Ref.\cite{Witten:1994tz}.

We now come to the action for the fundamental scalar supermultiplet in case of the dual or the complementary model. An integral over the analytic superspace is used to represent the corresponding action:
\begin{equation}\label{acX0}
\hat{S}_{\hat\Phi} = i\int d^2x d\hat{u}d^2{\hat\theta}^+_+ \hat\Phi^{+Y'}\partial_{++} \hat\Phi^+_{Y'}.
\end{equation}

Here we are using Galperin and Sokatchev's notation instead of Witten's notation.

Here the $SU(2)$ and Lorentz weights of the Grassmann measure are adjusted so as to effect that the action is both an $SU(2)$ and Lorentz
singlet. To obtain the component content of (\ref{acX0}) we substitute the
short form (\ref{eqX0}),(\ref{solu0}) of the expansion (\ref{X0}) in
(\ref{acX0}) and perform two more steps. First, we do the Grassmann
integral, i.e., we pick out only the $({\hat\theta}^+_+)^2$ terms. Then we do
the harmonic integral according to the rules, which amounts
to extracting the $SU(2)$ singlet part. The result is
\begin{equation}\label{comX0}
\hat{S}_{\hat\phi} = \int d^2x\; \left( \hat\phi^{A'Y'}\partial_{++}\partial_{--}\hat\phi_{A'Y'} +
\frac{i}{2}\hat\chi^{AY'}_-\partial_{++}
\hat\chi_{-AY'}\right).
\end{equation}
This action matches with the kinetic energy part of the action for $\hat\phi$ and $\hat\chi$ fields in Ref. \cite{Ali:2023csc}.

Now we come to the second supermultiplet for the complementary ADHM sigma model. In addition to the superfield listed in Eqn.(\ref{X0}) whose action is given in Eqn.(\ref{comX0}) the Complementary Model uses a second $(0,4)$ supermultiplet that exclusively includes chiral fermions. Its harmonic superspace description is actually quite parallel to what we saw for the original model in Sub-appendix \ref{free}. Consider the following real superfields that are {\it anti-commuting}.
\begin{equation}\label{Lam0}
\hat{\Lambda}^{a'}_+(x,\hat{\theta}^+,\hat{u}) = \hat\lambda^{a'}_+(x,\hat{u}) + \hat{\theta}^{+A}_+\hat{g}^{-a'}_{A}(x,\hat{u})
+i(\hat{\theta}^+_+)^2  \hat\sigma^{--a'}_{-}(x,\hat{u}).
\end{equation}
The action for them is
\begin{equation}\label{acL0}
S_{\hat\Lambda} = \frac{1}{2}\int d^2x d\hat{u} d^2\hat{\theta}^+_+ \;
\hat\Lambda^{a'}_+\hat{D}^{++} \hat\Lambda^{a'}_{+}.
\end{equation}
The range of the external index is
$a'=1,\cdots, n'+4k$ as an $SO(n'+4k)$ index. Integrating $\hat{D}^{++}$ in Eqn.
(\ref{acL0}) by parts changes the sign, but the superfields $\hat\Lambda^{a'}_+$
anticommute, so the trace $a'a'$ is symmetric.

Obtaining the component content of action in Eqn.(\ref{acL0}) is as easy as in the case of
the action in Eqn.(\ref{acX0}). First we do the Grassmann integral and find
\begin{equation}\label{comL0}
S_{\hat\Lambda} = \int d^2xd\hat{u} \; \left( \frac{i}{2}\hat\lambda^{a'}_+\partial_{--}
\hat\lambda^{a'}_{+}
+ i\hat\sigma^{--a'}_{-} \partial^{++}\hat\lambda^{a'}_+ + \frac{1}{4}
\hat{g}^{-a'A} \partial^{++}\hat{g}^{-a'}_{A} \right).
\end{equation}

Now following the route taken in Sub-appendix (\ref{free}) and Section 3.1 of Ref.\cite{Galperin:1994qn} we
obtain simply the action for $n'+4k$ free chiral fermions
\begin{equation}\label{chfe0}
S_{\hat\lambda} = \frac{i}{2}\int d^2x \;
\hat\lambda^{a'}_+(x)\partial_{--}\hat\lambda^{a'}_{+}(x).
\end{equation}
This is an example of an on-shell supermultiplet in which supersymmetry has a minor effect. The multiplet's off-shell variant  of (\ref{comL0}), which we wish to emphasize, calls for a {\it infinite number} of auxiliary fields. Galperin and Sokatchev offer a rationale in this regard in their Appendix A. The problem, as we understand, is the following. Since both original as well as the complementary models are $(0, 4)$ supersymmetric there fore the right moving fermions in the either model should not have super partners. But Witten assigns a superpartner $F^a$ to $\lambda^a$ to derive the potential for the scalar field in the $SU(2)$ case. A similar thing has to be done in the Complementary Model. We shall leave this issue here and refer to the rationalization by Galperin and Sokatchev mentioned above.

Finally, the Complementary Model too utilizes the so-called ``twisted" scalar multiplet, in
which the $SU(2)$ indices carried by the bosons and fermions are
interchanged as compared to the standard multiplet. Its
superspace description turns out to be unusual as in case of the Original Model. We need
a set of {\it anti-commuting abelian gauge} superfields
\begin{equation}\label{X1}
\hat X^{+Y}_+(x,\hat{\theta}^+,\hat{u}) = \hat\rho^{+Y}_+(x,\hat{u}) +
\hat\theta^{+A}_+\hat X^{Y}_{A}(x,\hat{u}) +i(\hat{\theta}^+_+)^2 \hat\psi^{-Y}_{-}(x,\hat{u}),
\end{equation}
satisfying the reality condition
$\widetilde{X^{+Y}_+}=\epsilon_{YZ}X^{+Z}_+$. Also $\hat\psi^{-Y}_{-}=\hat{u}^{-A'}\hat\psi_{-A'}^Y$.
Here $Y'=1,\ldots,2k'$ is an index of a new symplectic group $Sp(k')$, as
we shall see below.  It turns out that $k'$ is just the number of instantons in
the corresponding ADHM construction.
These superfields undergo the following abelian gauge transformations
\begin{equation}\label{gau0}
\delta\hat X^{+Y}_+ = \hat{D}^{++}\hat\omega^{-Y}_+
\end{equation}
with analytic parameters $\hat\omega^{-Y}_+(x,\hat{\theta}^+,\hat{u})$. As stressed, in  Ref.\cite{Galperin:1994qn} and reviewed in Sub-appendix \ref{free},  by Galperin and Sokatchev this gauge invariance has nothing to do with the target space
gauge group.  It is an artifact of the superspace description of the
multiplet.

If we compare the gauge invariance (\ref{gau0}) with the four-dimensional abelian gauge field $A_\mu(x)$, just as in case of the Original Model,  the role of the gauge invariance becomes clear. This field is known to contain two Poincar\'e group representations with spins 1 and 0. The spin 0 component can be removed in one of two ways. One method is to impose an irreducibility constraint, such as $\partial^\mu A_\mu =0$, while the alternative involves putting $A_\mu$ through gauge transformations, such as $\delta A_\mu = \partial_\mu
\hat\omega(x)$, and recording a gauge invariant action. Here, we note a similar phenomenon. Equivalent to Eqn.(\ref{irr0}) for the superfield in Eqn.(\ref{X0}) is the irreducibility condition $\partial^\mu A_\mu =0$. This condition has the result of making the superfield harmonically short, which results in the elimination of an unlimited number of additional degrees of freedom. Eqn.(\ref{gau0}) provides the analogue of the second gauge mechanism for $A_\mu$. The expansion of the parameter should be examined in order to comprehend what transpires.

\begin{equation}\label{om0}
\omega^{-Y}_+(x,\hat{\theta}^+,\hat{u}) = \hat\tau^{-Y}_+(x,\hat{u}) + \hat{\theta}^{+A'}_+
\hat{l}^{--Y}_{A'}(x,\hat{u})
+i(\hat{\theta}^+_+)^2  \hat\mu^{---Y}_{-}(x,\hat{u}).
\end{equation}
{}From Eqn.(\ref{X1}),  Eqn.(\ref{om0}) and using Eqn.(\ref{der0}), we get, for
instance, 
\begin{equation}
    \delta\hat\rho^+_+(x,\hat{u}) = \partial^{++}\hat\tau^-_+(x,\hat{u}). 
\end{equation}
Comparing
the harmonic expansions of $\hat\rho^+$ and $\hat\tau^-$, we easily see
that each term in the expansion of $\hat\rho^+$ has its counterpart in that of
$\hat\tau^-$. So, the component $\hat\rho^+$ can be completely gauged away.
Similar arguments show that the expansions of the parameters $\hat{l}^{--}$ and
$\hat\mu^{---}$ are a little ``shorter" than those of the fields $\hat{X}$ and
$\hat\psi^{-}$, correspondingly. What is missing is just the lowest order
term, the smallest $SU(2)'$ representations in each expansion. Thus, the
fields $\hat{X}^{Y}_{A}(x,\hat{u})$ and $\hat\psi^{-Y}_{-}(x,\hat{u})$ can also be gauged
away, except for the first terms in their harmonic expansions, the fields
$\hat{X}^{Y}_{A}(x)$ and $\hat{u}^-_{A'}\hat\psi^{A'Y}_{-}(x)$. The net result of all
this is the following ``short", i.e., irreducible harmonic superfield in
the {\it Wess-Zumino-type gauge}
\begin{equation}\label{WZ0}
\hat X^{+Y}_+(x,\hat{\theta}^+,\hat{u}) = \hat{\theta}^{+A}_+\hat{X}^{Y}_{A}(x)
+i(\hat{\theta}^+_+)^2  \hat{u}^{-}_{A'}\hat\psi^{YA'}_{-}(x).
\end{equation}
This is the precise meaning of the Complementary Model's ``twisted" multiplet. It is important to remember that this multiplet is off shell, much as the one that the superfield $\phi^{+Y'}$ in Eqn.(\ref{X0}) describes. The only difference between the two multiplets in terms of component fields is that the two automorphism groups $SU(2)$ are reversed. It is evident why their superspace descriptions differ, as we have seen, because we only have harmonic variables for one of the $SU(2)$ groups. We also want to emphasise that manifest supersymmetry is lost in the Wess-Zumino gauge (\ref{WZ0}).

We now need an action for the superfield $\hat X^+_+$. Following the discussion in Section 3.1 of the Ref.\cite{Galperin:1994qn} and Sub-appendix \ref{free} this is written as 
\begin{equation}\label{acP0}
\hat{S}_X=\int d^2x d^4\hat{\theta}_+ d\hat{u}_1d\hat{u}_2\; \frac{1}{\hat{u}^+_1\hat{u}^+_2}
\hat X^{+Y}_+(1)\partial_{++} \hat X^+_{+Y}(2).
\end{equation}

The discussion now is parallel to the one in Sub-appendix \ref{free} but still has to be repeated because of the duality.

The notation $\hat X^+_+(1) $ means that the analytic superfield is written
down in an analytic basis (\ref{bas0}) defined by the harmonic variable
$\hat{u}_1$; similarly, $\hat X^+_+(2) $ is given in another basis, defined by
$\hat{u}_2$.  Since both superfields appear in the same integral, they should be
written down in the same non-analytic basis $x_{\pm\pm},  
\hat{\theta}^{AA'}_+, \hat{u}$. This explains
why the Grassmann integral in (\ref{acP0}) is taken over the full superspace
and not over an analytic subspace, as in (\ref{acX0}) or (\ref{acL0}). Note
that, as always, we take care of matching $U(1)$ charges and Lorentz
weights (this accounts for the presence of $\partial_{++}$ in (\ref{acP0})).
Note also that the contraction ${}^{Y}{}_{Y} \equiv {}^{Y}\epsilon_{YZ}
{}^{Z}$ this time must be of the type $Sp(k)$. Indeed, interchanging 1 and 2 involves
integration by parts of $\partial_{++}$, flipping the odd superfields
$\hat X(1)$, $\hat X(2)$ and using $\hat{u}^+_1\hat{u}^+_2 = - \hat{u}^+_2\hat{u}^+_1$, so the fourth
antisymmetric factor $\epsilon_{YZ}$ restores the symmetry required by
the double harmonic integral.

The reason for the exotic form of (\ref{acP0}) is the gauge invariance
(\ref{gau0}). It works in the following way. Varying in (\ref{acP0}) with
respect to, e.g., $\hat X^+_+(1) $ and according to the transformation law
(\ref{gau0}) makes the harmonic derivative $\hat{D}^{++}_1$ appear under the
integral. Integrating it by parts, we see that it only acts upon the
harmonic distribution $(\hat{u}^+_1\hat{u}^+_2)^{-1}$ (since $\Phi^+_+(2) $ depends on
the second harmonic variable $\hat{u}_2$).  After this both superfields $\hat X^+_+$ become analytic with
respect to the same basis, i.e., they depend only on the two Grassmann
variables $\hat{\theta}^{+A}_+$. However, the Grassmann integral in (\ref{acP0})
is over $d^4\hat{\theta}$, so it gives 0. This establishes the gauge invariance
of the action (\ref{acP0}).

Due to its harmonic non-locality, the action (\ref{acP0}) may look strange but that is not the correct perception. In fact, all of the fields in the Wess-Zumino gauge (\ref{WZ0}) have brief harmonic expansions. When calculating the Grassmann integral, we find factors of $(\hat{u}^+_1\hat{u}^+_2)$  in the numerator that cancel out the singular denominator without forgetting the various arguments of $\Phi(1,2)$. The double harmonic integral then loses significance, and we discover
\begin{equation}\label{comP0}
S_{\hat X} = \int d^2x\; \left( \hat{X}^{YA}\partial_{++}\partial_{--}
\hat{X}_{YA} + \frac{i}{2}\hat\psi^{YA'}_-
\partial_{++}\hat\psi_{-YA'} \right).
\end{equation}
These are the kinetic energy terms of the action for the fields that are part of the twisted multiplet of the Complementary Model.

This completes our construction of the free supermultiplets that we need for the off-shell harmonic superspace formulation of Ali-Ilahi's complementary ADHM instanton sigma model.

\section{Dual ADHM Interaction}\label{dual-int}

Thus we have at our hands three supermultiplets in the dual harmonic superspace - the dual fundamental scalar superfield, the dual chiral superfield and the dual twisted scalar superfield. How to couple the  supermultiplets to get the interaction part of the complementary ADHM instanton sigma model in the harmonic superspace is the key issue now. We are solely interested in potential type couplings, i.e., without space-time derivatives, in keeping with Witten's original construction. The coupling ought to be managed by a mass-dimensioned parameter $m$. Examining the resulting theory in the limit $m\rightarrow\infty$ with the intention of demonstrating that it flows into a CFT is the goal. Complementary ADHM sigma model construction stipulates that the latter must have an unbroken $SU(2)$ invariance as opposed to the $SU(2)'$ symmetry of the Original Model. 

Corresponding discussion is parallel but dual to the discussions in Section 3.2 of Galperin and Sokatchev that has been reviewed in Sub-appendix \ref{oleg} of this note. The interaction terms of the action can be integrals over either the full dual
$(0,4)$ harmonic superspace $\int d^2x d\hat{u} d^4\hat{\theta}$ or the dual analytic
superspace $\int d^2x d\hat{u} d^2\hat{\theta}^+_+ $; they should involve a positive
power of $m$. To see how this can be arranged, let us first examine the
dimensions of our superfields $\hat\Phi^{+Y'}$, $\hat\Lambda_+^{a'}$ and $\hat X^{+Y}_+$.
The position of the physical spinors (dimension 1/2) $\hat\chi_-$ in
Eqn.(\ref{X0}), $\hat\lambda_+$ in Eqn.(\ref{Lam0}) and $\hat\psi_-$ in Eqn.(\ref{X1}) fixes
the dimensions $[\hat\Phi]=0$, $[\hat\Lambda] = 1/2, [\hat X] = -1/2$.
It is easy to see that the full superspace integrals are ruled out. Indeed,
the full measure has dimension $0$ and Lorentz weight $0$. The presence of
the mass in $m\int d^2x d^4\hat{\theta}$ requires at least a pair of $X^+_+$
superfields
($[(\hat X)^2]=-1$), but this is not  consistent with the Lorentz invariance.
Thus, we are left with analytic superspace
interaction terms only. The analytic measure $\int d^2x d\hat{u}
d^2\hat{\theta}^+_+ $ has dimension 1, Lorentz weight $-2$ and $U(1)$ charge
$-2$. There are only two possible coupling terms in the action, in
which the dimensions, charges and weights are matched (we
omit the indices):
\begin{eqnarray}
\hat{S}_{\hat X\hat\Lambda}&=&m\int d^2x d\hat{u} d^2\hat{\theta}^+_+\; X^+_+\hat\Lambda_+ \hat{v}^+(\hat\Phi^+,
\hat{u}),
\label{m0} \\
\hat{S}_{\hat X\hat X}&=&m^2\int d^2x d\hat{u} d^2\hat{\theta}^+_+\;  X^+_+X^+_+ \hat{t}(\hat\Phi^+,
\hat{u}).
\label{mm0}
\end{eqnarray}
Here $\hat{v}^+$ and $\hat{t}$ are arbitrary functions of the dimensionless and
weightless superfields $\hat\Phi^+$ and of the harmonic variables $\hat{u}$. Also in place of the mass parameter $\hat{m}$ we are using the same notation as for the Original Model, that is $m$. 

Let us first investigate the term in Eqn.(\ref{m0}). To get an
idea how it can be constructed we follow the same route that we took in Sub-appendix \ref{oleg} following Ref.\cite{Galperin:1994qn}. Thus we first take the simplified case, in which
$n'=0$ and the index $a'$ can be written as a pair of symplectic indices, $a'
= A'Y$. Then the charged object $\hat{v}^+$ can be the harmonic variable $\hat{u}^+_{A'}$
itself. Thus we come to the following coupling term
\begin{equation}\label{coup0}
S_{\hat X\hat\Lambda} =m \int d^2x d\hat{u} d^2\hat{\theta}^+_+ \;  X^{+Y}_+ \hat{u}^{+A'}
\hat\Lambda_{+A'Y}.
\end{equation}
The main problem now for us is to make sure that the
coupling in Eqn.(\ref{coup0}) is invariant under the
gauge transformation of Eqn.(\ref{gau0}). The way to do this is to make the 
superfield $\hat\Lambda_{+A'Y}$ to transform as follows
\begin{equation}\label{gauL0}
\delta \hat\Lambda_{+A'Y} =m \hat{u}^+_{A'} \hat\omega^-_{+Y}.
\end{equation}
One can see that the variation with respect to $\hat\Lambda$ in the kinetic term in Eqn.
(\ref{acL0}) compensates for that of $\hat X$ in Eqn.(\ref{coup0}), whereas
$\delta\hat\Lambda$ in Eqn.(\ref{coup0}) is annihilated by the harmonic $\hat{u}^+$
$(\hat{u}^{+A'}\hat{u}^+_{A'} = 0)$.

Just like in case of the Original Model this time too the second possibility of Eqn.(\ref{mm0}) to construct a coupling term is now
clearly ruled out by the gauge invariance stated in Eqn.(\ref{gau0}). This term appears in a fixed gauge.

Like the the case of the Original Model discussed in Ref.\cite{Galperin:1994qn} and Sub-appendix \ref{oleg} the coupling in Eqn.(\ref{coup0}) is just a mass term despite the existence of gauge invariance. Once again there are two ways to see that - either we use component form or work directly with the superfields. We shall take up the latter first.  

Decomposing the index $A'$ of $\hat\Lambda_{+A'Y}$ on the basis of $\hat{u}^{\pm A'}$:
\begin{equation}\label{deco0}
\hat\Lambda_{+A'Y} = \hat{u}^+_{A'}\hat\Lambda^-_{+Y} + \hat{u}^-_{A'} \hat\Lambda^+_{+Y}, \;\;\;
\hat\Lambda^-_{+Y} \equiv -\hat{u}^{-A'}\hat\Lambda_{+A'Y}, \;\;
\hat\Lambda^+_{+Y} \equiv \hat{u}^{+A'}\hat\Lambda_{+A'Y}.
\end{equation}
Then from Eqn.(\ref{gauL0}) we see that $\delta \Lambda^-_{+Y} =m
\hat\omega^-_{+Y}$,
which allows us to fix the following {\it supersymmetric gauge} (as opposed to
the non-supersymmetric Wess-Zumino gauge (\ref{WZ0}))
\begin{equation}\label{sugau0}
\hat\Lambda^-_{+Y}  = 0.
\end{equation}
Substituting this into Eqn.(\ref{coup0}) and the kinetic term of Eqn.(\ref{acL0}), we
obtain two action terms involving the remaining projection $\Lambda^+_{+Y'} $:
\begin{equation}\label{alg0}
\int d^2x d\hat{u} d^2\hat{\theta}^+_+ \; \left(-\frac{1}{2}
\Lambda^{+Y}_{+}\Lambda^+_{+Y} +
m \hat X^{+Y}_+ \hat\Lambda^+_{+Y}\right).
\end{equation}
Clearly, $\hat\Lambda^+_{+Y}$ can be eliminated from Eqn. (\ref{alg0}) by means of its
algebraic field equation $\hat\Lambda^+_{+Y} = m \hat X^+_{+Y'}$. Then we are left
only with the superfield $\hat X$ which has the action (recall Eqn. (\ref{acP0}))
\begin{eqnarray}\label{massac0}
\hat{S}_{\hat{X}} &=&i \int d^2x d^4\hat{\theta}_+ d\hat{u}_1d\hat{u}_2\; \frac{1}{\hat{u}^+_1\hat{u}^+_2}
\hat X^{+Y}_+(1)\partial_{++} \hat X^+_{+Y}(2) \nonumber\\
&{}&+ \frac{m^2}{2} \int d^2x d\hat{u}
d^2\hat{\theta}^+_+ \; \hat X^{+Y}_+\hat X^+_{+Y}.
\end{eqnarray}
The very appearance of Eqn.(\ref{massac0}) shows that it is the action for $2k$
massive superfields. We conclude that when the number of the multiplets
$\hat\Lambda^{a'}$ equals $4k$, the coupling term in Eqn.(\ref{coup0}) is in fact a mass
term for the $4k$ real bosons $\hat X^{AY}$ contained in $\hat X$ and for
the $4k$ pairs of right-handed fermions $\psi^{A'Y}_-$ from $\hat X$ and
left-handed ones $\hat\lambda^{A'Y}_+$ from $\hat\Lambda$.

We now come back to the general case when $n' \neq 0$. We can try to
generalize the coupling in Eqn.(\ref{coup0}) by replacing $\hat{u}^+_{A'}$ by a matrix
$\hat{v}^{+a'}_{Y}(\hat\Phi^+, \hat{u})$, satisfying the reality condition $\widetilde{\hat{v}^{+a'}_Y}
= \epsilon^{YZ}\hat{v}^{+a'}_Z$. It cannot depend on the superfields $\hat X$ or
$\hat\Lambda$
because of Lorentz invariance, but it can still be a function of the
weightless superfields $\hat\Phi^{+Y'}$ and of the harmonic variables. So, we
write down
\begin{equation}\label{int0}
\hat{S}_{int} = m \int d^2x d\hat{u} d^2\hat{\theta}^+_+ \;  \hat {X}^{+Y}_+ \hat{v}^{+a'}_{Y}(\hat\Phi^+,\hat{u})
\hat\Lambda_+^a.
\end{equation}

The problem now is how to make Eqn.(\ref{int0}) compatible with the gauge
invariance defined in Eqn.(\ref{gau}). Like in case of Eqn.(\ref{gauL0}), we can try
\begin{equation}\label{gauLv0}
\delta\hat\Lambda^a_{+} = m\hat{v}^{+a'}_{Y}(\hat\Phi^+,\hat{u})\hat\omega^{-Y}_+.
\end{equation}
It is not hard to see that this transformation compensates for
$\delta\hat{X}$ in Eqn.(\ref{int0}), provided the following two conditions hold:
\begin{equation}\label{c10}
\hat{v}^{+a'}_{Y}\hat{v}^{+a'}_{Z} = 0,
\end{equation}
\begin{equation}\label{c20}
\hat{D}^{++}\hat{v}^{+a'}_{Y}(\hat\Phi^+,\hat{u}) = 0.
\end{equation}

Like in case of the Original Model the condition (\ref{c20}) is very restrictive. Assuming that the
function $\hat{v}^{+a'}_{Y}(\hat\Phi^+,\hat{u})$ is regular, i.e., can be expanded in a Taylor
series in $\hat\Phi^+$ we get:
\begin{equation}\label{tay0}
\hat{v}^{+a'}_{Y}(\hat\Phi^+,\hat{u}) = \sum^{\infty}_{p'=0}\hat{v}^{(1-p')a'}_{Y{Y'}_1\cdots{Y'}_{p'}}(\hat{u})
\hat\Phi^{+{Y'}_1} \cdots \hat\Phi^{+{Y'}_{p'}}.
\end{equation}

Since $\hat\Phi^+$ satisfies the irreducibility condition in Eqn.(\ref{irr0}) therefore the Eqn.  (\ref{c20})
implies $\hat{D}^{++}\hat{v}^{(1-p')a'}_{Y{Y'}_1\cdots{Y'}_{p'}}(\hat{u}) = 0$, where $1-p'$ is the $U(1)$
charge. The only non-vanishing solution to
this is 
\begin{equation}\label{lin0}
\hat{v}^{+a'}_{Y}(\hat\Phi^+,\hat{u}) = \hat{u}^{+A'} \hat{N}^{a'}_{A'Y} + \hat{E}^{a'}_{YY'}\hat\Phi^{+Y'},
\end{equation}
where the matrices $\hat{N}^{a'}$ and $ \hat{E}^{a'}$ are constant. In other words, the matrix
$\hat{v}^{+a'}_{Y}(X^+,\hat{u})$ can at most depend linearly on $\hat\Phi^+$. If we put
$\hat{\theta}^+=0$, i.e., consider only the lowest components in the superfield
$\hat\Phi^+$ (see Eqn.(\ref{eqX0})), then Eqn.(\ref{lin0}) becomes
\begin{equation}\label{lin'0}
\hat{v}^{+a'}_{Y}(\hat\Phi^+,\hat{u})|_{\hat{\theta}=0} = \hat{u}^{+A'} (\hat{N}^{a'}_{A'Y}  +
\hat{E}^{a'}_{YY'} \hat\phi^{Y'}_{A'}) \equiv \hat{u}^{+A'}
A^{a'}_{A'Y}(\hat\phi).
\end{equation}
Where we have defined
\begin{equation}
 \hat{N}^{a'}_{A'Y}  +
\hat{E}^{a'}_{YY'} \hat\phi^{Y'}_{A'} = A^{a'}_{A'Y}(\hat\phi).  \label{linear6}
\end{equation}
Here in Eqn.(\ref{linear6}) we have recovered the form of the $A$-tensor in Eqn.(\ref{linear5}).

The other condition (\ref{c10}) is purely algebraic here too. Putting Eqn.(\ref{lin'0})
in it and removing the harmonic variables $\hat{u}^{+A'}\hat{u}^{+B'}$, we get
\begin{equation}\label{ADHM0}
A^{a'}_{(A'Y}A^{a'}_{B')Z} = 0, \;\; \mbox{i.e.,} \;\;\;\;
A^{a'}_{A'Y}(\hat\phi)A^{a'}_{B'Z}(\hat\phi) = \epsilon_{A'B'} \hat{R}_{YZ}(\hat\phi).
\end{equation}

What we have uncovered are the two defining criteria on the matrix $A$ that are utilized as a starting point in the ADHM construction for instantons. As referred to earlier, the matrices $\hat{N}$ and $\hat{E}$ must have the highest rank possible, in this case $4k$, in order to be used in the ADHM construction. This implies that all $4k$ right-handed chiral fermions in $\hat{X}^{+Y}_+$ are coupled with a subset $4k$ of left-handed ones from $\hat\Lambda^a_+$ and acquire mass in the context of the Complementary Model \cite{Ali:2023csc}. This sigma model corresponds to $k$ instantons in $R^{4k'}$ with gauge group $SO(n')$. 

At the end of their Section 3.3 
instanton parameter counting the ADHM case. We shall not repeat it here.

To sum up this section, we have demonstrated how $(0,4)$ supersymmetry may accurately pinpoint the nature of the interaction for the linear sigma model and align it with the foundation of the ADHM structures. We underline the importance of Eqn.(\ref{gau0}) in stabilizing this interaction and the abelian gauge invariance. Of course, the model we have uncovered is simply the Complementary Model in superspace.

\section{Dual Instanton Gauge Field}\label{oleg0}

In parallel with what we did in Sub-appendix \ref{oleg}, on the basis of the aforementioned matrix $A$, we can construct the instanton gauge field in case of the complementary ADHM instanton linear sigma model. This is possible in off-shell supersymmetric manner in the harmonic superspace.  That is what we shall do first and describe the component method in a cursory manner at the end of this discussion.   

In Section \ref{dual-int} we have demonstrated that the model describes $4k+4k$ free massive bosons and fermions when $n'=0$ and the coupling is simply given by Eqn. (\ref{coup0}). This holds true to some extent even for the interaction Lagrangian in Eqn.  (\ref{int0}) in place of the one given in Eqn. (\ref{coup0}). The more accurate statement is that there is a subset of $4k$ of the $n'+4k$ left-handed fermions $\hat\lambda_+^{a'}$ in $\hat\Lambda^{a'}_+$ that couple with the right-handed fermions in $X$ and become massive (along with the bosonsic superpartners). Remaining chiral fermions maintain their masslessness. The issue is how to diagonalize the action in order to distinguish between the massive and the massless modes.
We'll try a diagonalization that looks similar to Eqn. (\ref{deco0}).

As the first step let us complete the $2k\times (n'+4k)$ matrix
$\hat{v}^{+a'}_{Y}(\Phi^+,\hat{u})$ to a
full {\it orthogonal} matrix $\hat{v}^{{\tilde a}'a'}(\Phi^+,\hat{u})$, where the $n'+4k$
dimensional index ${\tilde a} = (+Y, -Y, i')$ and
$i'=1,\cdots, n'$ is an index of the group $SO(n')$. This time orthogonality
means the following
\begin{equation}\label{or0}
\hat{v}^{\tilde a'a'}\hat{v}^{\tilde b' a'} = \delta^{\tilde a'\tilde b'},
\end{equation}
where $\delta^{+Y, -Z} = - \delta^{-Y, +Z} = \epsilon^{YZ}, \;\;
\delta^{+Y, +Z}=\delta^{-Y, -Z}=\delta^{\pm Y, i'}=0 $.
Particularly, for $\tilde a' = +Y$ and $\tilde b' = +Z$ we obtain just
condition in Eqn. (\ref{c10}). Since $\hat{v}^{+a'}_{Y}$ is a function of $\Phi^{+Y'}$ and
$\hat{u}^\pm$, we expect the other blocks of $\hat{v}^{\tilde a'a'}$, namely $\hat{v}^{-Ya'}$
and $\hat{v}^{i'a'}$ to be such functions too. Of course, the fact that
$\hat{v}^{+a'}_{Y}$ must be a {\it linear} function of $\Phi^{+Y'}$ (see
Eqn. (\ref{lin0})) does not imply that the rest of $\hat{v}^{\tilde a'a'}$ are linear as
well. It is also clear that the new matrix blocks are not completely fixed
by Eqn. (\ref{or0}). The freedom consists of the subset of $SO(n'+4k)$
transformations acting on the index $\tilde a$ with analytic (i.e.,
functions of $\hat\Phi^{+Y'}$ and $\hat{u}^\pm$) parameters, which leave $\hat{v}^{+a'}_{Y}$
invariant.

With the help of the newly introduced matrix we can make a change of variables
from the superfield $\hat\Lambda^{a'}_+$ to $\hat\Lambda^{\tilde a'}_+ =\hat{v}^{\tilde
	a'a'}\hat\Lambda^{a'}_+$. Then the
gauge transformations of Eqn.(\ref{gauLv0}) can be
translated into
\begin{equation}\label{gauL-0}
\delta\hat\Lambda^{-Y}_+ = m\hat\omega^{-Y}_+, \ \ \ \delta\hat\Lambda^{+Y}_+ =
\delta\hat\Lambda^{i'}_+ = 0.
\end{equation}
As in the flat case of Section \ref{dual-int}, Eqn. (\ref{gauL-0}) allows us
to fix the supersymmetric gauge (cf. Eqn.(\ref{sugau0}))
\begin{equation}\label{sugauge0}
\hat\Lambda^{-Y}_+ = 0.
\end{equation}

Now we can switch to the new superfields $\Lambda^{\hat a}_+$ in the kinetic
term for $\hat\Lambda$ of Eqn. (\ref{acL0}) and the coupling term (\ref{int0}). It is
useful to introduce the following notation
\begin{equation}\label{VV0}
(\hat{V}^{++})^{\tilde a'\tilde b'} = \hat{v}^{\tilde a'a'}\hat{D}^{++} \hat{v}^{\tilde b'a'}.
\end{equation}
Then the terms of the Lagrangian containing $\hat\Lambda$ become
(in the gauge (\ref{sugauge0}))
\begin{eqnarray}\label{LLL0}
\hat{\cal L}^{++}_{++}(\hat\Lambda) &= &\frac{1}{2}
\hat\Lambda^{i'}_+[\delta^{i'j'} \hat{D}^{++} + (\hat{V}^{++})^{i'j'}]
\hat\Lambda^{j'}_+  \nonumber\\
&+& \hat\Lambda^{+Y}_+[\frac{1}{2}(\hat{V})_{YZ}\hat\Lambda^{+Z}_+ +
(\hat{V}^{++})^{-i'}_{Y}\hat\Lambda^{i'}_+ + mX^+_{+Y}],
\end{eqnarray}
where we have denoted 
\begin{equation}
    \hat{V}_{YZ} =(\hat{V}^{++})^{--}_{YZ}.
\end{equation}  

We see that
$\hat\Lambda^{+Y}_+$ enters without derivatives, so it can be eliminated from Eqn. (\ref{LLL0}). The result is
\begin{eqnarray}
\hat{\cal L}^{++}_{++}(\hat\Lambda)
&=& \frac{1}{2}\hat\Lambda^{i'}_+[\delta^{i'j'} \hat{D}^{++} + (\hat{V}^{++})^{i'j'}] \hat\Lambda^{j'}_+
\nonumber \\
&-& \label{LL0}
\frac{1}{2}[(\hat{V}^{++})^{-i'}_{Y} \hat\Lambda^{i'}_+ + m\hat{X}^+_{+Y}] (\hat{V}^{-1})^{YZ}
[(\hat{V}^{++})^{-j'}_{Z}\hat\Lambda^{j'}_+ + m\hat{X}^+_{+Z}].\nonumber\\
\end{eqnarray}
The complete action of the model is obtained by adding the kinetic terms for
$\hat{X}^+$ (\ref{acX0}) and $\hat\Phi^+_+$ (\ref{acP0}).

We can set the superfields $\hat\Lambda^{i'}_+$ to 0 and $\hat{V}_{YZ} = \epsilon_{YZ}$ to make contact with the free case of Eqn.(\ref{massac0}). Only the mass term for the superfields $\hat{X}$ (cf. Eqn. (\ref{massac0})) is left. This explains why the chiral fermions in the subset of superfields  that correspond to massive modes are $\hat\Lambda^{+Y}_+$. Once again we can simplify things more interestingly if we let the mass $m$ to infinity. The second component in Eqn.(\ref{LL0}) then becomes auxiliary and we can drop it. This is equivalent to suppressing the kinetic term for $X$. The straightforward outcome is what is left after that.
\begin{equation}\label{ADga0}
\hat{\cal L}^{++}_{++}(\hat\Lambda)|_{m\rightarrow\infty} = \hat\Lambda^{i'}_+[\delta^{i'j'}
\hat{D}^{++} +
(\hat{V}^{++})^{i'j'}] \hat\Lambda^{j'}_+,
\end{equation}
where
\begin{equation}\label{calV0}
(\hat{V}^{++})^{i'j'} = \hat{v}^{i'a'}(\hat\Phi^+,\hat{u})\hat{D}^{++}\hat{v}^{j'a'}(\hat\Phi^+,\hat{u}).
\end{equation}
There is a method for  decoding the field content of the action term in Eqn.(\ref{int0}). It entails locating the area of the component action when a composite gauge field is connected to massless chiral fermions with the symbol $\hat\lambda_+$. In order to achieve this, we must utilise the Wess-Zumino gauge of Eqn. (\ref{WZ0}) rather than the one that is obviously supersymmetric. To simplify our task we
shall keep only the relevant fields, i.e., the fermions in $\hat{X}^{+Y}_+$ and
$\hat\Lambda^{a'}_+$ and the bosons in $\hat{\Phi}^+$. Since $\hat\psi^{A'Y}_-$ is
accompanied by $(\hat{\theta}^+_+)^2$ in Eqn. (\ref{WZ0}),
the other superfields in Eqn. (\ref{int0}) contribute with their lowest order
components only,
i.e., $A^{a'}_{A'Y}$ from Eqn.(\ref{lin0}) and $\hat\lambda^{a'}_+$ from Eqn. (\ref{Lam0}).
This, together with the kinetic term for $\hat\Lambda^{a'}_{+}$ from Eqn. (\ref{comL0}), gives
\begin{equation}\label{fer0}
\hat{S} = \int d^2xd\hat{u} \; \left(\frac{i}{2}\hat\lambda^{a'}_+\partial_{--}\hat\lambda^{a'}_{+} +
i\hat\sigma^{--a'}_{-} \partial^{++}\hat\lambda^{a'}_+ + m \hat{u}^-_{A'}\psi^{A'Y}_- \hat{u}^+_{B'}
A^{a'B'}_{Y}(\hat\phi)\hat\lambda^{a'}_+\right).
\end{equation}
The important point here is that the Lagrange multiplier term with
$\hat\sigma^{--a'}_{-}$
has not changed, so we still have the field equation
$\partial^{++}\hat\lambda^{a'}_{+}(x,\hat{u}) = 0 \
\rightarrow \ \hat\lambda^{a'}_{+} = \lambda^a_{+}(x)$, like in the free case. Then
the harmonic integral in (\ref{fer0}) becomes trivial and we obtain the action
\begin{equation}\label{ferr0}
\hat{S} = \int d^2x \; \left( \frac{i}{2}\hat\lambda^{a'}_+\partial_{--}\hat\lambda^{a'}_{+}
- \frac{m}{2} \hat\psi^{A'Y}_- A^{a'}_{A'Y}(\hat\phi)\hat\lambda^{a'}_+\right).
\end{equation}

The subsequent steps are described by Complementary Model and we only sketch them. One
introduces an $n'\times (n'+4k)$ matrix $\hat{v}^{a'}_{i'}(\hat\phi)$, orthogonal to $A$,
$A^{a'}_{A'Y}\hat{v}^{a'}_{i'}=0$ and orthonormalized, $\hat{v}^{a'}_{i'}\hat{v}^{a'}_{j'} = \delta_{i'j'}$.
In a sense, this step is similar to the introduction of the matrix
$\hat{v}^{\tilde a'a'}$ in Eqn. (\ref{or0}). The aim is to diagonalize Eqn. (\ref{ferr0}), i.e., to
separate the massive fermions from the massless ones. The massless
fermions are just $\hat\lambda^{i'}_+ = \hat{v}^{i'a'}\hat\lambda^{a'}_+$.
After that one puts all massive fields to 0 (or, equivalently,
$m\rightarrow\infty$) and gets
\begin{equation}\label{fergau0}
\hat{S} =\frac{i}{2} \int d^2x \; \hat\lambda^{i'}_+(\delta^{i'j'}\partial_{--} +
\partial_{--}\hat\phi^{A'Y'} \hat{A}^{i'j'}_{A'Y'})
\hat\lambda^{j'}_+,
\end{equation}
where
\begin{equation}\label{ADHMgau0}
\hat{A}^{i'j'}_{A'Y'} = \hat{v}^{i'a'} \frac{\partial\hat{v}^{j'a'}}{\partial \hat\phi^{A'Y'}}.
\end{equation}
This is precisely the expression for the instanton field in the ADHM
construction in case of the Complementary Model.

\section{Conclusions}\label{conclusions}

The dual harmonic superspace formalism we developed in this note for off-shell (0,4) supersymmetry of complementary ADHM instanton linear sigma model is very technical in spite of being straightforward replication of the original formalism by Galperin and Sokatchev for the original  ADHM instanton sigma model of Witten. Like the component cases of the Original and Complementary Models as well as the off-shell formulation of Original Model by Galperin and Sokatchev the present off-shell model too flows to respective small N=4 superconformal field theories in the infrared limit Ref.\cite{Ademollo:1975an, Ademollo:1976pp, Ademollo:1976wv}. In this formalism a pair of  formidable constructions came together to make the analysis complex. These are ADHM construction  and the harmonic superspace construction. It would be instructive to develop a harmonic superspace formalism for the large N=4 supersymmetry. By large N=4 supersymmetry we mean a theory that would flow to large N=4 superconfromal symmetry in the infrared Ref.\cite{Sevrin:1988ew, Ivanov:1988rt, Ali:2000zu, Ali:2003aa}, that is, Ali and Salih's Complete ADHM instanton linear sigma model of Ref.\cite{Ali:2023icn}.

The $AdS_3$ superstrings are closely related to BTZ blackhole and hence to its stingy generalization\cite{Ali:1992mj}. It will be worthwhile to investigate that aspect in the context of ADHM instanton sigma models.

In Ref.\cite{Papadopoulos:2024uvi} Papdopoulos and Witten gave a direct proof of the fact that in two dimensions scale invariance implies conformal invariance. In Ref.\cite{Witten:2024yod} Witten investigated the relation of instantons to large $N=4$ algebra. It would be interesting to investigate connection of these findings with the issues discussed in this note (see also \cite{Ferko:2024uxi, Heydeman:2025fde, Murthy:2025moj}).

In Ref.\cite{Ali:2025jcu} we have collected the results of this note in brief.

{\it Acknowledgment : } AA and MI thank Professor DP Jatkar and Professor R. Gopakumar for hospitality. AA thanks Professor V. Ravindran for hospitality. MI, PPAS and SRT contributed to this work during their PhD work.
\appendix

\section{Yang-Mills Instantons and Sigma Models}\label{a}

Our objective in this note is to develop harmonic superspace covariant, that is off-shell, formalism  for Ali-Ilahi's complementary ADHM instanton sigma model  along the lines that was done for Witten's original ADHM instanton sigma model  by Galperin and Sokatchev. (For references see the introduction's part just before the plan of the paper.)

The subject matter of this note is technical. To make it easy to follow the narrative we include background material in the appendices. The material collected in the appendices is not pedagogical but of take-off nature. In no way our reviews and recapitulations serve to replace the original references but serve only as reminders of the background.

To follow that narrative we need some background on Yang-Mills instantons. This includes Christ, Stanton and Weinberg perspective as well as the Corrigan, Fairlie, Goddard and Templeton perspective on ADHM instanton construction   

These aspects are covered in the present appendix.
 
Organization of the present appendix is as follows.

The background on 't Hooft instantons is collected in sub-appendix \ref{thooft}.

This note is about linear sigma models. We do not need information on non-linear sigma models in this note. Yet non-linear sigma models are more familiar than the linear ones. Also the non-linear sigma models provide a comparative perspective. Thus initially we provide a perspective on non-linear sigma models. 

In Sub-appendix \ref{bosonic0} we recapitulate some essential features of bosonic non-linear sigma models.

Review of conventional superspace formalism for non-linear sigma models is presented in Sub-appendix \ref{fermionic}.

Callan, Harvey and Strominger solutions and sigma models are described in Sub-appendix \ref{chs0}.

In Sub-appendix \ref{csw} we first summarize the Christ-Stanton-Weinberg perspective of the ADHM instantons and then then Corrigan-Fairlie-Goddard-Templeton perspective in Sub-appendix \ref{cfgt}.  These two perspectives helped in uncovering the secrets of the ultra-brief ADHM construction of instantons, that is, self-dual Yang-Mills fields. These two sub-appendices should mitigate the fear of ADHM construction to some extent - just like the original refrences.

Both  Ali-Ilahi's ADHM instanton sigma model and  Witten's ADHM instanton sigma model are linear sigma models. To follow the ADHM instanton sigma models we need some background on linear sigma models. Linear sigma models are introduced in Appendix \ref{linear}.

\subsection{The 't Hooft Instanton}\label{thooft}

Yang-Mills instantons are solutions to self-dual Yang-Mills equations. We begin by the definition of 't Hooft instantons here and move onto ADHM instantons in a latter sub-appendix. In this Sub-appendix our main source is Ref.\cite{Callan:1991at}.

Let us take Euclidean Yang-Mills action
\begin{equation}
    S_E=\frac{1}{2g^2}\int d^4x F^{a\mu\nu}F^a_{\mu\nu}\label{yang-mills1}
\end{equation}
with 
\begin{equation}
    F^a_{\mu\nu}=\partial_\mu A^a_\nu - \partial_\nu A^a_\mu +f^{abc}A^b A^c\label{ymfst}
\end{equation}
and search for solutions that approach the pure gauge configuration at infinity
\begin{equation}
    A_\mu\longrightarrow g^{-1}\partial_\mu g.
\end{equation}
These configurations are labelled by the winding number of the map from $S^3$ into the the gauge group $G$ of $g$ evaluated at infinity.

The winding number $k$ is defined as
\begin{equation}
    k=\frac{1}{16\pi^2}\int d^4x F^{a\mu\nu}{\tilde F}^a_{\mu\nu}
\end{equation}
where the dual field strength tensor is defined as
\begin{equation}
    {\tilde F}^{a\mu\nu} =\frac{1}{2}\epsilon^{\mu\nu\lambda\sigma}F^a_{\lambda\sigma}.\label{dual0}
\end{equation}

Now we would like to have self-dual (or anti-self-dual) Yang-Mills solution corresponding to $SU(2)$ gauge group. 

\begin{equation}
    \tilde F^{a\mu\nu} =\pm F^{a\mu\nu}.\label{selfdual0}
\end{equation}

Self-dual solutions are called instantons and anti-self-dual solutions anti-instantons. The most general solution for $k$-instanton is the ADHM solution that has 8$k$ parameters. We shall talk about ADHM instantons later on. Here we begin 't Hooft's simple ansatz with 5$k$+3 parameters.

In this ansatz the gauge field is written as a derivative of a scalar field $\phi(x)$
\begin{equation}
    A_\mu=\bar\Sigma_\mu^\nu\partial_\nu \ln\phi(x)\label{ansatz0}
\end{equation}
where we have not written the $SU(2)$ indices. Here $\bar\Sigma_\mu^\nu$ is a two by two ant-symmetric matrix that is anti-self-dual with respect to the indices $\mu\nu$.

If we embed the $SU(2)$ gauge group in $SO(4)$ and denote $SO(4)$ indices by $m, n=1,\cdots, 4$ then a convenient representation for $\bar\Sigma_\mu^\nu$ is given by
\begin{equation}
    \bar\Sigma_{\mu\nu}^{mn}=\frac{1}{2}(\delta_{\mu\nu}^{mn}-\epsilon_{\mu\nu}^{mn})
\end{equation}
that is anti-self-dual in $\mu\nu$ as well as $mn$ indices.

If we substitute above ansatz in Eqn.(\ref{selfdual0}) we get a simple equation for $\phi(x)$
\begin{equation}
    \frac{1}{\phi(x)}\square\phi(x)=0.
\end{equation}

When $\phi(x)$ is non-singular then it is constant and then the gauge field vanishes because of (\ref{ansatz0}). On the other hand if $\phi(x)$ has singularities then these might cancel with those of $\square\phi(x)$. Such a solution looks like following
\begin{equation}
    \phi(x)=1+\sum_{i=1}^{k}\frac{\rho_i}{(x-x_i^0)^2}.
\end{equation}

Here 5$k$ parameters are $(x^0_{\mu i}, \rho_i)$ and three parameters are the overall displacements of the position.

With above specifications the $k=1$ instanton gauge field looks like as follows
\begin{equation}
    A_\mu=-2\rho^2\bar\sigma_{\mu\nu}\frac{x^\nu-x^{0\nu}}{(x-x^0)^2((x-x^0)^2+\rho^2)}.
\end{equation}

\subsection{Bosonic Sigma Models}\label{bosonic0} 

Above we have discussed solitons and instantons as solutions to field theories. Next we take up their realization as solutions first to bosonic string equations and then then to superstring equations.

Propagation of (super)strings on curved backgrounds is studied using the so called sigma models that can be linear or non-linear, chiral and non-chiral as well as bosonic or supersymmetric. We shall begin our journey by collecting the information about their field content and construction. Sigma models have their use in nuclear physics, particle physics, statistical mechanics and condensed matter physics. Needless to mention that our focus will be on their utilization to illustrate the dynamics of (super)strings.

Non-linear sigma models possess several very convenient properties. These are  renormalizable in a non-trivial way. These possess chiral fermions and hence provide a tool for realistic particle physics model building. Chiral sigma models possess anomalies and one can devise ways to cancel them. 

The action for bosonic string theory propagating in the gravitational ($G_{\mu\nu}$), ant-symmetric tensor ($B_{\mu\nu}$) and Fradkin-Tseytlin dilaton \cite{Fradkin:1985ys}($\Phi$) field backgrounds is given by the  non-linear sigma model (NLSM) on a two-dimensional surface $\Sigma$ with intrinsic metric $\gamma^{mn}$.
\begin{eqnarray}
        S_{NLSM} &=& S_G+S_B+S_{\Phi}\nonumber\\
    S_G&=&\frac{1}{4\pi\alpha'} \int_\Sigma d^2 x \sqrt{\gamma}\gamma^{mn}G_{\mu\nu}(X)\partial_mX^\mu\partial_nX^\nu\nonumber\\
    S_B&=&\frac{1}{4\pi\alpha'} \int_\Sigma d^2 x \epsilon^{mn}B_{\mu\nu}(X)\partial_mX^\mu\partial_nX^\nu\nonumber\\
    S_{\Phi}&=&\frac{\alpha'}{4\pi\alpha'}\int_\Sigma d^2 x \sqrt{\gamma}R^{(2)}\Phi(X).
    \label{action1}
\end{eqnarray} 
Here $d^2 x=d\sigma d\tau$ and the scalars $X^\mu(x)$'s, $\mu = 1, 2, \cdots, D$ map the string onto a $D$ dimensional spacetime $\mathcal{M}$ called the target manifold \cite{Callan:1985ia}. The dimensional coupling constant $\alpha'$ turns out to be the string tension. The essential consistency requirement for the above NLSM as a quantum field theory is that it be scale invariant. This in turn demands tracelessness of the  following world-sheet stress-energy tensor.
\begin{equation}
    T_{mn}=\frac{4\pi}{\sqrt{\gamma}}\frac{\delta S_{NLSM}}{\delta\gamma_{mn}}.
\end{equation}

The scale invariance is explicitly broken by the dilaton and implicitly by the gravitational and anti-symmetric background field. The expression for the world-sheet stress-energy tensor turns out to be
\begin{equation}
 2\pi T^n_n= \beta^\Phi\sqrt{\gamma}R^{(2)}+ \beta^G_{\mu\nu}\sqrt{\gamma}\gamma^{mn}\partial_mX^\mu\partial_nX^\nu + \beta^B_{\mu\nu}\sqrt{\gamma}\gamma^{mn}\partial_mX^\mu\partial_nX^\nu
\end{equation}
where the dilaton, gravitational field and the anti-symmetric tensor field beta functions, $\beta^\Phi$, $\beta^G_{\mu\nu}$ and $ \beta^B_{\mu\nu}$ respectively, are local functionals of $G_{\mu\nu}, B_{\mu\nu}$ and $\Phi$.

The beta functions can be calculated either in conformal or light-cone gauge. In the conformal gauge $\gamma_{mn}=e^{2\sigma}\delta_{mn}$ (here the conformal factor $\sigma$ should not be confused the string spatial coordinate). If we work in complex co-ordinates $z$ and $z'$ then world sheet metric becomes

\begin{equation}
\begin{bmatrix}
    0 &  1\\
    1 & 0
\end{bmatrix}.
\end{equation}

After performing all the index algebra in two dimensions and then continuing to the $D$ dimensions where $D$ is the dimension of the spacetime manifold $\mathcal{M}$, we get the following expression for the effective action.
\begin{eqnarray} 
S^D &=& S_G^D+S_B^D+S_{\Phi}^D\nonumber\\
     S_G^D&=&\frac{1}{4\pi\alpha'} \int d^2xe^{(D-2)\sigma}G_{\mu\nu}(X)\partial X^\mu\bar\partial X^\nu\nonumber\\
    S_B^D&=&\frac{1}{4\pi\alpha'} \int d^2xe^{(D-2)\sigma}B_{\mu\nu}(X)\partial X^\mu\bar\partial X^\nu\nonumber\\
    S_{\Phi}^D&=&-\frac{\alpha'}{4\pi\alpha'}\int d^2xe^{(D-2)\sigma} 4\partial\bar\partial\sigma.
    \label{action8}
\end{eqnarray} 

The expressions for the non-linear sigma model beta functions turn out to be
\begin{eqnarray}
   \beta^\Phi&=&\frac{D-26}{48\pi^2}+\alpha'\left[\frac{1}{16\pi^2}\left\{4(\nabla \Phi)^2-4\nabla^2\Phi- R +\frac{1}{12}H^2  \right\}+O(\alpha')\right], \nonumber\\
   \beta^G_{\mu\nu}&=& R_{\mu\nu}-\frac{1}{4}H_\mu^{\lambda\sigma}H_{\nu\lambda\sigma}+O(\alpha')\nonumber\\
   \beta^B_{\mu\nu}&=&\nabla_\lambda H^\lambda_{\mu\nu}-2(\nabla_\lambda\Phi)H^\lambda_{\mu\nu}+O(\alpha'). \label{beta1}
\end{eqnarray}
Here wee have defined the anti-symmetric field strength tensor $H_{\mu\nu\lambda}=3\nabla_{[\mu} B_{\nu\lambda ]}$ and $R_{\mu\nu}$ is the Ricci tensor.

These equations can be derived by the variation of the following action
\begin{equation}
 S^S_{eff}\propto   \int d^D\sqrt{G}e^{-2\Phi}\left\{ R+(\nabla\Phi)^2-\frac{1}{12}H^2\right\}.\label{action9}
\end{equation}
Here the superscript $S$ means that the action is written in string metric $G^S_{\mu\nu}=G_{\mu\nu}$.

If we define the Einstein metric 
\begin{equation}
    G_{\mu\nu}^E=e^{4\Phi/(D-2)}G_{\mu\nu}
\end{equation}
then the action (\ref{action9}) changes to 
\begin{equation}
    S^E_{eff}\propto   \int d^D\sqrt{G}\left\{ R-\frac{4}{D-2}(\nabla\Phi)^2-\frac{1}{12}e^{-8\Phi/(D-2)}H^2\right\}.\label{action10}
\end{equation}
This is called the Einstein metric effective action because it is closer the the Einstein-Hilbert action owing to the absence of the $e^{-2\Phi}$ factor before the scalar curvature term.

In case of heterotic strings we have to take care of the fermions and the Yang-Mills background and the effective action in $D=10$ is the $N=1$ supergravity and super Yang-Mills action 
\begin{eqnarray}
 S_{eff}^{het}&=&\int d^D\sqrt{G}\left\{ R-\frac{1}{D-2}(\nabla\Phi)^2-\frac{1}{12}e^{-8\Phi/(D-2)} H^2\right. \nonumber\\
 &+&\left.\frac{1}{2}\alpha'e^{-4\Phi/(D-2)}(R^2-tr F^2)\right\}.\label{action11}
\end{eqnarray}
Here $R^2=R_{\mu\nu\lambda\sigma}R^{\mu\nu\lambda\sigma}$ and the anti-symmetric tensor field strength has been shifted by an $\alpha'$ order term
\begin{equation}
    H^{new}=H^{old}+\frac{1}{8}\alpha'tr(F\wedge A)-\frac{1}{8}\alpha'tr(R\wedge \omega).
\end{equation}

\subsection{Conventional Superspace}\label{fermionic}

This note is about harmonic superspace off-shell formalism for Ali-Ilahi's complementary ADHM instanton sigma model. This raises two questions. First question concerns the conventional superspace and its use for off-shell formalism of sigma models. Second question concerns why are we focusing on harmonic superspace. In this Sub-appendix we shall focus upon the conventional superspace and its use for sigma models. Second question was already answered by Galperin and Sokatchev. We shall rehash their answer with some more details in Sub-appendix \ref{background}.

In this Sub-appendix we shall focus upon conventional superspace and its use in case of sigma models.

Witten's original ADHM instanton sigma model has $(0, 4)$ supersymmetry and so does Ali-Ilahi's complementary model. Because of this our natural concern will be sigma models with higher supersymmetries and their off-shell formalism.

We pick up our story with a paper by Howe and Papadopoulos \cite{Howe:1987qv}. Their focus is on conditions for non-linear sigma models in two dimensions to have $(p, q)$ supersymmetries. They show how to construct off-shell $(p, q)$ superfields for a wide class of models. 

Ref.\cite{Howe:1987qv} is a milestone in construction of supersymmetric non-linear sigma models. In this section we shall collect useful information from this paper ((see also \cite{Howe:1992tg, Howe:1988cj}). Adding supersymmetry to sigma models makes the structure of these models very rich. Thus though any manifold can accommodate a supersymmetry increasing the number of supersymmetries puts restrictions on the structure of the target manifold owing to complex structure being intimately related to it. For example  the target manifold must be a K\"ahler one for the sigma model to have $(2, 2)$  supersymmetry and a $(4, 4)$ supersymmetry of the sigma model demands that the target manifold be hyper-K\"ahler\cite{Alvarez-Gaume:1981exv}. In general sigma models can have different number of left and right moving supersymmetries $(p, q)$. The notation derives from the fact that left moving fields (functions of $z$) and right moving fields (functions of $\bar z$) are dynamically independent \cite{Callan:1991at}.  These models can have extended supersymmetry. They also carried forward their analysis to ultra-violet power counting rules for supersymmetric non-linear sigma models. They found that $(4, q)$ models with $0\leq q\leq 4$ are ultra-violet finite. The latter issues are not relevant for the present note. (However see \cite{Ali:2023zxt}).

The supersymmetry algebra in the general $(p, q)$ case is
\begin{eqnarray}
    \{Q^I_+, Q^J_+\} &=&2\delta^{IJ} P_+,\;  I, J=1,\cdots, p\nonumber\\
    \{Q^I_-, Q^J_-\} &=&2\delta^{IJ} P_-,\;  I, J=1,\cdots, q\nonumber\\
    \{Q^I_-, Q^J_-\} &=& 0\label{susy7}
\end{eqnarray}
where
\begin{equation}
    P_\pm=(P_0\pm P_1)/\sqrt{2}.
\end{equation}

Howe and Papadopoulos formulate the non-linear sigma models with higher supersysmmetries in terms of constrained superfields.

They begin by discussing $(p, 0)$ models in terms of $(1, 0)$ superfields. The action for the $(1, 0)$ sigma model is
\begin{equation}
    S=-i\int d^2xd\theta^+(g_{ij}+b_{ij})D_+\phi^i\partial_-\phi^j,\label{action16}
\end{equation}
with
\begin{eqnarray}
    D_+&=&\frac{\partial}{\partial\theta^+}+i\theta^+\partial_-\nonumber\\
    D_+^2&=&i\partial_-\nonumber\\
    \phi^i(x, \theta)&=&\phi^i(x)+\theta^+\lambda^i_+.
\end{eqnarray}

The condition for the action of Eqn.(\ref{action16}) to have $(2, 0)$ supersymmetry is
\begin{equation}
    \delta\phi^i=\zeta^+{J^i}_jD_+\phi^j\label{susy8}
\end{equation}
where ${J^i}_j$ is a $(1, 1)$ tensor on the target manifold $\mathcal M$ and $\zeta^+$ is the supersymmetry parameter. Closure of the symmetry algebra requires that $J$ be a complex structure on $\mathcal M$
\begin{equation}
    {J^i}_j{J^i}_k=\delta^i_k
\end{equation}
and vanishing of Nijenhuis tensor
\begin{equation}
    {N^i}_{jk}=0.
\end{equation}
If $g$ is a Hermitian metric then
\begin{equation}
    g_{kl}{J^k}_i{J^l}_j=g_{ij}
\end{equation}
implies and is implied by
\begin{equation}
    J_{ij}=-J_{ji}.
\end{equation}
In addition if $J$ is covariantly constant, that is,
\begin{equation}
    \nabla_i{J^j}_k=0
\end{equation}
then the action is invariant under supersymmetry transformations (\ref{susy8}).

If we take $\zeta$ in (\ref{susy8}) to be constant then the classical action is conformally invariant. Due to complex structures the dimension $n$ of the target maniford $\mathcal M$ is even.

For each additional supersymmetry there will be a complex structure
\begin{equation}
    J_r,~~~r=1, 2, \cdots
\end{equation}
obeying above conditions. Commutation of supersymmetry transformations leads to 
\begin{equation}
    \{J_r, J_s\}=0,~~r\neq s.
\end{equation}
From this we conclude that given two complex structures, $J$ and $K$, we get a third one
\begin{equation}
    L=JK.
\end{equation}
Interesting models have $(2, 0)$ and $(4, 0)$ supersymmetry. In the latter case the complex structures obey the relation
\begin{equation}
J_rJ_s=-\delta_{rs}+\epsilon_{rst}J_t.
\end{equation}

No auxiliary field is needed and the supersymmetry algebra closes without the use of equations of motion. Thus the $(1, 0)$ superfields can be converted into $(p, 0)$ superfields.

Since this note is about the harmonic superspace for $(4, 0)$ supersymmetry of Witten's original and Ali-Ilahi's complementary ADHM instanton sigma model we shall now include some more details of conventional superspace for $(2, 0)$ and $(4, 0)$ supersymmetry.

Let us discuss the case of conventional off-shell $(2, 0)$ superfield formalism. 

In this case we take two Grassmann coordinates : one is $\theta_0^+$ (the original one) and the other is $\theta_1^+$. The constraint on the $(2, 0)$ superfield is now
\begin{equation}
   D_{1+}\phi^i={J^i}_jD_{0+}\phi^j.\label{constraint0}
\end{equation}
This can be written in another form
\begin{equation}
    \bar\Delta_+\phi^i=i{J^i}_j\bar\Delta_+\phi^j,\label{constraint1}
\end{equation}
with
\begin{equation}
    \Delta_+=\frac{1}{2}(D_{0+}-iD_{1+}).\label{constraint2}
\end{equation}
Geometry of the target manifold $\mathcal M$ makes these consistent. The off-shell $(2, 0)$ superspace action is 
\begin{equation}
    S=-i\int d^2x\{(g_{ij}+b_{ij})D_{0+}\phi^i\partial\phi^j\},\label{action2}
\end{equation}
Which is the same as action (\ref{action16}) but with constrained $(2, 0)$ superfields.

This formalism can be extended to $(4, 0)$ constrained superfields with minimal effort. 

We define the covariant derivatives $D_{r+}$, $r=1, 2, 3$. The constraints are
\begin{equation}
   D_{r+}\phi^i={J^i_r}_jD_{0+}\phi^j,\label{constraint3}
\end{equation}
with
\begin{eqnarray}
    \Delta_{1+}&=&\frac{1}{2}(D_{0+}-iD_{1+}),\nonumber\\
    \Delta_{2+}&=&\frac{1}{2}(D_{2+}-iD_{3+}).\label{constraint4}
\end{eqnarray}
These can be rewritten as follows:
\begin{eqnarray}
\bar\Delta_{1+}\phi^i&=&i{J^i_1}_j\bar\Delta_{1+}\phi^j,\nonumber\\
\bar\Delta_{2+}\phi^i&=&i{J^i_1}_j\bar\Delta_{2+}\phi^j,\nonumber\\
\Delta_{1+}\phi^i&=&i{J^i_2}_j\bar\Delta_{2+}\phi^j.\label{constraint5}
\end{eqnarray}
The action is again given by (\ref{action2}) but this time interpreted in terms of $(4, 0)$ superfields. Because of the constraints (\ref{constraint0}) and (\ref{constraint3}) the components of the $(p, 0)$ superfields consist only of the fields $\phi^i(x)$ and $\lambda^i_x(x)$.

Howe and Papadopolus extended their formalism to $(1, 1)$ and $(p, q)$ superfields too but we shall leave that out in this note. In Ref.\cite{Howe:1988cj} Howe and Papadopolulos further analyzed the geometry of two dimensional non-linear sigma models. In this reference they analyzed the conditions for linearisability of the constraints on the superfields. The conditions involve generalized Nijenhuis tensors and curvature tensors constructed out of complex structures on target manifold.
In this reference we get glimpse of Yang-Mills structure in non-linear sigma models with higher supersymmetries. In the conformal limit, in our view, these will be the $SU(2)$ symmetries that we see for $N=3$ and small, middle and large $N=4$ superconformal algebras\cite{Ademollo:1976wv, Ivanov:1987mz, Ivanov:1988rt, Sevrin:1988ew, Ali:1993sd, Ali:2000zu, Ali:2000we, Ali:2003aa}.

\subsection{CHS Solutions and Sigma Models}\label{chs0} 

Now we collect some information about Callan, Harvey, Strominger instanton solutions followed by their sigma models. In Refs.\cite{Strominger:1986uh, Strominger:1990et, Callan:1991dj, Callan:1991ky, Callan:1991at} these authors found various classical stringy soliton solutions and studied supersymmetric instanton sigma models. Our focus will be on instantonic aspects. 

Callan, Harvey and Strominger have discussed stringy realizations of instanton solutions to supersymmetric field theories. They start with a discussion of solitonic solutions to superstrings in background fields.

Beta functions for superstrings propagating in massless background fields are the equations following from a master action that can be computed as a series in $\alpha'$. In case of heterotic strings this action is the same as for $D=10$, $N=1$ supergravity with super Yang-Mills (SYM) theory. This is given below
\begin{equation}
    S=\frac{1}{{\alpha'}^4}\int d^{10}\sqrt{-g}e^{-2\phi}(R+4(\nabla\phi)^2-\frac{1}{3}H^2-\frac{\alpha'}{30}{\rm Tr} F^2).
\end{equation}
This time the three form anti-symmetric tensor field strength is related to the  two-form potential by the equation
\begin{equation}
    H=dB+\alpha'(\omega_3^L(\Omega_-)-\omega_3^{YM}(A))+\cdots.
\end{equation}
Here $\omega_3^L$ is the Chern-Simons three form and we have
\begin{equation}
    dH=\alpha'({\rm Tr}R\wedge R-\frac{1}{30}{\rm Tr}F\wedge F)
\end{equation}
and  $\Omega_\pm$ is the non-Riemannian connection
\begin{equation}
    \Omega^{AB}_{\pm M}={\omega_M}^{AB}\pm {H_M}^{AB}.\label{ooh}
\end{equation}

This is where these authors take the following revolutionary step. Rather than solving the equations of motion directly they look for those backgrounds that annihilate the supersymmetric variations of fermions:
\begin{eqnarray}
    \delta\chi&=&F_{MN}\gamma^{MN}\epsilon,\nonumber\\
    \delta\lambda&=&\gamma^M\delta_M\phi-\frac{1}{6}H_{MNP}\gamma^{MNP},\nonumber\\
    \delta\psi_M&=&(\partial_M-\frac{1}{4}\Omega^{AB}_{-M}\gamma_{AB})\epsilon.\label{variations0}
\end{eqnarray}
Vanishing of these variations gives us first order partial differential equation. It is easier to construct ansatz to solve these rather than solving the original equations of motion that are second order coupled partial differential equations and more difficult to solve than Einstein field equations.

There are solitonic solutions to the vanishing of variations (\ref{variations0}). Analogy with the Dirac monopole in electromagnetic theory lead Callan, Harvey and Strominger to the five brane solution as dual solution to strings. This in turn leads to a connection with the D1-D5 system.

This five brane solution is related to instantons in four dimensions. We no longer have $H=dB$ but
\begin{equation}
    H=dB-\frac{\alpha'}{30}\omega_3^{YM}.\label{h0}
\end{equation}
There is a gravitational term too. From (\ref{h0}) we get
\begin{equation}
    dH=-\frac{\alpha'}{30}{\rm Tr}(F\wedge F).\label{dh0}
\end{equation}
The dual equation for Dirac monopole analogy is
\begin{equation}
    d^*H=0.\label{h1}
\end{equation}
In Eqn.(\ref{dh0}) Yang-Mills topological charge serves as the magnetic source term for $H$.

For the five brane solution the $SO(9, 1)\supset SO(5, 1)\oplus SO(4)$
positive chirality Majorana-Weyl spinor decomposes as
\begin{equation}
    16\rightarrow (4_+, 2_+))\oplus(4_-, 2_-).
\end{equation}
To make the supersymmetric variations the ansatz for the transverse metric and anti-symmetric tensor fields are
\begin{eqnarray}
    g_{\mu\nu}&=&e^{2\phi}\delta_{\mu\nu},~~~~ \mu, \nu=6,\cdots, 9\label{g0}\\
    H_{\mu\nu\lambda}&=&-{\epsilon_{\mu\nu\lambda}}^\sigma\partial_\sigma\phi\label{h2}
\end{eqnarray}
where $\phi$ is the dilaton field. The vierbein and the generalized spin connections are
\begin{eqnarray}
    e^m_\mu&=&\delta^m_\mu e^\phi,\label{vier0}\\
    \Omega_{\mu mn}&=&\delta_{m\mu}\partial_ n\phi-\delta_{n\mu}\partial_m\phi\mp{\epsilon_{\mu mn}}^\rho\partial_\rho\phi.
\end{eqnarray}
This ansatz leads to
\begin{equation}
  {\Omega_\mu}^{mn}\gamma_{mn}\epsilon_\eta=2({\gamma_\mu}^\rho\partial_\rho\phi)(1\pm\eta 1)\epsilon_\eta 
\end{equation}
where $\eta=\pm$ and $\Omega_\pm$ annihilates the $(4_+, 2_+)$ spinor.

Half of the supersymmetries are unbroken and the other half are associated with fermion zero modes bound to the soliton.

We still have to specify the dilaton. For this we begin with the curl of $H$ given by
\begin{equation}
    dH=-\frac{1}{2}*\square e^{-2\phi}=\alpha'(Tr R\wedge R-\frac{1}{30}Tr F\wedge F).\label{dh1}
\end{equation}

Perturbation theory arguments lead to the following expression for the dilaton field
\begin{equation}
e^{2\phi}=e^{2\phi_0}+8\alpha'\frac{x^2+2\rho^2}{(x^2+\rho^2)^2}+{\mathcal O}(\alpha'^2). \label{dilaton0}
\end{equation}

Callan, Harvey and Strominger call it the {\it gauge} solution. Corresponding instanton number and axion charge are
\begin{eqnarray}
    \nu&=&\frac{1}{480\pi^2}\int {\rm Tr}F\wedge F,\label{nu0}\\
    Q&=&-\frac{1}{2\pi^2}\int H\label{q0}
\end{eqnarray}
respectively.

The other solution discussed by Callan-Harvey-Strominger is called the {\it symmetric} solution. Its specifications are as follow
\begin{equation}  ds^2=e^{2\phi}\delta_{\mu\nu}dx^\mu dx^\nu+\eta_{\alpha\beta}dy^\alpha dy^\beta,\label{metric0}
\end{equation}
\begin{equation}  H_{\mu\nu\lambda}=-{\epsilon_{\mu\nu\lambda}}^\rho\partial_\rho\phi\label{anti0}
\end{equation}
and
\begin{equation}
    {F_{\mu\nu}}^{mn}={\tilde F_{\mu\nu}}^{mn}={R_{\mu\nu}(\Omega_-)}^{mn}.
\end{equation}

Our discussion in this Sub-appendix so far has been on solitonic solutions discovered by Callan, Harvey and Strominger. 

We now take up the task of covering some background on the sigma models investigated by the same authors.   

After going through bosonic string sigma models and corresponding effective actions and the equations of motion in Sub-appendix \ref{bosonic0} it is natural to to the supersymmetric case.  This is what we do next. The $(1, 0)$ superspace action for nonlinear sigma models is
\begin{equation}
    S_{HP1} = \int_\Sigma d^2xd\theta (g_{ij}+b_{ij})D_+\phi^i\partial_-\phi^j\label{action12}
\end{equation}
whose component form is
\begin{equation}
    S_{HP1}=\int_\Sigma d^2x [(g_{ij}+b_{ij}\partial_+\phi^i\partial_-\phi^ +\frac{i}{2}g_{ij}\lambda^i_+\nabla_-\lambda^j_+]\label{action13}
\end{equation}
with
\begin{equation}
    \nabla_-\lambda^i_+=\partial_-\lambda^i_++\Gamma^{+i}_{jk}\partial\phi^j_-\lambda^k_+
\end{equation}
and
\begin{equation}
    \Gamma^{\pm i}_{jk}=\Gamma^{i}_{jk}\pm\frac{1}{2}H^i_{jk},\;\;\;H^i_{jk}=3\partial_{[i}b_{jk]}
\end{equation}
where $\Gamma^{i}_{jk}$ is the Christoffel symbol.

Corresponding instanton and the sigma models were studied in detail by Callan-Harvey-Strominger in Refs.\cite{Strominger:1986uh, Strominger:1990et, Callan:1991dj, Callan:1991ky, Callan:1991at}. The 't Hooft instanton is a restricted construction while ADHM construction is a general construction.  In this sub-appendix we give a brief summary of the instanton sigma model constructed by Callan, Harvey and Strominger.
 
The supersymmetry algebra in the general case is defined in Eqn.(\ref{susy7}) and the equation after that.

The supersymmetric sigma models can either be discussed in component form or in terms of superfields living in superspace. 

For $N=p$ supersymmetry we need $p-1$ complex structures $J^r, r=1,2,\cdots, p-1$ with the property
\begin{equation}
    \{J^r, J^s\}=-\delta^{rs}.
\end{equation}
The complex structures on the left and right sectors are differentiated by $\pm$ signs and we have
\begin{equation}
    [J^{+r}, J^{-s}]=0.
\end{equation}

The two dimensional supersymmetric sigma model action for heterotic string that describes the dynamics of $D$ worldsheet bosons $X^M$, $D$ right moving fermions $\psi^M_R$ plus left moving fermions $\lambda_L^a$ which lie in a representation of $SO(32)$ or $E_8\times E_8$ is described by the action
\begin{eqnarray}
    \frac{1}{4\pi\alpha'}\int d^2\sigma\{G_{MN}(X)\partial_+X^M\partial_-X^N+2B_{MN}(X)\partial_+X^M\partial_-X^N\nonumber\\
    +iG_{MN}\psi^M_R{\mathcal D_-}\psi^N_R+i\delta_{ab}\lambda^a_L{\mathcal D_+}\lambda^b_L+\frac{1}{2}(F_{MN})_{ab}\psi^M_R\psi^N_R\lambda^a_L\lambda^b_L\}.\nonumber\\
    \label{action0}
\end{eqnarray}
Here
\begin{eqnarray}
    H&=&dB,\nonumber\\
    {\mathcal D_-}\psi^M_R&=&\partial_-\psi^M_R+{{\Omega_-N}^M}_A\partial_-X^N\psi^A_R,\nonumber\\
    {\mathcal D}_+\lambda^a_L&=&\partial_+\lambda^a_L+{{A_N}^a}_b\partial_+X^N\lambda^b_L.
\end{eqnarray}

The left-right symmetric sigma model corresponding to the five brane solution discussed in Sub-appendix \ref{chs0} has the action
\begin{eqnarray}
    \frac{1}{4\pi\alpha'}\int d^2\sigma\{G_{\mu\nu}(X)\partial_+X^\mu\partial_-X^\nu+2B_{\mu\nu}(X)\partial_+X^\mu\partial_-X^\nu\nonumber\\
    +iG_{\mu\nu}\psi^\mu_R{\mathcal D_-}\psi^\nu_R+iG_{\mu\nu}\lambda^\mu_L{\mathcal D_+}\lambda^\nu_L+\frac{1}{2}R(\Omega_+)_{\mu\nu\lambda\rho}\psi^\mu_R\psi^\nu_R\lambda^\lambda_L\lambda^\rho_L\}.\nonumber\\
    \label{action4}
\end{eqnarray}
There is an exchange symmetry here that depends upon the non-Riemannian relation
\begin{equation}
R(\Omega_+)_{\mu\nu\lambda\rho}=R(\Omega_-)_{\lambda\rho\mu\nu}
\end{equation}
which holds for (\ref{ooh}) when $dH=0$.

The basic $(1, 1)$ supersymmetry of the worldsheet model (\ref{action4}) is 
\begin{eqnarray}
\delta X^M&=&\epsilon_L\psi^M_R+\epsilon_R\psi^M_L,\nonumber\\
\delta\psi^A_L+\Omega_{+M}^A\delta X^M\psi^B_L&=&\partial X^A\epsilon_R+\cdots,\nonumber\\
\delta\psi^A_R+\Omega_{-M}^A\delta X^M\psi^B_R&=&\partial X^A\epsilon_L+\cdots.\label{susy0}
\end{eqnarray}
We are interested in extended supersymmetry. A second supersymmetry might have the following structure
\begin{eqnarray}
\hat\delta X^M&=&\epsilon_L{f_R(X)^M}_N\psi^N_R+\epsilon_R{f_L(X)^M}_N\psi^N_L,\nonumber\\
\hat\delta\psi^A_L+\Omega_{+M}^A\delta X^M\psi^B_L&=&-{f_L(X)^A}_B\partial X^B\epsilon_R+\cdots,\nonumber\\
\hat\delta\psi^A_R+\Omega_{-M}^A\delta X^M\psi^B_R&=&-{f_L(X)^A}_B\partial X^B\epsilon_L+\cdots.\label{susy1}
\end{eqnarray}
The tensors $f_{L, R}$ should be complex structures
\begin{equation}
    f^2_\pm=-1
\end{equation}
and covariantly constant with respect to corresponding connection
\begin{equation}
    {\mathcal D}^\pm_A{{f_\pm}^B}_C=\partial_A{{f_\pm}^B}_C+{\Omega_{AD}^{(\pm)}}^B{{f_\pm}^D}_C-{\Omega_{AC}^{(\pm)}}^D{{f_\pm}^B}_D=0.\label{cc0}
\end{equation}
For $(p, p)$ supersymmetry we have $p-1$ such complex structures that follow the Clifford algebra
\begin{equation}
    f^{(r)}_\pm f^{(s)}_\pm=-\delta_{rs}+\epsilon_{rst}f^{(t)}_\pm.\label{ca0}
\end{equation}
Each complex structure leads to a conserved current
\begin{equation}
 J_\pm^{(r)}={f_\pm^{(r)}}_{AB}\psi_\pm^A\psi^B_\pm   
\end{equation}
In case of (4, 4) supersymmetry these currents lead to an $SU(2)$ symmetry.

Let us define a tensor
\begin{equation}
    J_{AB}=-i\eta^\dagger\gamma_{AB}\eta
\end{equation}
where the four dimensional spinor $\eta$ has definite chirality, for example $\gamma_5\eta=\eta$, and has unit norm ($\eta^\dagger\eta=1$). It can be verified that it square to minus one
\begin{equation}
    {J_A}^B{J_B}^C=-{\delta_A}^C.
\end{equation}
With these we can construct three right and three left complex structures
\begin{eqnarray}
    J_1^+&=&\begin{pmatrix}
      i\sigma_2&0\\
      0&i\sigma_2
    \end{pmatrix},~~J_1^-=\begin{pmatrix}
       -i\sigma_2&0\\
      0&-i\sigma_2 
    \end{pmatrix},\nonumber\\
    J_2^+&=&\begin{pmatrix}
        0&1\\
        -1&0
    \end{pmatrix},~~J_2^-=\begin{pmatrix}
        0&-\sigma_3\\
        \sigma_3&0
    \end{pmatrix},\nonumber\\
    J_3^+&=&\begin{pmatrix}
        0&i\sigma_2\\
        i\sigma_2&0
    \end{pmatrix},~~J_3^-=\begin{pmatrix}
       0&-\sigma_1\\
       \sigma_1&0
    \end{pmatrix}.\label{cs1}
\end{eqnarray}

In reference \cite{Galperin:1995pq} Galperin and Sokatchev gave off-shell harmonic superspace formalism for 't Hooft instant non-linear sigma models of Callan, Harvey and Strominger \cite{Callan:1991at} described in this Sub-appendix.

\subsection{ADHM Instantons : CSW Perspective}\label{csw}  

Yang-Mills instantons have a daunting reputation because the impression is that these are solutions to second order coupled partial differential equations. Witten cleared  this confusion in Ref.\cite{Witten:1978qe} where he clarifies that the instantons are solutions to self-dual Yang-Mills equations that are first order partial differential equations. This was done immediate after ADHM's seminal paper \cite{Atiyah:1978ri}. 

Yet the mystery of ADHM instanton does not stop with this clarification. ADHM construction  is an ultra brief one. Plus it is incredibly linear construction. There is yet another connection of ADHM construction with complexity - Galperin and Sokatchev have made remarks about the connection of ADHM instantons with twistor construction. We shall not focus upon the twistor connection in this note and limit our explorations to field theory constructions. 

In this Sub-appendix we begin the task of making the ADHM construction accessible - with the help of some papers listed below.

Aspects of ADHM construction were disentangled mainly in two papers. One was by Christ-Stanton-Weinberg (CHW) \cite{Christ:1978jy} and the other by Corrigan-Fairlie-Goddard-Templeton (CFGT) \cite{Corrigan:1978ce}. In this sub-appendix  we shall review the aspects of the original ADHM instanton sigma model as expounded by Christ-Stanton-Weinberg. It is the place where we learn that we should not be intimidated by the algebro-geometric brevity of the original ADHM construction. He also learn that ADHM construction while being a solution to second order coupled partial differential equations the actual problem solved by ADHM is just the problem of inverting a non-square  matrix equation with quaternionic entries.

Though Galperin and Sokatchev do not focus on quaternionic connection of the ADHM construction of instantons Christ, Stanton and Weinberg do and so do. Thus we can say that the Galperin and Sokatchev approach and Christ-Stanton-Weinberg approach are complementary to each other.  For  Corrigan, Fairlie, Goddard and Templeton explanation refer to next Sub-appendix. We include some aspects of these two clarifications of the ADHM construction because we believe that this would add to the clarification of the construction.

In this sub-appendix we shall pick up the clarifications of the ADHM construction of instantons from Ref.\cite{Weinberg:2012pjx} rather than \cite{Christ:1978jy} (see also \cite{Tong:2005un}).

In Ref.\cite{Weinberg:2012pjx} the author first clarifies the ADHM construction first for the $SU(2)$ group and then he takes it up for the larger groups. We shall recapitulate the salient points here and believe that this adds to the exegesis.

To follow this exposition we need a brieff introduction to quaternions. Quaternions are generalizations of complex numbers involving three imaginary units $i, j$ and $k$. These anti-commute and obey
\begin{equation}
    i^2=j^2=k^2,~~~ijk=-1.\label{q1}
\end{equation}
An arbitrary quaternion $q$ is written as
\begin{equation}
   q=a+bi+cj+dk \label{q2}
\end{equation}
where $a, b, c$ and $d$ are real numbers. The quaternion is said to be real if only $a$ is non-zero. On the other hand if $a=0$ then it is purely imaginary.

The complex conjugate of above quaternion is defined by
\begin{equation}
    q^*=a-bi-cj-dk \label{q3}.
\end{equation}
Quaternion $q$ is called unit quaternion if it obeys
\begin{equation}
    q^*q=1.\label{q4}
\end{equation}

If we define
\begin{equation}
    \hat e_1=-i,\hat e_2=-j,\hat e_3=-k,\hat e_4=-1,
\end{equation}
then an arbitrary quaternion can be expanded as 
\begin{equation}
    q=q_r\hat e_r.
\end{equation}
With this convention the convention the quantity
\begin{equation}
    \chi_{rs}=\hat e_r\hat e_s^*-\hat e_s\hat e_r^*
\end{equation}
is self-dual. This can be proved by defining
\begin{eqnarray}
    e_r&=&i\sigma_r, r=1,2,3,\nonumber\\
    e_4&=&1_2,\nonumber\\
    e^\dagger_r&=&-i\sigma_r, r=1,2,3,\nonumber\\
    e^\dagger_4&=&1_2
\end{eqnarray}
where $\sigma_r$ are the Pauli matrices and using
\begin{eqnarray}
    \eta_{rs}&=&-i(e_re^\dagger_s-\delta_{rs}1_2),\nonumber\\
    \bar\eta_{rs}&=&-i(e^\dagger_re_s-\delta_{rs}1_2).
\end{eqnarray}

In view of the brevity of the ADHM construction the question arises about the expressions of the instanton gauge field as well as the field strength tensor in this construction. These are the issues spelt out in \cite{Christ:1978jy} and \cite{Weinberg:2012pjx}.

Let us begin with a $(k+1)\times k$ quaternionic matrix
\begin{equation}
    M=B-Cx
\end{equation}
with $B$ and $C$ being constant matrices while $x$ is a quaternionic representation of a point in Euclidean space
\begin{equation}
    x=x_r\hat e_r.
\end{equation}
Using a dagger for matrix transpose and quaternionic conjugation we get a $k\times k$ matrix
\begin{equation}
    R=M^\dagger M
\end{equation}
that is required to be real and invertible.

Then let us define a $k+1$ component column vector $N$ with the properties
\begin{equation}
    N^\dagger M=0
\end{equation}
and 
\begin{equation}
    N^\dagger N=I.\label{normalization0}
\end{equation}
Then the self-dual $SU(2)$ gauge field with instanton number $k$ is given by
\begin{equation}
    \mathcal{A}=\frac{1}{g}N^\dagger\partial_pN.\label{instanton0}
\end{equation}
Using Eqn. (\ref{normalization0}) the instanton expression can also be written as
\begin{equation}
    \mathcal{A}_p=-\frac{1}{g}(\partial_pN^\dagger)N.\label{instanton2}
\end{equation}
Since the expressions in Eqn. (\ref{instanton0}) and Eqn. (\ref{instanton2}) are minus conjugate of each other we conclude that the instanton gauge field $\mathcal{A}_p$ is purely imaginary. This can be expanded as
\begin{equation}
    \mathcal{A}_p=-\Sigma_{a=1}^{3}\frac{\hat e^a}{2}A_p^a.
\end{equation}
Corresponding self-dual field strength tensor is defined as
\begin{equation}
    \mathcal{F}_{pq}=\partial_p\mathcal{A}_q-\partial_q\mathcal{A}_p+g[\mathcal{A}_p, \mathcal{A}_q].\label{fst0}
\end{equation}
Or
\begin{equation}
    \mathcal{F}_{pq}=-\Sigma_{a=1}^{3}\frac{\hat e^a}{2}F_{pq}^a.\label{fst1}
\end{equation}
The algebraic proof of self-duality of the field strength tensor in Eqn.(\ref{fst0}) is given in Refs. \cite{Christ:1978jy, Tong:2005un} and \cite{Weinberg:2012pjx}.

It can be proved that the instanton number for above construction is $k$. Also the 't Hooft instanton with instanton number $k$ can be obtained in this way.

Weinberg also clarifies that the matrix $M$ is not uniquely determined by the gauge field. The gauge field remains unchanged under the transformations
\begin{eqnarray}
    M&\rightarrow& M'=SMT,\nonumber\\
    N&\rightarrow& N=SN.
\end{eqnarray}
Here $S$ are $x$-dependent matrices with $S$ a quaternionic matrix. This freedom can be used to put $C$ in the form
\begin{equation}
    C=\begin{pmatrix}
        0&0&\cdots&0\\
        1&0&\cdots&0\\
        0&1&\cdots&0\\
        \cdots&\cdots&\cdots&\cdots\\
        0&0&\cdots&1\\
    \end{pmatrix}.
\end{equation}

This does not eliminate all the freedom. Transformations with
the $(k+1)\times(k+1)$ matrix $S$ with
\begin{equation}
    S=\begin{pmatrix}
        u&0\\
        0&T^{-1}
    \end{pmatrix}
\end{equation}
with $u$ a unit quaternion and $T$ a $k\times k$ orthogonal matrix, is still allowed. 

There are some more specifications about the construction that can be made.

We shall leave out the counting of instanton parameters, specification of expressions for 't Hooft's SU(2) solution as well as the constructions for general groups.

\subsection{ADHM Instantons : CFGT Perspective}\label{cfgt}

ADHM construction is a complex one. Several people have clarified its various aspects for us. In this Sub-appendix we shall cover those aspects of ADHM construction that were clarified by Corrigan, Fairlie, Goddard and Templeton referred to above.

The explicit expression for instanton gauge field that we get out of Witten's original construction is closest to the one we have in the description given by Corrigan, Fairlie, Goddard and Templeton.

Atiyah-Drinfeld-Hitchin-Manin showed how to construct the general self-dual solution for an arbitrary compact classical group. Corresponding constant parameters satisfy certain quadratic constrains called the ADHM conditions. Their construction involves only elementary linear algebra and leads to expressions for gauge potentials rational functions of spatial coordinates.

The ADHM construction leads to the gauge potential of the form
\begin{equation}
    A_\alpha=-iv^+\partial_\alpha v\label{adhm1}
\end{equation}
where $v$ is a matrix function of spatial co-ordinates. The spatial co-ordinates might have real, complex or quaternionic entries depending on the group considered. Function $v$ obeys the normalization condition
\begin{equation}
    v^+v=1.\label{normalization}
\end{equation}
It also follows the linear conditions
\begin{equation}
    v^+(a_{iB}+b_{iA}x_{AB})=0.\label{linear0}
\end{equation}
Here $a_{iA}$ and $b_{iA}$ are the constant matrix parameters of the solution. These obey certain quadratic conditions.

The quaternionic representation of the position co-ordinate is
\begin{equation}
    x=x_0-i\mathbf{x}\cdot\mathbf{\sigma}.\label{quaternion1}
\end{equation}
The range of the index $i$ and the shapes and sizes of $v$, $a$ and $b$ depend on group and instanton number $k$. Equations (\ref{normalization}) and (\ref{linear0}) determine $v=v(x)$ up to an arbitrariness corresponding to gauge freedom.

Above description can be made more specific for specific symplectic, unitary and orthogonal groups. The specification for symplectic and unitary groups is particularly simple.

For symplectic groups, $Sp(n)$, the Eqn. (\ref{linear0}) takes the form
\begin{equation}
    v^+\Delta(x)=0\label{linear3}
\end{equation}
with
\begin{equation}
    \Delta_{\lambda i}(x)=a_{\lambda i}+b_{\lambda i}x
\end{equation}
where the ranges of the indices are $1\leq i\leq k$, $-(n-1)\leq\lambda\leq k$.

the compact form of the symplectic group $Sp(n)$ is the group of $2n$ dimensional unitary matrices in the form of $n\times n$ array of quaternions, quaternions being just the arbitrary multiples of $SU(2)$ matrices. Particularly the group $Sp(1)$, the group of unit quaternions is just the group $SU(2)$.

The quantities $\Delta, a$ and $b$ are $(n+k)\times k$ matrices of quaternions with the normalization condition given by the Eqn. (\ref{normalization}).

The quantities $a^+a, b^+b$ and $b^+a$ are $k\times k$ quaternion matrices. Their symmetry is sufficient for Eqn. (\ref{adhm1}) to be a self-dual solution to Yang-Mills equations. The solution is non-singular if $k$ columns of $\Delta$ are linear independent of each other for each value of $x$.

The transformation
\begin{equation}
    v(x)\rightarrow v'(x)=v(x)\gamma(x)
\end{equation}
corresponds to gauge invariance of $A_\alpha$. Here $\gamma(x)$ is an $n\times n$ unitary quaternionic matrix with arbitrary spatial dependence.

The description of ADHM instantons for unitary groups $U(n)$ is even simpler. This time $v$ is a $(n+2k)\times n$ complex matrix that satisfies Eqn. (\ref{linear3}). $\Delta=\Delta_{\lambda,iA}$ is a $(n+2k)\times 2k$ matrix of the form
\begin{equation}
    \Delta_A(x)=a_A+b_Bx_{BA}
\end{equation}
with $1\leq i\leq k$, $-(n-1)\leq\lambda\leq 2k$ and $A=1, 2$. Here $a_A$ and $b_A$ are complex $(n+2k)\times k$ matrices. Here too $v$ should satisfy the normalization condition in Eqn. (\ref{normalization}). This time the conditions on $a$ and $b$ are such that Eqn. (\ref{adhm1}) yields a self-dual solution are
\begin{equation}
a^+_Aa_B=\mu\delta_{AB},~~b^+_Ab_B=\nu\delta_{AB}, ~~\epsilon_{AB}b^+_Ba_C=\epsilon_{BC}a^+_Bb_A.
\end{equation}
where  $\mu$ and $\nu$ are $k\times k$ Hermitian matrices and
\begin{equation}
    \epsilon_{AB}=-\epsilon_{BA}, ~~\epsilon_{12}=1.
\end{equation}
Once again the condition for non-singularity of the resulting solution is that $2k$ columns of $\Delta$ be linearly independent for each $x$.

The covariant derivative is defined as 
\begin{equation}
    D_\alpha=\frac{\partial}{\partial x_\alpha}+v^+\frac{\partial v}{\partial x_\alpha}.
\end{equation}

The gauge invariant expression for the field strength tensor is
\begin{equation}
    iF_{\alpha\beta}=[D_\alpha, D_\beta]=P[\partial_\alpha P, \partial_\beta P]P
\end{equation}
with
\begin{equation}
    P(x)=v(x)v^+(x).
\end{equation}

In this note we are mainly concerned with the expressions in Eqn. (\ref{adhm1}) and and Eqn. (\ref{linear0}). For the proof of the self-duality of the field strength tensor we refer to Ref.\cite{Corrigan:1978ce}. Once again we shall not be concerned with the completeness of the ADHM solutions in spite of its importance.

\subsection{Linear Sigma Models}\label{linear}

This note is about off-shell harmonic superspace formalism of Ali-Ilahi's ADHM instanton sigma model that is complementary to Witten's original ADHM instanton sigma model in the sense of being related to it by an in principle simple duality. Like the original model the complementary model is a linear sigma model. The linear sigma models have a potential for the scalar fields. These  are less investigated than their non-linear counterparts. 

In this Sub-appendix we shall collect some background on these. Our choice of the material in this Sub-appendix is impressionist in view of the focus this note being about ADHM instanton sigma models and not the linear sigma models in general.

The simplest term in potential for a scalar field can be the mass term. The mass term for scalar fields occurs naturally in the process of   regularization of infrared divergences \cite{Alvarez-Gaume:1981exa}. The ultraviolet divergence in terms of the dimensional regularization parameter $\epsilon=n-2$ is  defined as follows:
\begin{equation}
   -i{\rm log}\Omega[\phi^1] =\frac{1}{2}\int d^2xC_{ab}\langle\xi^a(x_1) \xi^b(x_2)\rangle=\frac{1}{2}\int d^2xC_{ab}i\delta^{ab}\Delta_F(x_1-x_2).
\end{equation} 
After that we had computed  the Euclidean momentum integral 
\begin{eqnarray}
I&\equiv&i\Delta_F(0)\nonumber\\
&=&\frac{1}{(2\pi)^n}\int d^nk_E\frac{1}{k^2_E+\mu^2}\nonumber\\
&\rightarrow&\lim\limits_{\epsilon\to 0}\frac{-1}{2\pi\epsilon} + {\rm finite}.
\end{eqnarray}
Here the mass term helps us in the regularization of the integral and disappears in the process of evaluation.

In Ref.\cite{Alvarez-Gaume:1983uye} Alvarez-Gaum\'e and Freedman determined the most general potential terms compatible with $N=1, 2$ and $4$ supersymmetry in two dimensions. The geometry of the internal manifold plays a crucial role in these constructions. 

In case of bosonic sigma models the potential $V(\phi)$ can be an arbitrary function on the defining Riemannian manifold $\mathcal M.$ In case of $N=1$ supersymmetric sigma model we can define an arbitrary super-potential $W(\phi)$ but this still leaves the possibility of parity non-conserving terms if $\mathcal M$ posses Killing vectors.

It is well known that while there is no restriction on the geometry of the internal manifold in the $N=1$ case it has to K\"ahler in $N=2$ case and hyper K\"ahler in $N=4$ case. This puts additional constraints on the potential terms. These constraints involve the Killing vectors that are holomorphic with respect to the complex structures of $\mathcal M$. 

All these aspects were analyzed in above reference. There are two methods to obtain the potential terms.

In the first method a general ansatz about the structure of the Lagrangian is made and requirement of supersymmetry is used to determine the structure of the terms. This procedure gives the most general potential terms.

In the second method a modification of the superspace methods is used to determine the structure of the potential terms. This time we do not get the most general terms.

In these papers the torsion term was missing.
In Ref.\cite{Hull:1993ct, Papadopoulos:1994tn} Hull, Papdopoulos and Townsend determined the structures of scalar potentials for $(p, 0)$, (1, 1) and $(p, q)$ supersymmetric sigma models with torsion. The action for the bosonic sigma model and the second order differential equations of motion are determined by three tensors. First one is the metric tensor $g_{ij}$. The second one is three form tensor $H_{ijk}$ that depends on the two form anti-symmetric tensor $b_{ij}$. Finally we have the scalar potential $V(\phi)$.

In case of supersymmetric sigma models the scalars are promoted to scalar superfields ${\phi^i(x, \theta^+)}$; ${i=1, \cdots, D}$ and the spinor superfields ${\psi^a(x, \theta^+)}$; ${a=1, \cdots, n}$ are introduced. 

The action in this case is
\begin{equation}
    S=\int d^2\sigma d\theta^+\{D_+\phi^i\partial_=(g_{ij}+b_{ij})+i\psi^a_-\nabla_+\psi^b_-h_{ab}+ims_a\psi^a_-\}.
\end{equation}

Scalar superfields $\phi^i$ define maps from the $(1, 0)$ superspace $\Sigma^{(1, 0)}$ onto $\mathcal M$. The fermionic superfield $\psi^a$ are sections of vector bundle $S_-\otimes\phi^*\xi$. Also $s_a(\phi)$ is a section of $\xi$.

The component action can be obtained by standard methods after elimination of the auxiliary fields. The potential in this case turns out to be
\begin{equation}
    V(\phi)=\frac{1}{4}m^2h^{ab}s_as_b.
    \label{potential0}
\end{equation}
It is well known that the generic sections of vector bundles over compact manifolds have isolated zeros. Thus for many models of interest $s_a(\phi)$ have isolated zeros. The potential vanishes at these and we get classical vacua. 

For more than one zeros there are interpolating solitonic solutions \cite{Abraham:1992vb, Abraham:1992qv}. Subject matter of this Sub-appendix is linear sigma models in general. The overall note is about linear ADHM instanton sigma models. Instantons have an interpretation as interpolating solutions between different vacua.

These authors also presented an extended superfield formalism of the $(p, 0)$ models and demonstrated that $(1, 1)$ models can be obtained from the $(1, 1)$-superspace formulation of the gauged but massless $(1, 1)$ sigma model.

The $(p, q)$ supersymmetry algebras is generated by the generators $Q_+^I; I=1, \cdots, p$ and $Q_-^{I'}; I'=1, \cdots, q$ and looks like
\begin{equation}
    \{Q^I_+, Q^J_+\}=2\delta^{IJ}P_+,~~\{Q^{I'}_-, Q^{J'}_-\}=2\delta^{I'J'}P_-,~~\{Q^I_+, Q^{I'}_-\}=2Z^{II'}.
\end{equation}

The $(p, 0)$ case can be analyzed in two ways. Either we use $(1, 0)$ superspace or  the generalized superspace method developed in Refs.\cite{Howe:1988cj, Howe:1987qv}. When supersymmetries of both chiralities are present there is the possibilities of central charges in the supersymmetry algebra. This complicates the process of finding the potential. In presence of central charges the standard superfield method are ineffective and therefore the authors of Ref.\cite{Hull:1993ct} used the superfield methods of \cite{Gates:1983py} and the geometry of the supersymmetric gauged sigma models of Ref.\cite{Hull:1989jk, Hull:1991uw}

For $m\neq 0$ condition supersymmetry puts restrictions of the couplings $g_{ij}$, $b_{ij}$, $h_{ab}$ and $s_a$
For example Lie derivatives of both $g_{ij}$ and $H_{ijk}$ w.r.t. $X^i$ must vanish, that is,
\begin{equation}
    \nabla_{(i}X_{j)}=0,\label{liex}
\end{equation}
\begin{equation}
    X^kH_{ijk}=\partial_{[i}u_{j]}\label{lieh}
\end{equation}
where $u$ is a locally defined 1-form and 
\begin{equation}
    s_i=u_i-X_i.\label{sux}
\end{equation}
Here the globally defined tensor $X^i$ occurs in the supersymmetry transformation of $\phi^i$
\begin{equation}
    \delta_\xi\phi^i=D_+\zeta e^i_a\psi^a_-+m\zeta X^i(\phi).\label{susy6}
\end{equation}
Using Eqn.(\ref{sux}) in Eqn.(\ref{potential0}) the structure of the potential in this case looks like:
\begin{equation}
    V=\frac{m^2}{4}g^{ij}(u-X)_i(u-X)_j.
\end{equation}

In Ref.\cite{Papadopoulos:1993mf} Papadopoulos and Townsend generalized these results to general ($p$, $q$) supersymmetric models including the (2, 2) and (4, 4) models. As expected the focus in initial papers on supersymmetric linear sigma models is on their construction. Studies of their properties received attention latter.

In Ref.\cite{Witten:1993yc} Witten studied the phase transitions in $N=2$ theories very extensively in two dimensions as the parameters of the theories are varied. He used linear sigma models that gave him freedom to analyze the issues. Here we are informed that at low energies the linear sigma models reduce to non-linear ones. He found a natural relation between sigma models based on Calabi-Yau hyper surfaces in weighted projective spaces and Landau-Ginzburg models. His construction gave him the known results about this correspondence and facilitated the discovery of new models. The famed topology change observation makes an appearance in this work.

His note is about the phases of $N=2$ supersymmetric theories. In this paper he exploited an idea in two dimensions  that was familiar in four dimensions for $N=1$ supersymmetric theories. He began by describing the idea in four dimensional context. Consider renormalizable gauge theories constructed from vector (gauge) multiplets and charged chiral multiplets. These contain a Fayet-Iliopoulos interaction D-term if the gauge group has a $U(1)$ factor. The structure of this term in terms of the vector superfield $V$ is
\begin{equation}
    -r\int d^4xd^4\theta V.
\end{equation}

The expression for the potential energies of the scalar components $s_i$, $i=1, \cdots, n$ of the chiral multiplets $S_i$ of chanrge $n_i$ is
\begin{equation}
    U(s_i)=\frac{1}{2e^2}D^2+\sum_i\left(\frac{\partial W}{\partial s_i}\right)^2.\label{potential2}
\end{equation}
Here $e$ is the gauge coupling, $W$ the super potential and 
\begin{equation}
    D=-e^2\left( \sum_in_is_i^2-r\right).\label{d-term}
\end{equation}
The parameter $r$ corresponds to the possibility of adding a constant to the Hamiltonians. As $r$ is varied phase transitions occur in the system. Typically this leaves the supersymmetry intact but changes the pattern of the scalars. Witten uncovered a very rich structure of these phase transitions including, as mentioned above, examples of topology change.

In Ref.\cite{Witten:1994tz} Witten gave his construction if the linear ADHM instanton sigma model. With this the narrative of the present Sub-appendix merges with the main narrative of this note. We are brought back to the main task of this note - to develop a harmonic superspace formalism for Ali-Ilahi's linear ADHM instanton sigma model with $(0, 4)$ supersymmetry.

In Ref.\cite{Lambert:1995dp} Lambert discussed two loop renormalization of the massive $(p, q)$ sigma models. Quantization of Witten's linear ADHM instanton sigma model was discussed by Lambert in Ref.\cite{Lambert:1995hs}. 
In this reference Lambert calculated $\beta$-functions of general massive $(p, q)$ supersymmetric sigma models to two loop order using $(1, 0)$ superfields. He discussed finiteness of mass terms in relation to $(p, q)$ supersymmetry. He also calculated the effective potential in one-loop order and considered the possibility of perturbative supersymmetry breaking in massive theory. He also considered the effect of one-loop finite counter terms and the UV behaviour of off-sell $(p, q)$ supersymmetric models to all orders of perturbation theory.

\section{Harmonic Superspace}\label{b}

The burden of this note is the explicitly covariant formalism of Ali-Ilahi's ADHM instanton sigma model using harmonic superspace off-shell formalism. Ali-Ilahi's ADHM complementary instanton sigma model is related to Witten's original ADHM instanton sigma model by a simple duality. Both Witten's original and Ali-Ilahi's complementary models are in component form. The off-shell covariant superspace formalism for Witten's original model was done by Galperin and Sokatchev using harmonic superspace. To take up the task of off-shell harmonic superspace formalism of Ali-Ilahis's ADHM sigma model we need background both on harmonic superspace and on Galperin-Sokatchev formalism. Former is covered in this appendix. Material of this appendix is essential to follow Galperin-Sokatchev's harmonic superspace formalism that we recapitulate in next appendix, that is, Appendix \ref{c}. 

In Sub-appendix  \ref{background} we begin by answering the critical question as to why we need harmonic superspace for the off-shell supersymmetric formulation of the linear ADHM instanton sigma models. 

Basis for harmonic superspace is introduced in Sub-appendix \ref{basis0}. 

Various aspects of the harmonic variables are discussed in Sub-appendix \ref{calculus}.

For formulation of Ali-Ilahi's Complementary Model in an off-shell manner we shall need the dual harmonic superspace. This is developed in Section \ref{dual}.

\subsection{Background} \label{background}

In this Sub-appendix we cover the background to make the journey from conventional superspace to harmonic superspace.

In N = 1 supersymmetry (SUSY), the use of unconstrained superfields effectively describes all known theories. However, in $N = 2$ SUSY , this is not the case. The reason is the proliferation of fields when we move from $N=1$ to $N=2$ superspace. For Witten's original and Ali-Ilahi's complementary ADHM sigma models the situation is all the more critical because these have $(0, 4)$ supersymmetry. Clearly we have to go beyond $N=1$ supersymmetry. The super Yang-Mills theory, supergravity and matter hypermultiplet theory  have been formulated using component formulations as well as constrained superfields. We have covered some aspects of the constrained superfield formalism in Sub-appendix \ref{fermionic}. 

Now we present the argument as to why we move on from usual superspace to harmonic superspace.

In Sub-appendix \ref{fermionic} we described conventional off-shell $(p, 0)$ superfield formalism for non-linear sigma models. This raises a question concerning why should we still be bothering about a superspace covariant formalism? Reason is that the Howe-Papadopoulos formalism of Sub-appendix (\ref{fermionic}) starts with a supersymmetry and then adds $p-1$ more supersymmetries to get to $(p, 0)$ supersymmetry. This breaks the internal $SO(p)$ automorphism to $SO(p-1)$. Thus the $(4, 0)$ supersymmetry's $SO(4)$ automorphism of Witten and Ali-Ilahi's ADHM instanton sigma models will be broken to $SO(3)$. We do not know how to maintain the full $SO(4)$ automorphism in the conventional superfield formalism. Hence we need a formalism that respects the full $SO(4)$ automorphism.

This is why we are looking for harmonic superspace off-shell formalism.

Apart from this the harmonic superspace formalism has several other advantages  - clear understanding, technical convenience and drastic simplifications in the cancellations of ultraviolet divergences.  

At this point we should also cover another point : why do we go from component formalism to superspace formalism? The answer is that the superspace formalism is, though difficult to design, more elegant, efficient to use and economical.
   
\subsection{Basis for Harmonic Superspace} \label{basis0}

In this Sub-appendix we shall introduce the elements of harmonic superspace. This includes the harmonic variables and harmonic expansion. We shall illustrate it for gauge theories in case of Abelian gauge invariance.

The main reference for this Sub-appendix is \cite{Galperin:1984av}.

First of all we introduce the real or central basis and then a subspace of it. Then we introduce zwiebeins and the analytic subspace. Harmonic expansion is introduced for general superfields and then for gauge superfields.

All initial efforts to discover a superfield formulation for the N=2 theories used the real or the central basis given by
\begin{equation}
   (x^\mu, \theta_{\alpha A}, \bar{\theta}^{A}_{\alpha'}).\label{basis}
\end{equation}
Corresponding supersymmetry transformations are
\begin{eqnarray}\label{n=2ss}
    \delta x^\mu&=&i(\epsilon^{A}\sigma^{\mu}\bar{\theta}_{A}-\theta^{A}\sigma^{\mu}\bar{\epsilon}_{A}),\nonumber\\~~~\delta\theta_{\alpha A}&=&\epsilon_{\alpha A},\nonumber\\~~~\delta\bar{\theta}^{A}_{\dot{\alpha}}&=&\bar{\epsilon}^{A}_{\dot{\alpha}}.
\end{eqnarray}

The $N=2$ theories usually have $U(1)$ symmetry and in above basis this invariance comes from a latent $SU(2)$ symmetry.

The issue with the Eqn.(\ref{n=2ss}) and with any associated  theory is the presence of too many $\theta$'s, and as a result, the superfields defined in the equation have numerous unnecessary higher spin components. These higher spin fields have to be eliminated in an {\it ad hoc} but systematic manner. This is usually done in two different ways. First one is to use constraints and the second one is to use complex prepotentials.  Latter lacks clear geometrical interpretation because it uses large gauge  groups.

Alternatively, there exists a reduced subspace  where N = 2 supersymmetry can be achieved. This subspace is the following subset of Eqn.(\ref{n=2ss}).
\begin{equation}\label{redss}
    (\Tilde{x}^\mu,~~~(\theta_{\alpha 1}+i\theta_{\alpha 2})/\sqrt{2},~~~(\bar{\theta}_{\dot{\alpha }}^1+i\bar{\theta}^2_{\dot{\alpha}})/\sqrt{2}).
\end{equation}
The superfields  defined in this context are considerably shorter, enabling a more efficient formulation of the linearized theory of N = 2 supersymmetric Yang-Mills theory. However, this specific basis only allows for an O(2) or U(1) automorphism group, making it challenging to extend the gauge groups to the non-Abelian case.

The main advantages of the two approaches can be merged. This entails utilizing a superspace, similar to Eqn. (\ref{redss}), that possesses an explicit $U(1)$ symmetry, enabling the consideration of short superfields. Simultaneously, it is crucial to maintain the required $SU(2)$-invariance present in most theories. To achieve this, special objects called {\it zweibeins} $u^{+A}$, $u^{-}_{A}=\overline{(u^{+A})}$ are introduced.
\begin{equation}\label{zweib}
 u^{+A}u^{-}_{A}=1,~~~~~~   \begin{pmatrix}
u^{-}_{1} & u^{+}_{1} \\
u^{-}_{2} & u^{+}_{2} 
\end{pmatrix} \in SU(2).
\end{equation}
These objects possess an $SU(2)$ index $A$ and a $U(1)$ index $\pm$. By utilizing them, an $SU(2)$ isospinor, such as $\theta$, can be transformed into two separate and independent $U(1)$ entities.
\begin{equation}\label{uone}
\theta^{+}_{\alpha}=\theta^{A}_{\alpha}u^{+}_{A}, ~~~~\theta^{-}_{\alpha}=\theta^{A}_{\alpha}u^{-}_{A}.
\end{equation}
Additionally, the conversion can be reversed due to property (\ref{zweib}).
\begin{equation}\label{conuone}
\theta^{A}_{\alpha}=u^{+A}\theta^{-}_{\alpha}-u^{-A}\theta^{+}_{\alpha}.
\end{equation}
The subsequent progression involves recognizing that
\begin{equation}\label{subset}
     x^{\mu}_S = x^\mu-2i\theta^{(A}\sigma^\mu\bar{\theta}^{B)}u^{+}_{A}u^{-}_{B},~~~\theta^{+}_{\alpha},~~~\bar{\theta}^{+}_{\dot{\alpha}},~~~u^{\pm}_{A}
\end{equation}
  constitute a subset of Eqn. (\ref{n=2ss}) that remains invariant under N = 2 supersymmetry transformations. These are
\begin{eqnarray}\label{neq2susytr}
 \delta x^{\mu}_S &=&-2i(\epsilon^{A}\sigma^\mu\bar{\theta}^{+}+\theta^{+}\sigma^\mu\bar{\epsilon}^{A})u^{-}_{A},\nonumber\\
\delta\theta^{+}_{\alpha}&=&\epsilon^{A}_{\alpha}u^{+}_{A},~~~\delta \bar{\theta}^{+}_{\dot{\alpha}}=\bar{\epsilon}^{A}_{\dot{\alpha}}u^{+}_{A},~~~\delta u^{\pm}_{A}=0.    \end{eqnarray}
The subspace of N=2 superspace defined by Eqn.(\ref{subset}) is termed as the analytic subspace. It bears resemblance to equation (\ref{redss}) (which is derived by selecting specific values for $u^{+1}$ and $u^{+2}$ as $-1/\sqrt{2}$ and $-i/\sqrt{2}$, respectively). However, the distinction lies in the fact that in equation (\ref{subset}), the original N=2 supersymmetry with parameters $\epsilon^{A}$ (SU(2) spinors) is realized, whereas in equation (\ref{redss}), only its remaining $U(1)$ aspect is retained. 

Let's examine the analytic superfields that are defined using Eqn.(\ref{subset}).
\begin{eqnarray}\label{anasf}
\phi^{(q)}(x_S,\theta^{+}, \bar\theta^{+}, u^{\pm})&=&F^{(q)}(x_S, u^{\pm})+\theta^{\alpha+}\psi^{(q-1)}_{\alpha}(x_S, u^{\pm})\nonumber \\
&+&\bar{\theta}^{+}_{\dot{\alpha}}\bar{\varphi}^{\dot{\alpha}(q-1)}(x_S, u^{\pm})
+\theta^{+}\theta^{+}M^{(q-2)}(x_S, u^{\pm})\nonumber \\
&+&\bar{\theta}^{-}\bar{\theta}^{+}N^{(q-2)}(x_S, u^{\pm}) + \theta^{+}\sigma^{a}\bar{\theta}^{+}A_{a}^{(q-2)}(x_S, u^{\pm}) \nonumber \\
&+& \bar{\theta}^{+}\bar{\theta}^{+}\theta^{\alpha+}\xi^{q-3}_{\alpha}(x_S, u^{\pm})+\theta^{+}\theta^{+}\bar{\theta}^{+}_{\dot{\alpha}}\bar{\chi}^{\dot{\alpha}(q-3)}(x_S, u^{\pm}) \nonumber \\
&+& \theta^{+}\theta^{+}\bar{\theta}^{+}\bar{\theta}^{+}D^{(q-4)}                                        (x_S, u^{\pm}).
\end{eqnarray}
The entity  $\phi^{(q)}$ on the left hand side as a whole has $U(1)$ charge value of $q$. Its constituents possess charges that range from $q$ to $q - 4$, which relies on the charges carried by the $\theta^{+}$-terms within the decomposition. Each individual constituent also has the expansion 
\begin{equation}\label{fnewcor}
F^{(q)}(x_S, u^{\pm})=\sum_{n=0}^{\infty} f^{A_1\cdots A_{n+q}B_1\cdots B_n}(x_S)u^{+}_{(A_1}\cdots u^{+}_{A_{n+q}}u^{-}_{B_1}\cdots u^{-}_{B_n)}
\end{equation}
for $q\geq 1$.

The $f^{\cdots}$'s are symmetric irreducible representations of $SU(2)$ and singlets of $U(1)$. The involvement of $U(1)$ indices is solely through $u^+$ and $u^-$. This  will soon become apparent. 

Hence, we can work with superfields (representations) that have sufficiently short decompositions, similar to N = 1 superfields. This is the juncture where we would like to point out once again that when we go to higher supersymmetries there is a proliferation of extra field. The problem is to limit their numbers. The construction of harmonic superfields is one way of doing that. 

Although they are U(1) representations, their ordinary field components are representations of SU(2), ensuring that the SU(2) symmetry is preserved in a hidden manner. This situation bears some resemblance to the vierbein formalism in gravity, where the vierbeins $e^m_a$, enable the conversion of world vectors into Lorentz ones, allowing for a Lorentz covariant formulation without sacrificing general coordinate transformation invariance. This equivalence permits the description of linearly spinors as typical Lorentz objects, even though they are not general covariant. Likewise, with the assistance of the `zweibein' $u^{\pm}$, we can now express the theory in a U(1)-covariant manner while retaining the broader SU(2) symmetry.

We can now demonstrate that the analytic superfields equation (\ref{anasf}) serve as the appropriate entities for formulating all recognized $N = 2$ supersymmetric theories. To illustrate, consider the example of $N = 2$ supersymmetric gauge theory, where the dimensionless superfield, having   $(q = +2)$, $ V^{++}(x,  \theta^+, \bar{\theta}^-, u^{\pm})$ is well-suited to serve as the sole prepotential. In the case of Abelian theory, it undergoes the following gauge transformation: 
\begin{equation}\label{gaugetran}
\delta V^{++} = D^{++}\lambda. 
\end{equation}
In this context, $\lambda(x, \theta^+,\bar{\theta}^-,u^{\pm}) (q = 0)$ represents the dimensionless analytic parameter associated with the gauge group. (This field should not be confused with chiral fermion supermultiplet to be introduced latter). $D^{++}$ refers to one of the (super) covariant derivatives acting on the sphere $SU(2)/U(1)$. When this derivative is applied to an analytic superfield, it takes the following expression: 
\begin{equation}\label{deransf}
D^{++} = u^{+A}\frac{\partial}{\partial u^{-A}} -2 i\theta^{+}\sigma^\mu\bar{\theta}^+
\frac{\partial}{\partial x_S^\mu}.
\end{equation}
It is evident that $D^{++}\lambda$ shares the same property as $V^{++}$ in being an analytic superfield. By utilizing equations (\ref{anasf}), (\ref{fnewcor}), (\ref{gaugetran}), and (\ref{deransf}), it is straightforward to derive the Wess-Zumino-like gauge for $V^{++}$ given by the following expression:
\begin{eqnarray}\label{wzlike}
V^{++}(x_S, \theta^+,\bar{\theta}^-,u^{\pm})&=&\theta^{+}\theta^{+}[M(x_S)+iN(x_S)] + \bar{\theta}^{+} \bar{\theta}^{+} [M(x_S)-iN(x_S)] \nonumber \\
&+&i{\theta}^{+}\sigma^{a}\bar{\theta}^{+}A_{a}(x_S)
+\bar{\theta}^{+}\bar{\theta}^{+}\theta^{+\alpha}\psi^{A}_{\alpha}(x_S)u^{-}_{A}\nonumber \\
&+&\theta^{+}\theta^{+}\bar{\theta}^{+}_{\dot{\alpha}}\bar{\psi}^{\dot{\alpha}A}(x_S)u^{-}_{A}+  \theta^{+}\theta^{+}\bar{\theta}^{+}\bar{\theta}^{+}D^{(AB)}          (x_S)u^{-}_{A}u^{-}_{B}.\nonumber\\
\end{eqnarray}
The remaining gauge transformations solely impact the vector field $A_a$. It is evident that equation (\ref{wzlike}) precisely represents the field content found in $N = 2$ supersymmetric gauge theory.

At this point we can explain the gratifying aspect of the harmonic superspace where the infinite number of degrees of freedom of $V^{++}$, coming from Eqn. (\ref{fnewcor}), are killed by gauge transformations of Eqn. (\ref{gaugetran}). Both $V^{++}$ and $\lambda$ describe the same superspin 0 analytic scalar superfield. $V^{++}$ has super isospins $0, 1, 2, 3, \cdots$ while $\lambda$ has super isospins $ 1, 2, 3, \cdots$. It is precisely the super isospin $0$ that survives after the gauge transformations. The result is that while the harmonic superspace has infinite number of auxiliary and gauge degrees of freedom only a finite number of these survive.

The real power of this formalism becomes even more apparent in case of super Yang-Mills theories, supergravity and hyper multiplets. Since we shall not need corresponding machinery in the present note we shall not digress to describe those aspects. We shall not also cover the massive $N=2$ spin-1 theory where once again the infinite number of auxiliary degrees of freedom occur.

The main achievements of the harmonic superspace formalism are the following. The superspace of $N=2$ supersymmetry is extended and now includes the sphere $SU(2)/U(1)$. We then introduce harmonic coordinates and this allows us to get to a new analytic subspace of $N=2$ superspace. There is a gauge invariance that helps us in killing an infinite number of auxiliary and gauge degrees of freedom.

\subsection{Harmonic Calculus}\label{calculus}

In the last Sub-appendix we introduced the basics of the harmonic superspace and explained that there are infinite number of auxiliary and gauge degrees of freedom that are killed by gauge transformations leaving behind a finite number of physical fields.

In this Sub-appendix we shall describe additional aspects of the harmonic variables $u^\pm_A$ that we have already introduced above. We shall also introduce the concepts of analytic basis in $N=2$ superspace as well as the analytic superfield and many useful conventions and formulas.

We now take up the harmonic variables and their calculus. The key point is to convert the $SU(2)$ indices to $U(1)$ charge. This is done by the introduction of a zwiebein like variables $u_A^\pm$
\begin{eqnarray}
    \psi^\pm&=&\psi^Au^\pm_A,\nonumber\\
    \psi^A&=&u^{+A}\psi^--u^{-A}\psi^+,\nonumber\\
    u^-_A&=&\overline{(u^{+A})}.\label{harv}
\end{eqnarray}
Clearly the variables $u^\pm_A ~\epsilon~SU(2)$ converts the $SU(2)$ index $A$ of $\psi^A$ into $U(1)$ charges $\pm$. In other words $\psi^A$ are $SU(2)$ doublets and $\psi^\pm$ are corresponding $U(1)$ projections. The variables $u^{\pm A}$ are called the harmonic variables.

The harmonic variables have the following properties
\begin{eqnarray}
u^{+A}u^{-}_A&=&1,\nonumber\\
u^{+A}u^{+}_A&\equiv&u^{+A}\epsilon_{AB}u^{+B}=0,\nonumber\\
u^{-A}u^{-}_A&=&0.\label{harv0}
\end{eqnarray}

In the notation of Ref.\cite{Galperin:1984av} the ordinary complex conjugation is denoted by $(\bar{.})$ and the $(*)$ operation acts on $U(1)$ index only and thus we have
\begin{equation}
(u^{+A})^*=u^{-A},~~(u^-_A)^*=-u^+_A.\label{harv1}
\end{equation}
In this notation we have the following unusual property
\begin{equation}
   (u^\pm)^{**}=-u^\pm.\label{harv2}
\end{equation}

Here would like to point out that the * operation defined by above authors is not the standard one.

In the following we shall not use any specific parametrization of $SU(2)$. Its presence will be indicated through the harmonic variables only. 
Using these three differential operators can be constructed: 

\begin{equation}\label{threedefop}
D^{++} = u^{+A}\frac{\partial}{\partial u^{-A}},~~~    D^{--} = \overline{D^{++}},~~~ D^{0} = u^{+A}\frac{\partial}{\partial u^{+A}} - u^{-A}\frac{\partial}{\partial u^{-A}} 
\end{equation}

The derivative $D^{++}$ is different from the one in Eqn.(\ref{deransf}). The derivatives in Eqn.(\ref{threedefop}) are defined so as to preserve Eqn.(\ref{harv0}). The derivative in Eqn. (\ref{deransf}) is defined for analytic superfield. The $U(1)$ charge of the variables ${u}^{\pm}$ is clearly counted in by the third operator $D^{0}$ in Eqn.(\ref{threedefop}). As eigen-functions of this charge operator, harmonic functions are defined as follows:

\begin{equation}\label{harmfn}
D^{0}{f}^{q}({u}) = q{f}^{q}({u}) ,~~~ q=0,\pm 1, \pm 2,\cdots,
\end{equation}
whose harmonic expansion was given earlier in Eqn.(\ref{fnewcor}) as:
\begin{equation}\label{harmexp}
{f}^{q}({u})=\sum_{n=0}^{\infty}{f}^{A_1\cdots A_{n+q}B_1\cdots B_n}{u}^{+}_{(A_1}\cdots {u}^{+}_{A_{n+q}}{u}^{-}_{B_1}\cdots {u}^{-}_{B_n)}
\end{equation}

Here $q\geq 0$. There is an analogous  expression  for $q < 0$. This is spherical harmonic expansion of a square integrable function or tensor (for $q \ne 0$) on the sphere $S^2 \sim SU(2)/U(1)$. Here $f^{(q)}$ as a whole transforms as a representation of the U(1) group with charge $q$. The expansion coefficients $f^{\cdots}$ are $SU(2)$ tensors with isospin $n+\frac{q}{2}$.  The homogeneity criteria (\ref{harmfn}) can be used to obtain the coset $SU(2)/U(1)$. We want to underline that this method of representing the sphere is global, so we don't need to worry about the analytic qualities of our functions in various coordinate patches (as opposed to, say, using a complex parametrization).

The harmonic functions in Eqn. (\ref{harmexp}) can be made real with respect to the following particular conjugation for even values of the $U(1)$ charge $q$:
 \begin{equation}\label{partconj}
\widetilde{{f}^{AB\cdots}}=\overline{{f}^{AB\cdots}},~~~\widetilde{{u}^{\pm A}}={u}^{\pm}_A,~~~\widetilde{{u}^{\pm}_A}=-{u}^{\pm A}. 
 \end{equation}

This is combination of the antipodal map on $S^2$ and common complex conjugation, represented in terms of ${u}^{\pm}$.

If we define
\begin{equation}
    D^3=\frac{1}{2}D^0
\end{equation}
then we get the $SU(2)$ algebra of derivatives as:
\begin{equation}
   [D^{++}, D^{--}]=2D^3,\nonumber 
\end{equation}
\begin{equation}   
  [D^3, D^{++}]=D^{++}, [D^3, D^{--}]=-D^{--}.\label{dalgebra}  
\end{equation} 

Action of the derivatives on the harmonic variables is given below.
\begin{equation}
D^{++}u^{+A}=0,~~~D^{++}u^{-A}=u^{+A}  
\end{equation}
and
\begin{equation}
D^{-}u^{+A}=u^{-A},~~~D^{++}u^{-A}=0  
\end{equation}
as well as
\begin{equation}
    D^0u^{\pm A}=\pm u^{\pm A}.
\end{equation}

We can discuss the problem of solution to the differential equation
\begin{equation} D^{++}X^{(q)}=F^{(q+2)}\label{diffe}
\end{equation}
in $u^\pm$-calculus for some unknown $X$.

We begin with the homogeneous problem
\begin{equation}
 D^{++}X_0^{(q)}=0. \label{diffe0}
\end{equation}
Corresponding solution is
\begin{eqnarray}
X_0^{(q)}&=&X^{(A_1\cdots A_q)}u_{A_1}\cdots u_{A_q},\;\;q\geq 0\nonumber\\
X^{(q)}_0&=&0,\;\;{q< 0}.\label{soln}
\end{eqnarray}

The solution to the inhomogeneous equation now can be written as
\begin{eqnarray}
    X^{(q)}&=&(1/D^{++})F^{(q+2)}+X_0^{(q)},q\geq 0\nonumber\\
    X^-&=&(1/D^{++})F^{(1)},\;\;q=-1.\label{soln0}
\end{eqnarray}
For $q\leq -1$ the solution exists only for special $F^{(q+2)}$s. He we have used the following partial solution to Eqn.(\ref{diffe})
\begin{equation}
\frac{1}{D^{++}}F^{(q+2)}=\sum_{n=0}^{\infty}\frac{1}{n+1}f^{\cdots 2n+q+2\cdots}(u^+)^{n+q+1} (u^-)^{n+1}.  \label{diffe3} 
\end{equation}

Here are two formulas that will be used in the following to wrap up our brief introduction to the harmonic calculus. The simple rule gives the definition of a harmonic integral.

\begin{equation}\label{harmint}
\int d{u} 1=1,~~\int d{u} {u}^{+}_{(A_1}\cdots{u}^{+}_{A_{p}}{u}^{-}_{B_1}\cdots{u}^{-}_{B_r)} ~\text{for}~~\text{for} ~p+r>0.
\end{equation}

In the expansion of the integrand, the harmonic integral essentially projects out the singlet portion. Integration by parts for the derivatives in Eqn. (\ref{threedefop}) is compliant with the aforementioned integration rule. Additionally, one can can prove the following identity:

\begin{equation}\label{identity}
D^{++}_1 \frac{1}{{u}^{+}_1{u}^{+}_2}=\delta^{+,-}({u}_1,{u}_2),
\end{equation}
where ${u}^{+}_1{u}^{+}_2 \equiv {u}^{+A}_1{u}^{+}_{2A}$ and $\delta^{+,-}(u_1,u_2)$ is a harmonic delta function. This  is equivalent to
\begin{equation}\label{delta-rule}
    \frac{\partial}{\partial \bar z}=\pi\delta(z).
\end{equation}

In central basis the $N=2$ superspace is parametrized by the coordinates
\begin{equation}
    z^M=(x^\mu, \theta_{\alpha A}, \bar\theta^A_{\dot\alpha}=\overline{(\theta_{\alpha A})}).
\end{equation}
Corresponding supersymmetry transformations are given by the Eqn.(\ref{n=2ss}).

The coordinates of the analytic basis include the harmonic variables
\begin{equation}
    Z^M_S=(x^\mu_S, \theta^+_\alpha, \bar\theta^+_{\dot\alpha}, \theta^-_\alpha, \bar\theta^-_{\dot\alpha},u^\pm_A).
\end{equation}
Where  we have defined
\begin{eqnarray}
    \label{def}
    x^\mu_S&=&x^\mu-2i\theta^{(A}\sigma^\mu{\bar\theta}^{B)}u^+_Au^-_B,\nonumber\\
    \theta^\pm_\alpha&=&\theta^A_\alpha u^\pm_A,\nonumber\\
    {\bar\theta}^\pm_\alpha&=&{\bar\theta}^A_\alpha u^\pm_A.
\end{eqnarray}
Corresponding supersymmetry transformations are given below:

\begin{eqnarray}\label{neq2susytr1}
 \delta x^{\mu}_S &=&-2i(\epsilon^{A}\sigma^\mu\bar{\theta}^{+}+\theta^{+}\sigma^\mu\bar{\epsilon}^{A})u^{-}_{A},\nonumber\\
\delta\theta^{+}_{\alpha}&=&\epsilon^{A}_{\alpha}u^{+}_{A},~~~\delta \bar{\theta}^{+}_{\dot{\alpha}}=\bar{\epsilon}^{A}_{\dot{\alpha}}u^{+}_{A},\nonumber\\   
\delta\theta^{-}_{\alpha}&=&\epsilon^{A}_{\alpha}u^{-}_{A},~~~\delta \bar{\theta}^{-}_{\dot{\alpha}}=\bar{\epsilon}^{A}_{\dot{\alpha}}u^{-}_{A},~~~\delta u^{\pm}_{A}=0.
\end{eqnarray}

\section{Witten's Original Model}\label{c}

To formulate Ali-Ilahi's ADHM instanton sigma model in an off-shell covariant harmonic superspace we need background on Witten's original ADHM instanton sigma model. Ali-Ilahi model is complementary to Witten's original model in the sense of being dual to it. In this Section we shall summarize Witten's construction of the original ADHM instanton linear sigma model on the basis of  \cite{Witten:1994tz}. 

We do so by listing the field contents and the actions for the the model including the Yukawa interaction (that eventually  incorporates the Higgs mechanism for fermions). We then introduce  the tensor $C^{a}_{AA'}$ that is involved in the Yukawa interactions as well as the ADHM conditions. Latter follow when we demand $N=4$ supersymmetry.

Latter on, following Witten, we introduce the tensor $B^{a}_{AY'}$. After introduction of the basis for massless right moving fermions Witten's construction leads us to the Corrigan, Fairlie, Goddard and Templeton form of the ADHM instanton $A_{ijAY}$. Linear sigma models are characterized by the potential terms for the fields, as already explained in Sub-appendix \ref{linear}. We describe the route to get to the potential $V_{W}(X,\phi)$ for Witten's original model in case of $SU(2)$ gauge group and one instanton. The structure of corresponding moduli space and corresponding mystery is discussed after that. 

The duality between the original and complementary models will be pointed out in the next section. This duality is a key discovery in the scheme of things that involve instanton sigma models, both 't Hooft and ADHM, the N=4 superconformal symmetries, both small and large, as well as the $AdS_3$ superstrings\cite{Ali:2024amc}.

To start, we introduce the action of Witten's original ADHM sigma model. This model comprises two sets of bosons: $X^{AY}$ with indices $A = 1,2$ and $Y = 1,2,...,2k$, totaling $4k$ bosons; and $\phi^{A'Y'}$ with indices $A' = 1,2$ and $Y' = 1,2,...,2k'$, totaling $4k'$ bosons. The corresponding right-handed superpartners are $\psi_{-}^{A'Y}$ for $X^{AY}$ and $\chi_{-}^{AY'}$ for $\phi^{A'Y'}$. 

The fields $X^{AY}$ and $\phi^{A'Y'}$ obey the following reality conditions
\begin{equation}
X^{AY}=\epsilon^{AB}\epsilon^{YZ}X_{BZ},\;\;\phi^{A'Y'}=\epsilon^{A'B'}\epsilon^{Y'Z'}\phi_{B'Z'}.\label{real}
\end{equation}

The multiplet $(X, \psi)$ is called the fundamental scalar multiplet and the multiplet $(\phi, \chi)$ as the twisted scalar multiplet.

Additionally, there exist left-handed fermions denoted as $\lambda_{+}^{a}$, where $a$ ranges from $1$ to $n$. The indices are manipulated using specific antisymmetric tensors: $\epsilon^{AB}$ and $\epsilon_{AB}$ for $SU(2)$, $\epsilon^{A'B'}$ and $\epsilon_{A'B'}$ for $SU(2)'$, $\epsilon^{YZ}$ and $\epsilon_{YZ}$ for $Sp(2k)$, and $\epsilon^{Y'Z'}$ and $\epsilon_{Y'Z'}$ for $Sp(2k')$.

The action for the original model can be expressed as follows:
\begin{equation}\label{action14}
S_W = S^{kin} + S^{int}
\end{equation}
where the kinetic term $S^{kin}$ and the interaction term $S^{int}$ are given by:
\begin{eqnarray}\label{action15}
 S^{kin} &=& \int d^2 \sigma ( \epsilon_{AB} \epsilon_{YZ} \partial_{-}  X^{AY}\partial_{+} X^{BZ}+ i\epsilon_{A^{\prime}B^{\prime}}\epsilon_{YZ}\psi_-^{A^{\prime}Y} \partial_{+} \psi_-^{B^{\prime}Z}  \nonumber \\
&+&\epsilon_{A^{\prime}B^{\prime}}\epsilon_{Y^{\prime}Z^{\prime}}\partial_{-}\phi^{A^{\prime}Y^{\prime}}\partial_{+} \phi^{B^{\prime}Z^{\prime}}
+i\epsilon_{AB}\epsilon_{Y^{\prime}Z^{\prime}}\chi_-^{AY^{\prime}}\partial_{+} \chi_{-}^{BZ^{\prime}} \nonumber \\&+&i\lambda_{+}^{a} \partial_{-} \lambda_+^{a})
\end{eqnarray}
\begin{eqnarray}\label{action3}
    S^{int}&=&-\frac{i}{2}m\int d^2 \sigma\lambda^{a}_{+}[(\epsilon^{BD} \frac{\partial C^{a}_{BB'}}{\partial X^{DY}}\psi^{B'Y}_{-} + \epsilon^{B'D'}\frac{\partial C^{a}_{BB'}}{\partial \phi^{D'Y'}}{\chi}^{BY'}_{-})\nonumber \\&-& \frac{im}{4} \epsilon^{AB}\epsilon^{A'B'}C^a_{AA'}C^a_{BB'})].
\end{eqnarray}
These terms are subject to certain conditions described by equations:
\begin{eqnarray}\label{adhm5}
\frac{\partial C_{AA'}^a}{\partial X^{B Y}}
+\frac{\partial C_{BA'}^a}{\partial X^{A Y}}
=0=  \frac{\partial C_{AA'}^a}{\partial \phi^{B'Y'}}
+\frac{\partial C_{AB'}^a}{\partial \phi^{A'Y'}},
\end{eqnarray}
\begin{equation}\label{adhm2}
\sum_{a}(C^{a}_{AA'}{C}^{a}_{BB'}+{C}^{a}_{BA'}{C}^{a}_{AB'})=0
\end{equation}
where the tensor $C^{a}_{AA'}$ satisfies the simple field dependence:
\begin{equation}\label{generalform1}
C^{a}_{AA'}=M^{a}_{AA'}+\epsilon_{AB} N^{a}_{A'Y}X^{BY}+\epsilon_{A'B'}D^{a}_{AY'}\phi^{~B'Y'}+\epsilon_{AB}\epsilon_{A'B'}E^{a}_{YY'}X^{BY}\phi^{~B'Y'}.
\end{equation}
The tensors  $M^{a}_{AA'}$, $N^{a}_{A'Y}$, $D^{a}_{AY'}$ and $E^{a}_{YY'}$ in Eqn.(\ref{generalform1}) are tensors with constant values. 

Here we also have introduced the right moving fermions $\lambda^a, a=1, \cdots, n$.

The model exhibits supersymmetry, which is characterized by the following transformations:
\begin{eqnarray}\label{susytransform}
\delta X^{AY} & = & i\epsilon_{A'B'} \eta_+^{AA'}\psi_-^{B'Y}, ~~~~ \delta \psi_-^{A'Y} = \epsilon_{AB} \eta_+^{AA'}\partial_{-}X^{BY} , \cr
~\delta \phi^{A'Y'} & = &i\epsilon_{AB} \eta_+^{AA'}\chi_-^{BY'}, ~~~~~~ \delta \chi_-^{A'Y} = \epsilon_{A'B'} \eta_+^{AA'}\partial_{-}\phi^{B'Y'} , \cr
\delta\lambda^a_+ &= &\eta_+^{AA'} C_{AA'}^a.
\end{eqnarray}
Here, $\eta_+^{AA'}$ represents an infinitesimal anti-commuting parameter.

The tensor $C^a_{AA'}$ is linear both in $X$ and $\phi$. In Witten's original ADHM sigma model, a specific choice is made for the tensors such that $M^{a}_{AA'}=N^{a}_{A'Y}=0$. As a result, the tensor $C^{a}_{AA'}$ simplifies to:
\begin{eqnarray}\label{caaa}
C^{a}_{AA'}&=&\epsilon_{A'B'}(D^{a}_{AY'}+\epsilon_{AB}E^{a}_{YY'}X^{BY})\phi^{B'Y'}=\epsilon_{A'B'}B^{a}_{AY'}(X)\phi^{B'Y'}\nonumber\\
&=&B^{a}_{AY'}(X){\phi_{A'}}^{Y'}.
\end{eqnarray}
Here, $B^{a}_{AY'}(X)$ is a linear function of $X$
\begin{equation}
 B^{a}_{AY'}(X)=D^{a}_{AY'}+\epsilon_{AB}E^{a}_{YY'}X^{BY}=D^{a}_{AY'}+E^{a}_{YY'}{X_A}^Y\label{linear1}   
\end{equation}
since the tensors $D^{a}_{A'Y}$ and $E^{a}_{YY'}$ are constant.

The genesis of the ADHM instanton in Witten's derivation is as follows. The potential $V_W$ for Witten's linear sigma model vanishes for $\phi=0$. Corresponding space of vacua, the moduli space, is ${\mathcal M} = {\mathcal R}^{4k}$ is parametrized $X$.

In this case the structure of the Yukawa couplings is very simple :
\begin{equation}
    \sum_a\lambda^a_+B^a_{AY'}\chi^{AY'}_-.\label{yukawa0}
\end{equation}
The fermion mass comes from Yukawa coupling and from above structure it is clear that the fermionic partners $\psi_-$ of $X$ are all massless. Assuming the number $n$ of $\lambda^a_+$ to be more than $4k'$, the number of components of $\chi^{AY'}$, generically all components of latter will get mass. By supersymmetry their partners too, $\phi$,  will get masses. This will leave some of the $\lambda^a_+$'s massless. The number of massless components is
\begin{equation}
    N=n-4k'.\label{massless}
\end{equation}
We choose $v^a_i$ as the basis for massless components of $\lambda^a_+$, with $i=1, \cdots, N.$  These obey the orthonormality
\begin{equation}
\sum_av^a_iv^a_j=\delta_{ij}.\label{ortho0}
\end{equation}

From the structure of Eqn.(\ref{yukawa0}) we get
\begin{equation}
\sum_{a=1}^nv^a_iB^a_{AY'}=0.\label{ortho1}
\end{equation}
Setting the massive modes to zero we get
\begin{equation}
\lambda^a_+=\sum_{i=1}^{N}v^a_i\lambda_{+i}
\end{equation}
where $\lambda_{+i}$'s are the massless right moving fermions.

From the kinetic energy term for the right moving fermions we get the crux of the instanton construction
\begin{eqnarray}
   \frac{i}{2}\int d^2 x\sum_{i,j,a}^{N,N,n} (v^a_i\lambda_{+i})\partial_-(v^a_j\lambda_{+j})\nonumber\\
   =\frac{i}{2}\int d^2 x\sum_{i,j}^{N,N}\{\lambda_{+i}(\delta_{ij}\partial_++\partial_-X^{AY}A_{ijAY})\lambda_{+j}\},\label{aijay2}
\end{eqnarray}
with the resulting instanton having the following expression:
\begin{equation}\label{aijay}
A_{ijAY}=\sum_{a=1}^{n=N-4k'}v^{a}_{i}\frac{\partial v^{a}_{j}}{\partial X^{AY}}.
\end{equation}
Which agrees with the expression that comes from Corrigan, Fairlie, Goddard and Templeton construction \cite{Corrigan:1978ce}.

The construction of the potential has been done by Witten in the following way. The right moving fermions are part of a superfield $\Lambda^a$ with superpartner $F^a$:
\begin{equation}
    \Lambda^a=\lambda^a+\theta F^a.
\end{equation}
The expression for the potential energy is
\begin{equation}
    V_W=\frac{1}{2}\sum_a F^aF^a.
\end{equation}
Now comparing the supersymmetric variations of $\lambda^a$
\begin{eqnarray}
    \delta\lambda^a=\eta F^a,\label{susy2}\\
    \delta\lambda^a=\eta_+^{AA'}C^a_{AA'}\label{susy3}
\end{eqnarray}
we define
\begin{equation}
    \eta^{AA'}=\eta c^{AA'}
\end{equation}
with the $c$-numbers $c^{AA'}$ normalized as
\begin{equation}
    \epsilon_{AB}\epsilon_{A'B'}c^{AA'}c^{BB'}=1.
\end{equation}
In this case the potential becomes
\begin{equation}
    V_W=\frac{1}{2}\sum_aC^aC^a
\end{equation}
with
\begin{equation}
   C^a=c^{AA'}C^a_{AA'}. 
\end{equation}

In the case of $SU(2)$ and one instanton, that is for $k=k'=1$, the right moving fermions obey the reality conditions
\begin{eqnarray}
\lambda^{AY'}_+&=&\epsilon^{AB}\epsilon^{Y'Z'}\bar\lambda_{+BZ'},\nonumber\\
\lambda^{YY'}_+&=&\epsilon^{YZ}\epsilon^{Y'Z'}\bar\lambda_{+ZZ'}\label{real2}
\end{eqnarray}
and the tensors $C^a_{AA'}$ take the form
\begin{eqnarray}
    {C^{YY'}}_{BB'}&=&{X_B}^Y{\phi_{B'}}^{Y'},\nonumber\\
    {C^{AY'}}_{BB'}&=&\frac{\rho}{\sqrt{2}}{\delta^A}_B{\phi_{B'}}^{Y'}.
\end{eqnarray}
Now if we define
\begin{eqnarray}
X^2&=&\epsilon_{AB}\epsilon_{YZ} X^{AY}X^{BZ},\nonumber\\
\phi^2&=&\epsilon_{A'B'}\epsilon_{Y'Z'}\phi^{A'Y'}\phi^{B'Z'}\label{square0}
\end{eqnarray}
then the potentials is given by:
\begin{equation}\label{potential}
V_W = \frac{m^2}{8}(X^{2}+\rho^{2})\phi^{2}.
\end{equation} 

Here $\rho$ is the size of the instanton. Clearly the instanton solution breaks the conformal invariance. It is believed that the model will flow to an $N=4$ superconformal theory in the infrared.

This is the time to analyze the structure of the moduli space of Witten's original ADHM instanton sigma model and see the mystery for ourselves.

Moduli state is the state of ground states and is parameterized by scalar quantities. From potential (\ref{potential}) it is clear that the moduli space of Witten's original model is given by $\phi=0$ and any $X$. This is the moduli space $\mathcal M=\mathcal R^{4k}$ that we mentioned earlier.

The mystery is in case of the small instanton where we take $\rho\rightarrow 0$. In this case the potential becomes $V_W=\frac{1}{8}X^2\phi^2$. This time an additional branch of moduli space opens up that is defined by $X=0$ and any $\phi$. We call it $\mathcal M'=\hat{\mathcal R}^{4k'}$. In Refs. \cite{Witten:1994tz} and \cite{Witten:1995zh} the origin of this branch was a mystery.

The origin of this branch of the moduli space was clarified in Ref.\cite{Ali:2023csc}. This lead to the resolution of this three decades old mystery. We shall present the corresponding argument in the next section.

Galperin and Sokatchev have their own perspective on Witten's original ADHM instanton linear sigma model. We summarize this in the Sub-appendix \ref{perspective0}.

\section{Ali-Ilahi's Complementary  Model}\label{ali-ilahi}\label{d}

In this Appendix we shall summarize Ali-Ilahi's construction of an ADHM instanton sigma model on the basis of the paper \cite{Ali:2023csc}. This model is complementary to Witten's original model in the sense that it is dual to Witten's model. The definition of the duality is interwoven into the narrative of this section. 

The starting fields, their supersymmetry transformations and the kinetic energy part of the action for Ali-Ilahi's complementary ADHM instanton sigma model is same as that of Witten's original model. Hence in this section we begin with the interaction part of the action containing the Yukawa and potential terms. In this model the roles of the fundamental scalar multiplet $(X, \psi)$ and the twisted scalar multiplet  $(\phi, \chi)$ multiplet are interchanged. This is the beginning of the duality between the Original Model and the Complementary Model.
 
In parallel with the original model we introduce the tensor $\hat C^{a'}_{AA'}$ that follows the ADHM conditions due to $N=4$ supersymmetry in this case. It occurs in the Yukawa terms and the potential term. Till this moment even the interaction part of the action looks like the one in Witten's original model. The difference comes later on when we interchange the roles of the fundamental and twisted multiplets.

The ADHM conditions in the complementary model are similar to those in the original model.

We also introduce the tensor $A^{a'}_{A'Y}$. This is dual to the tensor $B^a_{AY'}$ of Witten's original model. 

We then examine the moduli space of the Complementary Model and the analysis of the Yukawa terms leads in this case too an ADHM instanton in the form described by Corrigan, Fairlie, Goddard and Templeton $A_{i'j'A'Y'}$. This instanton is complementary to the instanton of Witten's Original Model.

Linear sigma models are characterized by the potential terms for the fields. We describe the route to get to the potential $V_{C}(X,\phi)$ for Ali-Ilahi's Complementary Model in case of $SU(2)$ gauge group and one instanton. This is the place where we analyze the moduli space of the complementary model and in the process we come to a very simple resolution of the mystery that Witten encountered in the original case.

What would be Galperin and Sokatchev's take on the Complementary Model? This question is answered in Section \ref{perspective}.
 
The complementary model follows the same supersymmetry transformation laws as in Eqs. (\ref{susytransform}) except the last equation, that in complementary case becomes $\delta \hat{\lambda}^a_+ = \eta_+^{AA'} \hat{C}_{AA'}^a$. Moreover, the corresponding interaction part of the action in Eqn. (\ref{action3}), conditions  in Eqs. (\ref{adhm2}), (\ref{adhm3})  and general form of the Eqn. (\ref{generalform1}) respectively for complementary model are given as 
\begin{eqnarray}\label{action6}
    \hat S^{int}&=&-\frac{i}{2}m\int d^2 \sigma\hat\lambda^{a^{\prime}}_{+}[(\epsilon^{BD} \frac{\partial \hat C^{a^{\prime}}_{BB'}}{\partial X^{DY}}\psi^{B'Y}_{-} + \epsilon^{B'D'}\frac{\partial \hat C^{a'}_{BB'}}{\partial \phi^{D'Y'}}{\chi}^{BY'}_{-})\nonumber \\&-& \frac{im}{4} \epsilon^{AB}\epsilon^{A'B'}\hat C^{a'}_{AA'}\hat C^{a'}_{BB'})]
\end{eqnarray}
\begin{eqnarray}\label{adhm3}
\frac{\partial \hat C_{AA'}^{a'}}{\partial X^{B Y}}
+\frac{\partial \hat C_{BA'}^{a'}}{\partial X^{A Y}}
=0=  \frac{\partial \hat C_{AA'}^{a'}}{\partial \phi^{B'Y'}}
+\frac{\partial \hat C_{AB'}^{a'}}{\partial \phi^{A'Y'}},
\end{eqnarray}
\begin{equation}\label{adhm4}
\sum_{a'}(\hat C^{a'}_{AA'}\hat C^{a'}_{BB'}+\hat C^{a'}_{BA'}\hat C^{a'}_{AB'})=0
\end{equation}
where $\hat C^{a'}_{AA'}$ is
\begin{equation}\label{generalform2}
\hat C^{a'}_{AA'}=\hat M^{a'}_{AA'}+\epsilon_{AB} \hat N^{a'}_{A'Y}X^{BY}+\epsilon_{A'B'}\hat D^{a'}_{AY'}\phi^{~B'Y'}+\epsilon_{AB}\epsilon_{A'B'}\hat E^{a'}_{YY'}X^{BY}\phi^{~B'Y'}.
\end{equation}

Here one possible doubt has to be cleared. Superficially it looks like that we are dealing with a superficial duality in which we have the transformations $(X, \psi)\leftrightarrow (\phi, \chi)$, $k\leftrightarrow k'$, $C\leftrightarrow \hat C$, $(a, i)\leftrightarrow (a', i')$. This is a red herring because the duality involves interchange of $k\rightarrow k'$ that involves exchange of target space dimensions with instanton number. This is highly non-trivial transformation.

In the Complementary Model the choice of tensors is different from the Original Model. In the Complementary Model we take $\hat M^{a'}_{AA'}=\hat D^{a'}_{AY'}=0$, leading to the reduction of the tensor $\hat C^{a'}_{AA'}$ to:
\begin{equation}\label{cnex}
\hat C^{a'}_{AA'}=\epsilon_{AB}(\hat N^{a'}_{A'Y}+\epsilon_{A'B'}\hat E^{a'}_{YY'}\phi^{B'Y'})X^{BY}=\epsilon_{AB}A^{a'}_{A'Y}(\phi)X^{BY}.
\end{equation}
In this case, $A^{a'}_{A'Y}(\phi)$ is linear in $\phi$ 
\begin{equation}
 A^{a'}_{A'Y}(\phi)=\hat N^{a'}_{A'Y}+\epsilon_{A'B'}\hat E^{a'}_{YY'}\phi^{B'Y'}\label{linear2}   
\end{equation}
since $\hat N^{a'}_{A'Y}$ and $\hat E^{a'}_{YY'}$ are constant.

The genesis of the ADHM instanton in Ali-Ilahi's model parallels that of Witten's model. The potential $V_C$ for Complementary Model vanishes for $X=0$. Corresponding space of vacua, the moduli space, is ${\mathcal M} = {\mathcal R}^{4k'}$ is parametrized $\phi$.

In this case too the structure of the Yukawa couplings is very simple :
\begin{equation}  \sum_a\lambda^{a'}_+A^{a'}_{A'Y}\psi^{A'Y}_-.\label{yukawa1}
\end{equation}

Here the right moving fermions in the complementary model are being called $\hat\lambda^{a'}_+$. The fermion mass comes from Yukawa coupling and from above structure it is clear that the fermionic partners $\chi_-$ of $\phi$ are all massless. Assuming the number $n'$ of $\hat\lambda^{a'}_+$ to be more than $4k$, the number of components of $\psi^{AY'}$, generically all components of latter will get mass.  By supersymmetry their partners too, $X$ will get masses. This will leave some of the $\hat\lambda^{a'}_+$'s massless. The number of massless components is
\begin{equation}
    N'=n'-4k.\label{massless1}
\end{equation}
We choose $\hat v^{a'}_{i'}$ as the basis for massless components of $\hat\lambda^{a'}_+$, with $i'=1, \cdots, N'.$  These obey the orthonormality
\begin{equation}
\sum_{a'}\hat v^{a'}_{i'}\hat v^{a'}_{j'}=\delta_{i'j'}.\label{ortho2}
\end{equation}

From the structure of Eqn.(\ref{yukawa1}) we get
\begin{equation}
\sum_{a'=1}^{n'}\hat v^{a'}_{i'}A^{a'}_{A'Y}=0.\label{ortho3}
\end{equation}
Setting the massive modes to zero we get
\begin{equation}
\hat\lambda^{a'}_+=\sum_{i'=1}^{N'}\hat v^{a'}_{i'}\hat\lambda_{+i'}
\end{equation}
where $\hat\lambda_{+i}$'s are the massless right moving fermions.

From the kinetic energy term for the right moving fermions we get the instanton construction
\begin{eqnarray}
   \frac{i}{2}\int d^2 x\sum_{i',j',a'}^{N',N',n'} (\hat v^{a'}_{i'}\hat\lambda_{+i'})\partial_-(\hat v^{a'}_{j'}\hat\lambda_{+j'})\nonumber\\
   =\frac{i}{2}\int d^2 x\sum_{i',j'}^{N',N'}\{\hat\lambda_{+i'}(\delta_{i'j'}\partial_++\partial_-\phi^{A'Y'}\hat A_{i'j'A'Y'})\hat\lambda_{+j'}\},\label{aijay3}
\end{eqnarray}
with the resulting instanton having the following expression:
\begin{equation}\label{aijay4}
\hat A_{i'j'A'Y'}=\sum_{a'=1}^{n'=N'-4k}\hat v^{a'}_{i'}\frac{\partial\hat v^{a'}_{j'}}{\partial\phi^{A'Y'}}.
\end{equation}
Which is again the the Corrigan-Fairlie-Goddard-Templeton construction \cite{Corrigan:1978ce} as in case of the Witten's original ADHM instanton sigma model.

The construction of the potential parallels the procedure in case Witten's original model. The right moving fermions are part of a superfield $\hat\Lambda^{a'}$ with superpartner $\hat F^{a'}$:
\begin{equation}
   \hat \Lambda^{a'}=\hat\lambda^{a'}+\theta \hat F^{a'}.
\end{equation}
The expression for the potential energy for the complementary model  is
\begin{equation}
    V_C=\frac{1}{2}\sum_{a'}\hat F^{a'}\hat F^{a'}.
\end{equation}
Now comparing the supersymmetric variations of $\hat\lambda^{a'}$
\begin{eqnarray}
    \delta\hat\lambda^{a'}=\eta \hat F^{a'},\label{susy4}\\
    \delta\hat\lambda^{a'}=\eta_+^{AA'}\hat C^{a'}_{AA'}\label{susy5}
\end{eqnarray}
we define
\begin{equation}
    \eta^{AA'}=\eta c^{AA'}
\end{equation}
with the $c$-numbers $c^{AA'}$ normalized as
\begin{equation}  \epsilon_{AB}\epsilon_{A'B'}c^{AA'}c^{BB'}=1.
\end{equation}
In this case the potential becomes
\begin{equation}
    V_C=\frac{1}{2}\sum_{a'}\hat C^{a'}\hat C^{a'}
\end{equation}
with
\begin{equation}
   \hat C^{a'}=c^{AA'}\hat C^{a'}_{AA'}. 
\end{equation}

In the case of $SU(2)$ and one instanton, that is for $k=k'=1$, the right moving fermions obey the reality conditions
\begin{eqnarray}
\hat\lambda^{A'Y}_+&=&\epsilon^{A'B'}\epsilon^{YZ}\bar{\hat\lambda}_{+B'Z},\nonumber\\
\hat\lambda^{YY'}_+&=&\epsilon^{YZ}\epsilon^{Y'Z'}\bar{\hat\lambda}_{+ZZ'}\label{real6}
\end{eqnarray}
and the tensors $\hat C^{a'}_{AA'}$ take the form
\begin{eqnarray}
    {C^{YY'}}_{BB'}&=&{X_B}^Y{\phi_{B'}}^{Y'},\nonumber\\
    {C^{A'Y}}_{BB'}&=&\frac{\omega}{\sqrt{2}}{\delta^{A'}}_{B'}{H_{B}}^{Y}.
\end{eqnarray}
Now if we define
\begin{eqnarray}
X^2&=&\epsilon_{AB}\epsilon_{YZ} X^{AY}X^{BZ},\nonumber\\
\phi^2&=&\epsilon_{A'B'}\epsilon_{Y'Z'}\phi^{A'Y'}\phi^{B'Z'}\label{square1}
\end{eqnarray}
then the potentials for this special case of com[lementary model is given by:
\begin{equation}\label{potential1}
V_C = \frac{m^2}{8}(X^{2}+\omega^{2})\phi^{2}.
\end{equation} 

Now it is time to analyze the structure of the moduli space of Ali-Ilahi's complementary ADHM instanton sigma model as given by (\ref{potential1}).

The moduli space is given by $\phi=0$ and any $X$ and it is $\mathcal M'=\hat{\mathcal R}^{4k'}$. 

This is complementary or dual to the moduli space of Witten's Original Model.

This is the same branch that opens up in the small instanton limit of Witten's original model. In the small instanton case of the present, that is Complementary, Model the moduli space of the original model opens up for $X=0$ and any $\phi$ and it is $\mathcal M={\mathcal R}^{4k}$. This resolved the mystery of the moduli space of Witten's original model.

\section{Witten's Model in Harmonic Superspace}\label{e}

Our aim in this note is to develop off-shell harmonic superspace formalism for Ali-Ilahi's complementary ADHM instanton sigma model in parallel with what Galperin and Sokatchev did for Witten's original ADHM instanton sigma model. We summarize latter in this Appendix. 

To give off-shell harmonic superspace formalism for Witten's original ADHM instanton sigma model Galperin and Sokatchev have their own perspective on it. We summarize that perspective in Sub-appendix \ref{perspective0}.

The harmonic superspace was originally proposed for $N=2$ supersymmetry. Thus harmonic superspace has to be extended for $(0, 4)$ supersymmetry because Witten's original ADHM instanton sigma model happens to be $(0, 4)$ sueprsymmetric. The $(0, 4)$ supersymmetry in the context of harmonic superspace is introduced in Appendix \ref{04susy}. 

Witten used two super-multiplets for the construction of his ADHM instanton sigma model - scalar and twisted scalar super-multiplets. Galperin and Sokatchev introduced corresponding super-multiplets. Free super-multiplets for Witten's model are introduced in Sub-appendix \ref{free}. Here, following Galperin and Sokatchev, we also introduce the third supermultiplet - the chiral fermion superfield.

It is obvious that the main structure of the Witten's original and Ali-Ilahi's complementary ADHM instanton sigma models is concerned with the interactions. The interaction in the complementary model are complementary to the ones in the original model in the sense of interchange of scalar and twisted scalar multiplets. We summarize the interactions and the description of the instanton gauge field for the original model in harmonic superspace formalism in Appendix \ref{oleg} following Galperin-Sokatchev formalism. The procedure is straightforward but technically involved.

\subsection{Galperin-Sokatchev Perspective of Witten's Model}\label{perspective0}

In Appendix \ref{c}  we have given our overview of Witten's original ADHM instanton sigma model. In Appendix \ref{d} we did the same for Ali-Ilahi's complementary ADHM instanton sigma model.

Galperin-Sokatchev gave us the of-shell, that is superspace, formalism for Witten's construction using harmonic superspace. To do this they needed and provided their own perspective on Witten's model. We recapitulate that perspective in this Sub-appendix.

Galperin and Sokatchev avoided use of quaternionic notation to describe ADHM construction. They opted for an alternative approach such that it avoids the use of quaternionic notation.

To give the Galperin-Sokatchev perspective on Witten's original ADHM instanton sigma model we shall use a notation that is closer to Witten's notation.

The initial step in the ADHM construction for the case of $SO(n)$ involves the rectangular matrix $B^{a}_{AY'}(X)$, that is linear in $X$, with indices $a = 1, \cdots, n+4k'$, $Y' = 1, \cdots, 2k'$ (where $n$ and $k'$ are positive numbers) and $A$ being the same $Sp(1)$ index as mentioned earlier. This matrix must satisfy the following reality condition:
 
\begin{equation}\label{real0}
\overline{B^a_{AY'}}=\epsilon^{AB}\epsilon^{Y'Z'}B^a_{BZ'}.
\end{equation}

As per Eqn.(\ref{linear1}) it consists of tensors $D^{a}_{AY'}$ and $E^{a}_{YY'}$ with the expression
\begin{equation}
 B^{a}_{AY'}(X)= D^{a}_{AY'}+E^{a}_{YY'}{X_A}^Y\label{linear4}   
\end{equation}

The tensor $B^a_{AY'}$ must satisfy the algebraic constraint:
\begin{equation}\label{real1}
B^{a}_{AY'}(X)B^{a}_{BZ'}(X)=\epsilon^{AB}R_{Y'Z'}(X),
\end{equation}
here $R_{Y'Z'}$ is an invertible antisymmetric $2k'\times 2k'$ matrix.

The matrices $D^{a}_{AY'}$ and $E^{a}_{YY'}$ should have maximal rank. These conditions are essential for the successful implementation of the ADHM construction in the case of $SO(n)$.

To obtain the instanton field the real rectangular matrix denoted as $v^a_{i}$, introduced above, is required, where $i$ is the $SO(n)$ index. The matrix satisfies two important properties given in Eqs. (\ref{ortho0}) and (\ref{ortho1}).

By employing these properties, the $SO(n)$ gauge field, Eqn.(\ref{aijay}),  can be straightforwardly expressed as follows:
\begin{equation}\label{aijay0}
A_{ijAY}=v^a_i\partial_{AY}v^a_j.
\end{equation}
Here
\begin{equation}
   \partial_{AY}=\frac{\partial}{\partial X^{AY}}. 
\end{equation}

It also corresponds to an instanton solution with finite action and instanton number $k'$. 

The outlined technique reveals that the matrices $B^{a}_{AY'}$ and $v^a_i$ have lots of arbitrariness. The index $a$ has  global $SO(n+4k')$ transformation freedom that leaves (\ref{aijay0}) invariant. Also the index $Y'$ has  global $GL(2k', C)$ transformation invariance.  The reality condition, Eqn.(\ref{real0}), reduces it to $GL(k', C)$ invariance. The index $i$ is the $SO(n)$ Yang-Mills local gauge invariance index.

The $SO(n+4k') \times GL(k,Q)$ freedom is the one that is relevant for counting the instanton parameters. This can be used to put the tensors $D^{a}_{AY'}$ and $E^{a}_{YY'}$ into canonical form 
 
\begin{equation}\label{canon1}
 D^a_{AY'}\ \rightarrow \ \left(\begin{array}{c} b_{4k' \times n}
\\ d_{4k' \times 4k' }
\end{array}\right), \ \ \
E^a_{YY'} \ \rightarrow \ \left(\begin{array}{c} 0_{4k'\times n} \\
1_{4k'\times 4k'}
\end{array} \right).
\end{equation}
The matrices $b$ and $d$ are subject to algebraic constraints derived from Eqn. (\ref{real1}). Together with the remaining elements in the instanton solutions, these symmetry transformations result in the appropriate count of independent parameters.  For more details on this demanding task we refer to \cite{Christ:1978jy, Weinberg:2012pjx}.

\subsection{Case of (0,4) Supersymmetry}\label{04susy}

In Appendix \ref{b} we have covered the background on harmonic superspace in some detail. That formalism was for $N=2$ supersymmetry. To formulate Witten's ADHM instanton sigma models in off-shell superspace we need harmonic superspace for $(0, 4)$ superspace. Galperin and Sokatchev discussed this in Section 2 of their narrative. We take up a review of that in this Sub-appendix.

Harmonic variables are useful for the analysis of $(0,4)$ supersymmetric theories because of the existence of three complex structures.

The $(0,4)$ world sheet superspace is described using the coordinates $x_{++}$, $x_{--}$ and $\theta^{AA'}_+$. To avoid confusion the harmonic $U(1)$ charges are expressed as upper indices while the Lorentz ($SO(1,1)$) weights as lower indices. The Grassmann variables $\theta^{AA'}_+$ carry doublet indices of 
the  automorphism group of $(0,4)$ supersymmetry,
\begin{equation}
    SO(4)\sim SU(2)\times SU(2)'
\end{equation}
and they also meet the reality condition 
\begin{equation}
\overline{{\theta}^{AA'}} =
\epsilon_{AB}\epsilon_{A'B'} {\theta}^{BB'}.
\end{equation}
 
The counterparts of the supersymmetry generators, the spinor covariant derivatives are defined as
\begin{equation}\label{D}
{D}_{-AA'} = \frac{\partial}{\partial{\theta}^{AA'}_+} + i{\theta}_{+AA'}
\frac{\partial}{\partial x_{++}}
\end{equation}
which satisfy the superalgebra with $(0,4)$ supersymmetry
\begin{equation}\label{susy}
\{ {D}_{-AA'}, {D}_{-BB'}\} = 2i\epsilon_{AB}\epsilon_{A'B'} \partial_{--}.
\end{equation} 

Defining  
\begin{equation}\label{D+}
{D}^+_{-A'} \equiv {u}^{+A}{D}_{-AA'},
\end{equation}
we can rewrite Eqn.(\ref{susy}) as follows
\begin{equation}\label{++}
\{ {D}^+_{-A'}, {D}^+_{-B'}\} = 0.
\end{equation}
The torsion term from the right side of Eqn.(\ref{susy}) disappears after this projection.

The Grassmann analytic 	superfields can be defined by the equation
\begin{equation}\label{ga}
{D}^+_{-A'}\Phi(x, {\theta}, {u}) = 0.
\end{equation}
We now define the {\it analytic basis} for the harmonic superspace, which has been enlarged by the addition of harmonic variables.
\begin{equation}\label{bas}
x_{S++} = x_{++} + i{\theta}^{AA'}_+{\theta}^{B}_{+A'} {u}^+_{(A}{u}^-_{B)},
\ \ x_{--}, \ \ {\theta}^{\pm A'}_{+}= {u}^\pm_A {\theta}^{AA'}_+, \ \ {u}^\pm.\end{equation}

The derivative $D^+_{-A'}$ is just a partial one
\begin{equation}
    D^+_{-A'} =
\frac{\partial}{\partial\theta^{-A'}_+}.
\end{equation}
The Grassmann analyticity criterion,  Eqn.(\ref{ga}),  can thus be resolved in the following manner:
\begin{equation}\label{asf}
D^+_{-A'}\Phi(x,{\theta},{u}) = 0 \ \ \Rightarrow \ \ \Phi = \Phi(x_{S++},
x_{--}, {\theta}^+_+, {u}).
\end{equation}
In the meantime the harmonic derivative ${D}^{++}$ receives a vielbein term:
\begin{equation}\label{der}
D^{++} = {u}^{+A}\frac{\partial}{\partial {u}^{-A}} + i{\theta}^{+A'}_+{\theta}^+_{+A'}
\frac{\partial}{\partial x_{S++}}.
\end{equation}
We also have
\begin{equation}\label{hgr}
[{D}^{++}, {D}^+_{-A'}] = 0.
\end{equation}

The analytic superfields defined in Eqn.(\ref{asf}) has a non-vanishing $U(1)$ harmonic charge $q$ short Grassmann expansion,
\begin{equation}\label{expa}
\Phi^q(x,{\theta}^+,\hat{u}) = \phi^q(x, {u}) + {\theta}^{+A'}_+\xi^{q-1}_{-A'}(x, {u})
+({\theta}^+_+)^2 f^{q-2}_{--}(x, {u}),
\end{equation}
where $({\theta}^+_+)^2 \equiv {\theta}^{+A'}_+{\theta}^+_{+A'}\;$. The coefficients in Eqn.(\ref{expa}) are harmonic-dependent fields (remember that all of the terms in Eqn.(\ref{expa}) conserve the total $U(1)$ charge $q$). In addition, we should note that the harmonic analytic superfields Eqn.(\ref{expa}) can occasionally be made real in the sense of the special conjugation Eqn.(\ref{partconj}).

Here Galperin and Sokatchev remind us that they have harmonized only one $SU(2)$ subgroup of the $SO(4)$ automorphism group and left the other, $SU(2)'$, unaffected. This is because   $SU(2)'$ does not belong to the $(0,4)$ superconformal group while $SU(2)$ does. The action in the model below is $SU(2)'$ invariant, not $SU(2)$ invariant. It therefore has the proper symmetries to flow to a $(0,4)$ CFT in the infrared. Here we would like to remind that in the Complementary Model the formalism is $F=SU(2)$ invariant and there is no $SU(2)'=F'$ invariance.

In other words in case of Ali-Ilahi's complementary ADHM instanton sigma model the role of the $SU(2)$ and $SU(2)'$ subgroups of the $SO(4)$ automorphism group of $(0, 4)$ supersymmetry are interchanged.

This is what we shall do while formulating the off-shell harmonic superspace for Ali-Ilahi's complementary model.

We discuss the $(0, 4)$ supersymmetry for Ali-Ilahi's ADHM instanton sigma model in Section \ref{04susy0}.

\subsection{Free Supermultiplets}\label{free}

In his construction of the original ADHM instanton sigma model Witten used three supermultiplets. These were in component form. To formulate Witten's  model in harmonic superspace three corresponding  free harmonic superfields are required. In this Sub-appendix we recapitulate the description of the corresponding free superfields constructed by Galperin and Sokatchev in their harmonic superspace formalism of Witten's model. They collected these superfields in their Section 3.1. 

The first supermultiplet employed by Witten in component form and cast in superfield form by Galperin and Sokatchev is the fundamental scalar superfield containing the coordinates $X^{AY}$ of the Euclidean target space $R^{4k}$. The second free superfield contains the right moving chiral fermions $\lambda^a$. Finally we have the third scalar superfield called the twisted scalar superfield containing the scalars $\phi^{A'Y'}$. These scalars contribute to the construction of $k'$-instantons. We shall take up these free superfields one at a time.

The Euclidean space $R^{4k}$ is the target space in which the Yang-Mills fields will be defined and the first supermultiplet $(X, \psi)$ contains its coordinates $X^{AY}$. The analytic superfields can be used to describe it in harmonic superspace (cf. Eqn. (\ref{expa})) as
\begin{equation}\label{X}
X^{+Y}(x, {\theta}^+, {u}) = X^{+Y}(x, {u}) + i{\theta}^{+A'}_+\psi^Y_{-A'}(x, {u})
+({\theta}^+_+)^2 f^{-Y}_{--}(x, {u}).
\end{equation}

This is the fundamental scalar superfield. The component fields are real in the sense of the conjugation defined in Eqn.(\ref{real}) that can be expressed as follows:
\begin{equation}\label{Xreal}
\widetilde{X^{+Y}} = \epsilon_{YZ} X^{+Z}.
\end{equation}
Harmonic fields have  infinite harmonic expansion on $S^2$. We are looking for those supermultiplets that have a finite number of physical fields. It turns out that by enforcing the following harmonic irreducibility condition, we can truncate the harmonic expansions
\begin{equation}\label{irr}
D^{++} X^{+Y} = 0.
\end{equation}
Finding the component solution to this constraint is not difficult.  Using the
analytic basis form of ${D}^{++}$ in Eqn. (\ref{der}), for $X^{+Y}(x, {u})$ we find
($\partial^{++}$ denotes the purely harmonic part of (\ref{der}))
\begin{equation}\label{eqX}
\partial^{++} X^{+Y}(x,{u}) = 0 \ \ \Rightarrow \ \ X^{+Y}(x,{u}) =
{u}^+_A X^{AY}(x),
\end{equation}
which follows from the harmonic expansion  for $q=+1$. Similarly, for
the other two fields in Eqn. (\ref{X}) we get
\begin{equation}\label{solu}
\psi^Y_{-A'}(x,{u}) = \psi^Y_{-A'}(x), \ \ \ f^{-Y}_{--}(x,{u}) = -i {u}^-_A
\partial_{--}X^{AY}(x).
\end{equation}

Here we would like to clarify once again that, for example, in field $f^{-Y}_{--}$ the $Y$ index is the $Sp(2k)$ index, upper $-$ index is the $U(1)$ charge and the lower indices $--$ are the Lorentz ones. The component fields are real because of Eqn. (\ref{Xreal}),
$\overline{X^{AY}}=
\epsilon_{AB}\epsilon_{YZ}X^{BZ}, \; \overline{\psi^{A'Y}}=
\epsilon_{A'B'}\epsilon_{YZ}\psi^{B'Z}$. The harmonic dependence of the fields  was therefore changed to a linear dependence as a result of the constraint. We may conclude that the fields are in  the form an off-shell $(0,4)$ supersymmetric multiplet since the constraint  is clearly supersymmetric and does not need equations of motion for the component fields. It matches with the field content in Ref.\cite{Witten:1994tz}.

Now come to the action for the fundamental scalar supermultiplet. An integral over the analytic superspace is used to represent the corresponding action:
\begin{equation}\label{acX}
S_X = i\int d^2x d{u}d^2{\theta}^+_+ \; X^{+Y}\partial_{++} X^+_Y.
\end{equation}

Here the $SU(2)$ and Lorentz weights of the Grassmann measure are adjusted so as to effect that the action is both an $SU(2)$ and Lorentz singlet. Here we have switched from Witten's notation   to that of Galperin and Sokatchev. To obtain the component content of Eqn.(\ref{acX}) they substitute the
short form Eqs.(\ref{eqX}) and (\ref{solu}) of the expansion in Eqn.(\ref{X}) in Eqn.
(\ref{acX}) and perform two more steps. First, they do the Grassmann
integral, i.e., they pick out only the $({\theta}^+_+)^2$ terms. Then they do
the harmonic integral according to the rules, which amounts
to extracting the $SU(2)$ singlet part. The result is
\begin{equation}\label{comX}
S_X = \int d^2x\; \left( X^{AY}\partial_{++}\partial_{--} X_{AY} +
\frac{i}{2}\psi^{A'Y}_-\partial_{++}
\psi_{-A'Y}\right).
\end{equation}
In this action, like Ref.\cite{Witten:1994tz}, we have $4k$ real scalars $X^{AY}$ and their chiral spinor superpartners $\psi^{A'Y}_-$. Clearly the fields are free and massless and  the multiplet in Eqn. (\ref{comX}) corresponds to a flat target space $R^{4k}$.  

Now they come to the second supermultiplet - the multiplet that contains the right moving chiral fermions. As a superfield they consider the following real and  {\it anticommuting} fields.
\begin{equation}\label{Lam}
\Lambda^a_+(x,{\theta}^+,{u}) = \lambda^a_+(x,{u}) + {\theta}^{+A'}_+g^{-a}_{A'}(x, {u})
+i({\theta}^+_+)^2  \sigma^{--a}_{-}(x,{u}).
\end{equation}
The action for them is
\begin{equation}\label{acL}
S_\Lambda = \frac{1}{2}\int d^2x d{u} d^2{\theta}^+_+ \;
\Lambda^{a}_+{D}^{++} \Lambda^a_{+}.
\end{equation}
The external index
$a=1,\cdots,n+4k'$ is an $SO(n+4k')$ one. Integrating ${D}^{++}$ in Eqn. 
(\ref{acL}) by parts changes the sign, but the superfields $\Lambda^a_+$
anticommute, so the trace $aa$ is symmetric.

Obtaining the component content of Eqn. (\ref{acL}) is as easy as in the case of
the scalar supermultiplet in Eqn.(\ref{acX}). First we do the Grassmann integral and find
\begin{equation}\label{comL}
S_\Lambda = \int d^2xd{u} \; \left( \frac{i}{2}\lambda^{a}_+\partial_{--}
\lambda^a_{+}
+ i\sigma^{--a}_{-} \partial^{++}\lambda^{a}_+ + \frac{1}{4}
g^{-aA'} \partial^{++} g^{-a}_{A'} \right).
\end{equation}
 
  The field
$\sigma^{--a}_{-}$ serves as a Lagrange multiplier for the harmonic
condition $\partial^{++}\lambda^a_+(x, {u}) =0$ which makes $\lambda^a_+$
harmonic independent.  The harmonic-dependent field $g^{-aA'}(x, {u})$ is
 auxiliary too. Eliminating both auxiliary harmonic fields, we
obtain simply the action for $n+4k'$ free chiral fermions
\begin{equation}\label{chfe}
S_\Lambda = \frac{i}{2}\int d^2x \;
\lambda^{a}_+(x)\partial_{--}\lambda^a_{+}(x).
\end{equation}
This is an example of an on-shell supermultiplet in which supersymmetry has a minor effect. The multiplet's off-shell variant  in Eqn. (\ref{comL}), calls for a {\it infinite number} of auxiliary fields. The usual {\it no go} counting arguments of can be used to demonstrate that the number of chiral fermions must be an integer multiple of four under the assumption that there are a {\it finite} number of auxiliary fields.
Additionally, the natural $SO(n+4k')$ symmetry of the free action is necessarily violated for multiplets with a finite number of auxiliary fields. A naive conclusion would be that an off-shell representation of $(0,4)$ supersymmetry cannot contain any number of chiral fermions \cite{Gates:1994bu}. The harmonic formalism, however, gets around this by utilizing an infinite number of auxiliary fields. Here we skip the detailed argument and instead refer the reader to the argument given in Appendix A of Galperin and Sokatchev in this regard.

Finally Witten's original model utilizes the so-called ``twisted" scalar multiplet, in
which the $SU(2)$ indices carried by the bosons and fermions are
interchanged as compared to the standard multiplet. Its
superspace description, constructed by Galperin and Sokatchev, turns out to be quite unusual. This time we need
a set of {\it anti-commuting abelian gauge} superfields
\begin{equation}\label{Phi}
\Phi^{+Y'}_+(x,{\theta}^+,{u}) = \rho^{+Y'}_+(x,{u}) +
\theta^{+A'}_+\phi^{Y'}_{A'}(x,{u}) +i({\theta}^+_+)^2  \chi^{-Y'}_{-}(x,{u}),
\end{equation}
satisfying the reality condition
$\widetilde{\Phi^{+Y'}_+}=\epsilon_{Y'Z'}\Phi^{+Z'}_+$ . Also 
\begin{equation}
\chi^{-Y'}_{-}=u^{-A}\chi_{-A}^{Y'}.
\end{equation}

Here $Y'=1,\ldots,2k'$ is an index of the symplectic group $Sp(k')$. We have earlier alluded to the fact that it turns out that $k'$ is just the number of instantons in
the corresponding ADHM construction.
These superfields undergo the following abelian gauge transformations
\begin{equation}\label{gau}
\delta \Phi^{+Y'}_+ = {D}^{++} \omega^{-Y'}_+
\end{equation}
with analytic parameters $\omega^{-Y'}_+(x, {\theta}^+, {u})$. Galperin and Sokatchev 
stress that this gauge invariance has nothing to do with the target space
gauge group.  It is an artifact of the superspace description of the
multiplet.

If we compare the gauge invariance in Eqn. (\ref{gau}) with the four dimensional abelian gauge field $A_\mu(x)$, the role of the gauge invariance may become clearer. This field is known to contain two Poincar\'e group representations with spins 1 and 0. The spin 0 component can be removed in one of two ways. One method is to impose an irreducibility constraint, such as $\partial^\mu A_\mu =0$, while the alternative involves putting $A_\mu$ through gauge transformations, such as $\delta A_\mu = \partial_\mu
\omega(x)$, and recording a gauge invariant action. Here, we note a similar phenomenon. Equivalent to the condition in Eqn. (\ref{irr}) for the superfield $X^+$ is the irreducibility condition $\partial^\mu A_\mu =0$. This condition has the result of making the superfield harmonically short, which results in the elimination of a (unlimited number of additional degrees of freedom). Eqn. (\ref{gau}) provides the analogue of the second gauge mechanism for $A_\mu$. The expansion of the parameter should be examined in order to comprehend what transpires.

The result is the following short superfield representation in Wess-Zumino gauge:
\begin{equation}\label{om}
\omega^{-Y'}_+(x, {\theta}^+,{u}) = \tau^{-Y'}_+(x, {u}) + {\theta}^{+A'}_+
l^{--Y'}_{A'}(x,{u})
+i({\theta}^+_+)^2  \mu^{---Y'}_{-}(x, {u}).
\end{equation}
{}From Eqs.(\ref{Phi}) and (\ref{om}) and using Eqn.(\ref{der}), we get, for
instance, $\delta\rho^+_+(x, {u}) = \partial^{++}\tau^-_+(x, {u})$. Comparing
the harmonic expansions of $\rho^+$ and $\tau^-$, we easily see
that each term in the expansion of $\rho^+$ has its counterpart in that of
$\tau^-$. So, the component $\rho^+$ can be completely gauged away.
Similar arguments show that the expansions of the parameters $l^{--}$ and
$\mu^{---}$ are a little ``shorter" than those of the fields $\phi$ and
$\chi^{-}$, correspondingly. What is missing is just the lowest order
term, the smallest $SU(2)$ representations in each expansion. Thus, the
fields $\phi^{Y'}_{A'}(x, {u})$ and $\chi^{-Y'}_{-}(x, {u})$ can also be gauged
away, except for the first terms in their harmonic expansions, the fields
$\phi^{Y'}_{A'}(x)$ and ${u}^-_A\chi^{Y'A}_{-}(x)$. The net result of all
this is the following ``short", i.e., irreducible harmonic superfield in
the  Wess-Zumino-type gauge
\begin{equation}\label{WZ}
\Phi^{+Y'}_+(x, {\theta}^+, {u}) =  {\theta}^{+A'}_+\phi^{Y'}_{A'}(x)
+i({\theta}^+_+)^2  u^{-}_A \chi^{Y'A}_{-}(x).
\end{equation}
This is exactly the content of the original model's ``twisted" multiplet. It is important to remember that this multiplet is off shell just like the one that the superfield $X^+$ describes. The only difference between the two multiplets in terms of component fields is that the two automorphism groups $SU(2)$ and $SU(2)'$ are reversed. It is evident why their superspace descriptions differ, as we have seen, because we only have harmonic variables for one of the $SU(2)$ groups. Galperin and Sokatchev point out  that the manifest supersymmetry is lost in the Wess-Zumino gauge of Eqn. (\ref{WZ}).

The superfield $\Phi^+_+$ needs a gauge invariant action, which we must find next.
Galperin and Sokatchev extract it from  the gauge superfield action for $N=2,D=4$.
 It has the following form:
 
\begin{equation}\label{acP}
S_\Phi =i \int d^2x d^4{\theta}_+ d{u}_1d{u}_2\; \frac{1}{{u}^+_1{u}^+_2}
\Phi^{+Y'}_+(1)\partial_{++} \Phi^+_{+Y'}(2).
\end{equation}
The notation $\Phi^+_+(1) $ means that the analytic superfield is written
down in an analytic basis of Eqn. (\ref{bas}) defined by the harmonic variable
${u}_1$; similarly, $\Phi^+_+(2) $ is given in another basis, defined by
${u}_2$.  Since both superfields appear in the same integral, they should be
written down in the same non-analytic basis $(x_{\pm\pm}, 
{\theta}^{AA'}_+, {u})$. This explains
why the Grassmann integral in Eqn.(\ref{acP}) is taken over the full superspace
and not over an analytic subspace, as in Eqn.(\ref{acX}) or Eqn.(\ref{acL}). Note
that, as always, one takes care of matching $U(1)$ charges and Lorentz
weights (this accounts for the presence of $\partial_{++}$ in Eqn.(\ref{acP})).
Note also that the contraction ${}^{Y'}{}_{Y'} \equiv {}^{Y'}\epsilon_{Y'Z'}
{}^{Z'}$ must be of the type $Sp(k')$. Indeed, interchanging 1 and 2 involves
integration by parts of $\partial_{++}$, flipping the odd superfields
$\Phi(1)$, $\Phi(2)$ and using ${u}^+_1{u}^+_2 = - {u}^+_2{u}^+_1$, so the fourth
antisymmetric factor $\epsilon_{Y'Z'}$ restores the symmetry required by
the double harmonic integral. The issue of the symmetry of this kinetic term is
important and the authors come back to it later on.

The reason for the strange form of Eqn. (\ref{acP}) is the gauge invariance in Eqn. (\ref{gau}). It works in the following way. Varying in the action in Eqn. (\ref{acP}) with
respect to, e.g., $\Phi^+_+(1) $ and according to the transformation law
in Eqn. (\ref{gau}) makes the harmonic derivative ${D}^{++}_1$ appear under the
integral. Integrating it by parts, we see that it only acts upon the
harmonic distribution $({u}^+_1{u}^+_2)^{-1}$ (since $\Phi^+_+(2) $ depends on
the second harmonic variable ${u}_2$).  After this both superfields $\Phi^+_+$ become analytic with
respect to the same basis, i.e., they depend only on the two Grassmann
variables ${\theta}^{+A'}_+$. However, the Grassmann integral in Eqn. (\ref{acP})
is over $d^4{\theta}$, so it gives 0. This establishes the gauge invariance
of the action in Eqn.(\ref{acP}).

Due to its harmonic non-locality, the action in Eqn.(\ref{acP}) may appear to be peculiar, however this is not so. In fact, all of the fields in the Wess-Zumino gauge of Eqn.(\ref{WZ}) have brief harmonic expansions. When calculating the Grassmann integral, we find factors of $({u}^+_1{u}^+_2)$  in the numerator that cancel out the singular denominator without forgetting the various arguments of $\Phi(1,2)$. The double harmonic integral then loses significance, and we discover
\begin{equation}\label{comP}
S_\Phi = \int d^2x\; \left( \phi^{A'Y'}\partial_{++}\partial_{--}
\phi_{A'Y'} + \frac{i}{2}\chi^{AY'}_-
\partial_{++} \chi_{-AY'} \right),
\end{equation}
which is the action for the twisted scalar multiplet used in Witten's original model.

\subsection{Original Interaction in Superspace}\label{oleg}

In this Sub-appendix we shall summarize the aspects of the interaction of Witten's original ADHM instanton sigma model as described by Galperin and Sokatchev's in harmonic superspace off-shell formalism.

In Sub-appendix \ref{free} we compiled the covariant superfields needed for Witten's original ADHM instanton sigma model. These were constructed by Galperin and Sokatchev. How to couple the aforementioned free multiplets to get Witten's interaction is the key issue now. We shall consider only the potential type couplings and ignore derivatives in parallel to what Witten investigated.   The coupling ought to be managed by a mass-dimensioned parameter $m$. Examining the resulting theory in the limit $m\rightarrow\infty$ with the intention of demonstrating that it flows into a CFT is the goal. In the Original Model the resulting theory should have an unbroken $SU(2)'$ invariance. 

The interaction terms of the action can be written in two ways : either as integrals over either the full
$(0,4)$ harmonic superspace $\int d^2x d{u} d^4{\theta}$ or the analytic
superspace $\int d^2x d{u} d^2{\theta}^+_+ $; they should involve a positive
power of $m$. To see how this can be arranged, let us first examine the
dimensions of our superfields $X^{+Y}$, $\Lambda_+^a$ and $\Phi^{+Y'}_+$.
The position of the physical spinors (dimension 1/2) $\psi_-$ in the superfield given in Eqn.
(\ref{X}), $\lambda_+$ in the superfield in the Eqn.(\ref{Lam}) and $\chi_-$ in the sueprfield in Eqn.(\ref{Phi}) fixes
the dimensions of these superfields as $[X]=0$, $[\Lambda] = 1/2$ and $ [\Phi] = -1/2$.
It is easy to see that the full superspace integrals are ruled out. Indeed,
the full measure has dimension $0$ and Lorentz weight $0$. The presence of
the mass in $m\int d^2x d^4{\theta}$ requires at least a pair of $\Phi^+_+$
superfields
(since we have $[\Phi^2]=-1$), but this is not  consistent with the Lorentz invariance.
Thus, we are left with analytic superspace
interaction terms only. The analytic measure $\int d^2x d{u}
d^2{\theta}^+_+ $ has dimension 1, Lorentz weight $-2$ and $U(1)$ charge
$-2$. There are only two possible coupling terms in the action, in
which the dimensions, charges and weights are matched (we
omit the indices):
\begin{eqnarray}
S_{\Phi\Lambda}&=&m\int d^2x d{u} d^2{\theta}^+_+\; \Phi^+_+\Lambda_+ {v}^+(X^+,
{u}),
\label{m} \\
S_{\Phi\Phi}&=&m^2\int d^2x d{u} d^2{\theta}^+_+\;  \Phi^+_+\Phi^+_+ {t}(X^+,
u).
\label{mm}
\end{eqnarray}
Here ${v}^+$ and ${t}$ are arbitrary functions of the dimensionless and
weightless superfields $X^+$ and of the harmonic variables ${u}$.

Let us first investigate the term in Eqn.(\ref{m}). To get an
idea how it can be constructed, the authors first take a simplified case, in which
$n=0$ and the index $a$ is written as a pair of symplectic indices, $a
= AY'$. Then the charged object ${v}^+$ is taken as the harmonic variable ${u}^+_A$
itself. Thus they come to the following coupling term
\begin{equation}\label{coup}
S_{\Phi\Lambda} =m \int d^2x d{u} d^2{\theta}^+_+ \;  \Phi^{+Y'}_+ {u}^{+A}
\Lambda_{+AY'}.
\end{equation}
The most important point now is to make sure that the
coupling in Eqn.(\ref{coup}) is invariant under the
gauge transformation of Eqn.(\ref{gau}). To this end the  the
superfield $\Lambda_{+AY'}$ must also transform as follows
\begin{equation}\label{gauL}
\delta \Lambda_{+AY'} =m {u}^+_A \omega^-_{+Y'}.
\end{equation}
Indeed, the variation with respect to $\Lambda$ in the kinetic term of Eqn.
(\ref{acL}) compensates for that of $\Phi$ in Eqn.(\ref{coup}), whereas
$\delta \Lambda$ in Eqn.(\ref{coup}) is annihilated by the harmonic ${u}^+$
$({u}^{+A}{u}^+_A = 0)$.

The second possibility of Eqn.(\ref{mm}) to construct a coupling term is now
clearly ruled out by the gauge invariance mentioned in Eqn.(\ref{gau}). However the term of the type in Eqn.(\ref{mm}) appears in a  fixed gauge.

Surprisingly, the coupling in Eqn.(\ref{coup}) is just a mass term despite the existence of gauge invariance. This can be seen in two ways. One method is to look at the component Lagrangian, while the other is to directly use superfields. We put off doing the former until we get to the $n\neq 0$ interaction. We begin with the superfield approach  which is actually lot simpler. 

Decomposing the index $A$ of $\Lambda_{+AY'}$ on the basis of ${u}^{\pm A}$

\begin{equation}\label{deco}
\Lambda_{+AY'} = {u}^+_A \Lambda^-_{+Y'} + {u}^-_A \Lambda^+_{+Y'}, \;\;\;
\Lambda^-_{+Y'} \equiv -{u}^{-A}\Lambda_{+AY'}, \;\;
\Lambda^+_{+Y'} \equiv {u}^{+A}\Lambda_{+AY'}.
\end{equation}
Then from Eqn.(\ref{gauL}) we see that $\delta \Lambda^-_{+Y'} =m
\omega^-_{+Y'}$,
which allows us to fix the following supersymmetric gauge, as opposed to
the non-supersymmetric Wess-Zumino  gauge of Eqn. (\ref{WZ}),
\begin{equation}\label{sugau}
\Lambda^-_{+Y'}  = 0.
\end{equation}
Substituting this into Eqn.(\ref{coup}) and the kinetic term of Eqn.(\ref{acL}), we
obtain two action terms involving the remaining projection $\Lambda^+_{+Y'} $:
\begin{equation}\label{alg}
\int d^2x d{u} d^2{\theta}^+_+ \; \left(-\frac{1}{2}
\Lambda^{+Y'}_{+}\Lambda^+_{+Y'} +
m \Phi^{+Y'}_+ \Lambda^+_{+Y'}\right).
\end{equation}
Clearly, $\Lambda^+_{+Y'}$ can be eliminated from the action in Eqn.(\ref{alg}) by means of its
 equation of motion $\Lambda^+_{+Y'} = m \Phi^+_{+Y'}$. Then we are left
only with the superfield $\Phi$ which has the action (recall Eqn.(\ref{acP}))
\begin{eqnarray}\label{massac}
S_\Phi &=&i \int d^2x d^4{\theta}_+ d{u}_1d{u}_2\; \frac{1}{{u}^+_1{u}^+_2}
\Phi^{+Y'}_+(1)\partial_{++} \Phi^+_{+Y'}(2) \nonumber\\
&{}&+ \frac{m^2}{2} \int d^2x d{u}
d^2{\theta}^+_+ \; \Phi^{+Y'}_+\Phi^+_{+Y'}.
\end{eqnarray}
The very appearance of Eqn.(\ref{massac}) shows that it is the action for $2k'$
massive superfields. We conclude that when the number of the multiplets
$\Lambda^a$ equals $4k'$, the coupling term of Eqn.(\ref{coup}) is thus a mass
term for the $4k'$ real bosons $\phi^{A'Y'}$ contained in $\Phi$ and for
the $4k'$ pairs of right-handed fermions $\chi^{AY'}_-$ from $\Phi$ and
left-handed ones $\lambda^{AY'}_+$ from $\Lambda$.

Let us now come back to the general case when $n \neq 0$. We can try to
generalize the coupling in Eqn.(\ref{coup}) by replacing ${u}^+_A$ by a matrix
$v^{+a}_{Y'}(X^+,u)$, satisfying the reality condition $\widetilde{v^{+a}_{Y'}}
= \epsilon^{Y'Z'}v^{+a}_{Z'}$. It can not depend on the superfields $\Phi$ or
$\Lambda$
because of Lorentz invariance, but it can still be a function of the
weightless superfields $X^{+Y}$ and of the harmonic variables. So, the authors
write down
\begin{equation}\label{int}
S_{int} = m \int d^2x d{u} d^2{\theta}^+_+ \;  \Phi^{+Y'}_+ v^{+a}_{Y'}(X^+,{u})
\Lambda_+^a.
\end{equation}

The main question is how to make the action in Eqn.(\ref{int}) compatible with the gauge
invariance in Eqn.(\ref{gau}). Like in Eqn.(\ref{gauL}), we can try
\begin{equation}\label{gauLv}
\delta \Lambda^a_{+} = mv^{+a}_{Y'}(X^+,{u})  \omega^{-Y'}_+.
\end{equation}
It is not hard to see that this transformation compensates for
$\delta\Phi$ in Eqn.(\ref{int}), provided the following two conditions hold:
\begin{equation}\label{c1}
v^{+a}_{Y'}v^{+a}_{Z'} = 0,
\end{equation}
\begin{equation}\label{c2}
{D}^{++} v^{+a}_{Y'}(X^+,{u}) = 0.
\end{equation}

Condition in Eqn.(\ref{c2}) is very restrictive. Indeed, assume that the
function $v^{+a}_{Y'}(X^+,{u})$ is regular, i.e., can be expanded in a Taylor
series in $X^+$:
\begin{equation}\label{tay}
v^{+a}_{Y'}(X^+,{u}) = \sum^{\infty}_{p=0}v^{(1-p)a}_{Y'Y_1\cdots Y_p}({u})
X^{+Y_1} \cdots X^{+Y_p}.
\end{equation}

Since $X^+$ satisfies the irreducibility condition of Eqn. (\ref{irr}) therefore the equation  (\ref{c2})
implies ${D}^{++}v^{(1-p)a}_{Y'Y_1\cdots Y_p}({u}) = 0$, where $1-p$ is the $U(1)$
charge. The only non-vanishing solution to
this is 
\begin{equation}\label{lin}
v^{+a}_{Y'}(X^+, {u}) = {u}^{+A} D^a_{AY'} + E^a_{YY'} X^{+Y},
\end{equation}
where the matrices $D^a$ and $E^a$ are constants. In other words, the matrix
$v^{+a}_{Y'}(X^+, {u})$ can at most depend linearly on $X^+$. If we put
${\theta}^+=0$, i.e., consider only the lowest components in the superfield
$X^+$ (see Eqn.(\ref{eqX})), then Eqn.(\ref{lin}) becomes
\begin{equation}\label{lin'}
v^{+a}_{Y'}(X^+,{u})|_{{\theta}=0} = {u}^{+A} (D^a_{AY'}  +
E^a_{YY'} X^{Y}_A) \equiv {u}^{+A}
B^a_{AY'}(X).
\end{equation}
Where we have defined
\begin{equation} \label{linear7}
D^a_{AY'}  +
E^a_{YY'} X^{Y}_A=B^a_{AY'}(X).
\end{equation}
Here we have recovered Eqn.(\ref{linear4}) in Eqn.(\ref{linear7}). 

The other condition of Eqn.(\ref{c1}) is purely algebraic. Putting Eqn.(\ref{lin'})
in it and removing the harmonic variables ${u}^{+A}{u}^{+B}$, we get
\begin{equation}\label{ADHM}
B^a_{(AY'}B^a_{B)Z'} = 0, \;\; \mbox{i.e.,} \;\;\;\;
B^a_{AY'}(X)B^a_{BZ'}(X) = \epsilon_{AB} R_{Y'Z'}(X).
\end{equation}
Here we have recovered and Eqn.(\ref{real1}) in Eqn.(\ref{ADHM}). This is a check on the veracity of the superfield construction of the interaction part of the action.

The two defining criteria on the matrix $B$, utilised as a starting point in the ADHM construction for instantons, are exactly what we have got. As referred to earlier, the matrices $D$ and $E$ must have the highest rank possible, in this case $4k'$, in order to be used in the ADHM construction. This implies that all $4k'$ right-handed chiral fermions in $\Phi^{+Y'}_+$ are coupled with a subset  $4k'$ of left-handed ones from $\Lambda^a_+$ and acquire mass in the context of the Original Model \cite{Witten:1994tz}. The sigma model that was initially proposed in \cite{Witten:1994tz} and discussed in this Sub-appendix corresponds to $k'$ instantons in $R^{4k}$ with gauge group $SO(n)$. 

At this point and at the end of Section 3.3 the authors make some remarks about which we make a brief comment at the end of Section \ref{dual-int}.

\subsection{Original Instanton Gauge Field}\label{original-gauge}
 
The authors have discussed Witten's original ADHM instanton sigma model in harmonic superspace in three main stages. First of all they discussed the free superfields and then the interaction. Finally they discussed the mathematical construction of the ADHM instanton gauge field.
This last item is what we now summarize in this Sub-appendix. 

The expression for the ADHM instanton gauge field will be written in terms of the matrix $B$ introduced above. Following the authors we shall first discuss the harmonic superspace formalism and then write down the component expression that matches with the expressions in Refs.\cite{Christ:1978jy} and \cite{Corrigan:1978ce}. 

First we shall take up the superfield formalism for the ADHM instanton gauge field.
The model describes $4k'+4k'$ free massive bosons and fermions when $n=0$ and the coupling is simply given by Eqn. (\ref{coup}), as we have proved at the beginning of Sub-appendix \ref{oleg}. This holds true to some extent even when the interaction Lagrangian in Eqn. (\ref{int}) is replaced by the one in Eqn. (\ref{coup}). The more accurate statement is that there is a subset of $4k'$ of the $n+4k'$ left-handed fermions $\lambda_+$ in $\Lambda^a_+$ that couple with the right-handed fermions in $\Phi$ and acquire mass (along with their bosonic superpartners from $\Phi$). Remaining chiral fermions maintain their masslessness. The issue is how to diagonalize the action in order to distinguish between the massive and the massless modes.
We'll try a diagonalization that looks similar to the one in Eqn.(\ref{deco}).

As the first step let us complete the $2k'\times (n+4k')$ matrix
$v^{+a}_{Y'}(X^+,{u})$ to a
full {\it orthogonal} matrix $v^{\tilde aa}(X^+,{u})$, where the $n+4k'$
dimensional index $\tilde a $ takes the values $ (+Y', -Y', i)$ and
$i=1,\cdots, n$ is an index of the group $SO(n)$. Orthogonality
means
\begin{equation}\label{or}
v^{\tilde aa} v^{\tilde b a} = \delta^{\tilde a\tilde b},
\end{equation}
where $\delta^{+Y', -Z'} = - \delta^{-Y', +Z'} = \epsilon^{Y'Z'}, \;\;
\delta^{+Y', +Z'}=\delta^{-Y', -Z'}=\delta^{\pm Y', i}=0 $.
In particular, for $\tilde a = +Y'$ and $\tilde b = +Z'$ we obtain just
condition Eqn.(\ref{c1}). Since $v^{+a}_{Y'}$ is a function of $X^{+Y}$ and
${u}^\pm$, we expect the other blocks of $v^{\tilde aa}$, namely $v^{-Y'a}$
and $v^{ia}$ to be such functions too. Of course, the fact that
$v^{+a}_{Y'}$ must be a linear function of $X^{+Y}$ (see
Eqn.(\ref{lin0})) does not imply that the rest of $v^{\tilde aa}$ too are linear as
well. It is also clear that the new matrix blocks are not completely fixed
by Eqn. (\ref{or}). The freedom consists of the subset of $SO(n+4k')$
transformations acting on the index $\tilde a$ with analytic (i.e.,
functions of $X^{+Y}$ and ${u}^\pm$) parameters, which leave $v^{+a}_{Y'}$
invariant.

With the help of the newly introduced matrix we can make a change of variables
from the superfield $\Lambda^a_+$ to $\Lambda^{\tilde a}_+ =v^{\tilde
	aa}\Lambda^a_+$. Then the gauge transformation of Eqn.(\ref{gauLv}) can be
translated into
\begin{equation}\label{gauL-}
\delta\Lambda^{-Y'}_+ = m\omega^{-Y'}_+, \ \ \ \delta\Lambda^{+Y'}_+ =
\delta\Lambda^{i}_+ = 0.
\end{equation}
As in the flat case of Sub-appendix \ref{oleg}, Eqn. (\ref{gauL-}) allows us
to fix the supersymmetric gauge (cf. Eqn. (\ref{sugau}))
\begin{equation}\label{sugauge}
\Lambda^{-Y'}_+ = 0.
\end{equation}

Now we can switch to the new superfields $\Lambda^{\tilde a}_+$ in the kinetic
term for $\Lambda$ in Eqn.(\ref{acL}) and the coupling term of Eqn.(\ref{int}). It is
useful to introduce the following notation
\begin{equation}\label{VV}
(V^{++})^{\tilde a\tilde b} = v^{\tilde aa}{D}^{++} v^{\tilde ba}.
\end{equation}
Then the terms of the Lagrangian containing $\Lambda$ become
(in the gauge given in Eqn.(\ref{sugauge}))
\begin{eqnarray}\label{LLL}
{\cal L}^{++}_{++}(\Lambda) &= &\frac{1}{2}
\Lambda^i_+[\delta^{ij} {D}^{++} + (V^{++})^{ij}]
\Lambda^j_+  \nonumber\\
&+& \Lambda^{+Y'}_+[\frac{1}{2}(V)_{Y'Z'} \Lambda^{+Z'}_+ +
(V^{++})^{-i}_{Y'} \Lambda^{i}_+ + m\Phi^+_{+Y'}],
\end{eqnarray}
where we have denoted 
\begin{equation}
    V_{Y'Z'} =(V^{++})^{--}_{Y'Z'}.
\end{equation}  We see that
$\Lambda^{+Y'}_+$ enters without derivatives, so it can be eliminated from
Eqn.(\ref{LLL}) using its equation of motion. The result is
\begin{eqnarray}
{\cal L}^{++}_{++}(\Lambda)
&=& \frac{1}{2} \Lambda^i_+[\delta^{ij} {D}^{++} + (V^{++})^{ij}]  \Lambda^j_+
\nonumber \\
&-& \label{LL}
\frac{1}{2}[(V^{++})^{-i}_{Y'} \Lambda^{i}_+ + m\Phi^+_{+Y'}] (V^{-1})^{Y'Z'}
[(V^{++})^{-j}_{Z'} \Lambda^{j}_+ + m\Phi^+_{+Z'}].\nonumber\\
\end{eqnarray}
The complete action of the model is obtained by adding the kinetic terms for
$X^+$ given in Eqn.(\ref{acX}) and for $\Phi^+_+$ given in Eqn.(\ref{acP}).

We can set the superfields $\Lambda^i_+$ to 0 and $V_{Y'Z'} = \epsilon_{Y'Z'}$ to make contact with the free case of Eqn.(\ref{massac}). Only the mass term for the superfields $\Phi$ is left. This explains why the chiral fermions in the subset of superfields  that correspond to massive modes are $\Lambda^{+Y'}_+$ . We can simplify things more interestingly if we let the mass $m$ to to go to infinity. The second component in Eqn.(\ref{LL}) then becomes auxiliary, and we can drop it. This is equivalent to suppressing the kinetic term for $\Phi$. The straightforward outcome is what is left after that.
\begin{equation}\label{ADga}
{\cal L}^{++}_{++}(\Lambda)|_{m\rightarrow\infty} = \Lambda^i_+[\delta^{ij}
{D}^{++} +
({V}^{++})^{ij}]  \Lambda^j_+,
\end{equation}
where
\begin{equation}\label{calV}
({ V}^{++})^{ij} = v^{ia}(X^+,{u}){D}^{++}v^{ja}(X^+,{u}).
\end{equation}
  
There is a method for  decoding the field content of the action term in Eqn.(\ref{int}) in component form. It entails extraction of the part of the  component action when a composite gauge field is connected to massless chiral fermions with the symbol $\lambda_+$. In order to achieve this, we must utilize the Wess-Zumino gauge of Eqn. (\ref{WZ}) and not the explicitly supersymmetric expression. To simplify our task we
shall keep only the relevant fields, i.e., the fermions in $\Phi^{+Y'}_+$ and
$\Lambda^a_+$ and the bosons in $X^+$. Since $\chi^{AY'}_-$ is
accompanied by $({\theta}^+_+)^2$ in Eqn.(\ref{WZ}),
the other superfields in Eqn. (\ref{int}) contribute with their lowest order
components only,
i.e., $\Delta^a_{AY'}$ from Eqn. (\ref{lin}) and $\lambda^a_+$ from Eqn. (\ref{Lam}).
This, together with the kinetic term for $\Lambda^a_{+}$ in Eqn. (\ref{comL}), gives
\begin{equation}\label{fer}
S = \int d^2xd{u} \; \left(\frac{i}{2}\lambda^{a}_+\partial_{--}\lambda^a_{+} +
i\sigma^{--a}_{-} \partial^{++}\lambda^a_+ + m {u}^-_A\chi^{AY'}_- {u}^{+B}
B^{a}_{BY'}(X)\lambda^a_+\right).\end{equation}
The important point here is that the Lagrange multiplier term with
$\sigma^{--a}_{-}$
has not changed, so we still have the field equation
$\partial^{++}\lambda^a_{+}(x,\hat{u}) = 0 \
\rightarrow \ \lambda^a_{+} = \lambda^a_{+}(x)$, like in the free case. Then
the harmonic integral in Eqn.(\ref{fer}) becomes trivial and we obtain the action
\begin{equation}\label{ferr}
S = \int d^2x \; \left( \frac{i}{2}\lambda^{a}_+\partial_{--}\lambda^a_{+}
- \frac{m}{2} \chi^{AY'}_- B^a_{AY'}(X)\lambda^a_+\right).
\end{equation}

The subsequent steps are parallel to what Witten did  and we only sketch them. One
introduces an $n\times (n+4k')$ matrix $v^a_i(X)$, orthogonal to $B$,
$B^a_{AY'}v^a_i=0$ and orthonormalized, $v^a_iv^a_j = \delta_{ij}$.
In a sense, this step is similar to the introduction of the matrix
$v^{\tilde aa}$ in Eqn.(\ref{or}). The aim is to diagonalize Eqn.(\ref{ferr}), i.e., to
separate the massive fermions from the massless ones. The massless
fermions are just $\lambda^i_+ = v^{ia}\lambda^a_+$.
After that one puts all massive fields to 0 (or, equivalently,
$m\rightarrow\infty$) and obtains
\begin{equation}\label{fergau}
S =\frac{i}{2} \int d^2x \; \lambda^i_+(\delta^{ij}\partial_{--} +
\partial_{--}X^{AY} A^{ij}_{AY})
\lambda^j_+,
\end{equation}
where
\begin{equation}\label{ADHMgau}
A^{ij}_{AY} = v^{ia} \frac{\partial v^{ja}}{\partial X^{AY}}.
\end{equation}
This is precisely the expression for the instanton field in the ADHM
construction.

\end{document}